\documentclass[letterpaper]{article}
\usepackage{graphicx}
\usepackage{color,bm,amsmath,hyperref,ulem}
\usepackage[square,numbers,sort&compress]{natbib}
\usepackage[most]{tcolorbox}
\usepackage[figuresright]{rotating}
\usepackage[]{subfig}
\usepackage{supertabular}
\usepackage{booktabs}
\usepackage{commath}
\usepackage{amsfonts}

\def\doubleunderline#1{\underline{\underline{#1}}}

\title{A catalog of pressure drop, deformation profile and tube laws for slender hyperelastic tubes conveying Newtonian flow at steady stat}

\author{Vishal Anand\aff{1}
  \corresp{\email{anand32@purdue.edu}}
  }

\author{Vishal Anand  \footnote{To whom the correspondence should be addressed. \href{mailto:anand32@purdue.edu}{\texttt{anand32@purdue.edu}}}; \\
\textit{Davidson School of Chemical Engineering, Purdue University,}\\ \textit{West Lafayette, Indiana 47907, USA}}
\begin{document}

\maketitle

\begin{abstract}
Slender tubes constituted of hyperelastic materials undergoing large deformations and conveying inertialess flow of Newtonian fluids at steady state are a model representations of complex systems in both biomechanics and bio-inspired technology. In this paper, we undertake a systematic study of this system. The tube's hyperelastic behavior is modeled by five (5) different constitutive laws: neo Hookean, Mooney Rivlin, Fung, Gent and Ogden. Invoking the principles of finite elasticity, we delineate the local pressure - deformation relationship for the tube for each of the hyperelastic models. The structural mechanical field of the tube is then coupled with fluid flow field, which is simplified using the tenets of lubrication approximation. The resultant fluid-structure interaction problem then throws up a cohort of interesting results including the deformation and pressure profile across the tube, and tube laws for five ($5$) hyperelastic models. Analysing these results, we posit that the behavior of neo Hookean and Mooney Rivlin tubes us converse of Fung's and Gent's tubes; the latter show a strain hardening behavior whilst the former exhibit a strain softening one. A sensitivity analysis which underscores the relative importance of geometrical nonlinearity in the problem then follows.
\end{abstract}

\section{Introduction}
\label{Sec:Introduction}
An inflation problem in mechanics is one in which a structure undergoes transverse deformation due to a prescribed pressure load \cite{ADG_Toroid,Selvadurai2006DeflectionsMembrane}. The prescribed pressure load is known \textit{apiori} and is therefore not subject to investigation during the analysis. Therefore, the inflation problem is generally analysed within the narrow purview and under the broad lens  of structural mechanics; indeed, the plethora of research literature pertaining to inflation problem  may serve as a useful microcosm of vast and varied field of structural mechanical principles and concepts \cite{Amabili_Book,Amabili_Review} . 

The research landscape of the inflation problem is patterned with diverse themes, categories and specializations.
For one, the inflation problem may be categorized by the shape of the geometry being inflated namely, circular\cite{ADG_Cicular_Dynamic,Circular_Hyper_JSV,ADG_PreStretched_Circular} , torus \cite{ADG_Toroid}, cylindrical \cite{ADG_Cited_Axisymmetric,Hyper_Torsion_Azimu_PreStress,PreStressed_Thick_Instability_Cylinder}, spherical \cite{Akkas_Sphere_Instability_Dynamic,Needleman_76_SphericalBallons}, rectangular \cite{BresLasky2,Breslavsky_plate1}. Both static \cite{ADG_PreStretched_Circular} and dynamic \cite{Akkas_Sphere_Instability_Dynamic,ADG_Cicular_Dynamic} inflation have been investigated with aplomb; and the concomitant complexities in the physical behavior including instabilities \cite{Akkas_Sphere_Instability_Dynamic,PreStressed_Thick_Instability_Cylinder}, failure \cite{Ren_Same_2008_Failure}, wrinkling \cite{Wrinkling_Toroid} have been delineated in detail. Another key differentiator is the material and geometrical characteristics of the structure; both hyperelastic \cite{Circular_Hyper_JSV,Needleman_76_SphericalBallons} and linearly elastic models \cite{TWK59,reddy07} with \cite{ADG_PreStretched_Circular} and without pre-stress have been analysed. Finally, the direction (azimuthal \cite{Hyper_Torsion_Azimu_PreStress}, torsional, normal \cite{Ren_Same_2008_Failure}) and type of load introduces yet another variant in the analysis of the inflation problem. 

Displaying a veritable mosaic of multi hued variations, it is no wonder then that the humble inflation problem has inspired a bevy of tailor made mathematical techniques and solution strategies, both analytical methods, for simple geometries where the final shape is already assumed \cite{ADG_Cicular_Dynamic,Mihai2015}, and rigorous numerical procedures \cite{Holzapfel_96,Bonnet_FEM,FSI_Hyper_Review_Num_Tezduyar}. A notable addition to the latter category is the meshless, numerical technique named local models method (LMM) invented by Amabili and co-workers\cite{Breslavsky_plate1,BresLasky2,Bredasvsky_3} which accounts for both geometrical and material nonlinearities without making any assumptions about the final deformed shape.

As mentioned earlier, the analysis of the inflation problem, predicates upon the assumption that the pressure load on the structure is known and prescribed. This assumption effectively treats the pressure load as 'black box' and no further investigations into its origin and evolution is warranted. However, this assumption is violated for \textit{fluid structure interactions (FSI)} problems \cite{C21}. Here, like inflation problems, the pressure load acting on the structure perpetuates a stress and deformation field in the structure. However, unlike inflation problem, the deformation of structure alters the fluid domain and the flow being conveyed therein, thereby a chain of two way coupled fluid structure interactions between fluid and solid domains is established. Therefore, unlike the inflation problem, a FSI problem is a multiphysics problem, where both fluid and solid domains must be analysed in tandem with information exchange at their boundaries being incorporated through suitable boundary conditions.

The additional complexities introduced by the unresolved fluid mechanical domain into the system render the solution of FSI  markedly challenging vis-a-vis the corresponding inflation problem. These complexities include the rheological behavior of the fluid (Newtonian \cite{CCSS17,AMC20}, shear thinning \cite{AC18b,ADC18}, viscoelastic \cite{RAB21,VAN22}), the compressible \cite{AC20,EJG18} or incompressible nature of the pressure-density formulation, the inclusion or neglection of inertial terms \cite{IWC20,WC21}, and the consideration of higher order effects like instability \cite{WC22} and turbulence \cite{RB20}.

Research into FSI problem of tubes is a time honored subject in both biomechanics \cite{SM04,SM05,GJ04} and microfluidics \cite{C21}, with emerging niches like soft robotics \cite{EG14} providing promising avenues of exciting research in this domain. However, this research has mostly been limited to the domain of small deformations and linearly elastic or linearly viscoelastic materials. The FSI of hyperelastic materials undergoing large deformation in dynamic contact with a fluid flow has large been a hitherto unexplored domain. In this paper, we consider the FSI problem of a slender, flexible microtube conveying Newtonian flow at steady state.  The microtube is constituted of hyperelastic material and undergoes large deformation; both geometrical and material nonlinearities are accounted for. To cover a wide spectrum of incompressible hyperelastic materials ranging from rubber like materials to soft tissues, we consider five different models of hyperelasticity namely, neo Hookean, Mooney Rivlin, Gent, Fung and $\text{Ogden}_{3}$ model \cite{Mihai2015,Gent_Original,Gent96,F93,Ogden_main}.  First, we analyse the structural mechanical aspect of the problem in isolation in Sec.~\ref{sec:StructMech} and prescribe the closed form expressions connecting local pressure and local deformation of the hyperelastic microtube for all five (5) models. Next, we widen our purview and delineate the fluid mechanical aspect of the problem in Sec.~\ref{sec:FluMech}. Predicated upon assumptions of inertialess, axisymmetric  flow of Newtonian fluid, the governing equations for fluid flow are linearised and solved to yield expressions for velocity field and flow rate in terms of the deformed radius. The flow and deformation fields are coupled in Sec.\ref{sec:coupling} and the closed form, analytical solutions of deformation profile and the pressure profile are obtained for the five hyperelastic models of the slender tube conveying Newtonian flow. The results are analysed in Sec.~\ref{sec:Results} along with the a discussion of the relative importance of geometrical and material nonlinearities. Conclusions and scope of future work are discussed in Sec.~\ref{sec:summ}. 


\section{Problem Description}
\label{sec:Problem_Description}
\begin{figure}
\centering
  \includegraphics[width=0.8\linewidth]{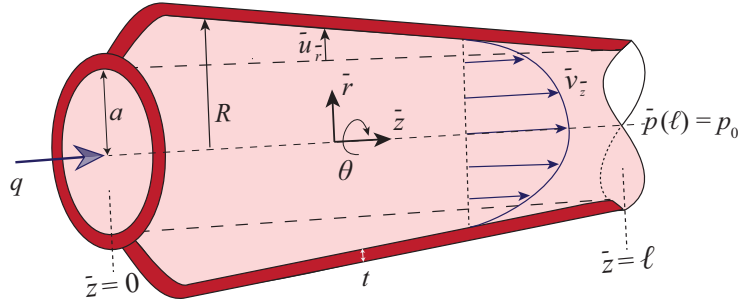}
\caption{Schematic of the system.  [Reproduced and adapted with permission from ``On the Deformation of a Hyperelastic Tube Due to Steady Viscous Flow Within,'' Vishal Anand, Ivan C.\ Christov, \textit{Dynamical Processes in Generalized Continua and Structures}, Advanced Structured Materials \textbf{103}, pp.~17--35, doi:10.1007/978-3-030-11665-1\_2.  \textcopyright\ Springer Nature 2019].}
\label{fig:Schematic}
\end{figure}

The schematic of the problem is shown in Fig. \ref{fig:Schematic}. The apparatus consists of a thin, slender cylindrical tube constituted of a hyperelastic material conveying Newtonian fluid at steady state. The origin of the cylindrical coordinate system is kept at the center of the inlet section of the tube. The $\bar{z}$ direction refers to the axial/flow wise direction of the tube while radial $\bar{r}$ and the circumferential $\bar{\theta}$ directions are defined conventionally. The tube is clamped at both its inlet $\bar{z}= 0$ and outlet $\bar{z} =\ell$. In the undeformed state, the tube has internal radius $a$ and thickness ${t}$. Steady, Newtonian flow at a fixed flow rate ${q}_0$ is imposed at the inlet of the flow. The hydrodynamic pressure $\bar{p}(\bar{z})$ exerted by the fluid on the tube wall perpetuates stress and deformation field inside the structure, leading to the deformed radius of the tube $\bar{R}=a +\bar{u}_{\bar{r}}(\bar{z})$, where $\bar{u}_{\bar{r}}(\bar{z})$ is the radial deformation field. The deformation of the tube wall in turn alters the flow field inside the tube and therefore the system behaves akin to a two way coupled  FSI system. The goal of this paper is then to delineate the structural and flow response of the system and ultimately obtain the quantitative  results for the pressure and deformation profiles. Both geometrical and material nonlinearities of the structure are accounted for by allowing the tube to undergo large deformation and by modeling the constitutive behavior of the structure with a bevy of hyperelastic models.

\section{Dimensionless variables}

We introduce the following dimensionless terms:
\begin{equation}
    r = \bar{r}/a, \quad
    z = \bar{z}/\ell, \quad
    u_r =\bar{u}_{\bar{r}}/a , \quad
    R = \bar{R}/a = 1+u_r \quad
    v_{z} = \bar{v}_{\bar{z}}/\mathcal{V}_z, \quad
    p = \bar{p}/\mathcal{P}_c, \quad q =\bar{q}/{q}_0
\label{eq:nd_vars_tube}
\end{equation}

A few pertinent comments are in order. First we note that the deformation scales as the radius of the undeformed tube $a$. This choice of scale of deformation contradicts the characteristic deformation observed in linearly elastic tubes (\cite{AC18b}); this difference in the characteristic deformation is attributed to the geometrical nonlinearity of the current problem, where the deformation need not be small compared to the tube dimensions, unlike the case of linearly elastic tubes where the assumption of small strains necessitates the characteristic deformation to be much smaller than the tube radius (or any other tube geometrical parameters).

Secondly, we have introduced the characteristic scale of axial velocity only, but refrained from introducing the scale of radial and swirl velocities. This omission is attributed to the fact, as discussed later in the Sec.\ref{sec:FluMech}, the velocity field is predominantly unidirectional to the leading order in $\frac{a}{\ell}$; the radial velocity is negligibly small while the swirl velocity is identically zero.

Finally, there are three scales from the fluid mechanical aspect of the problem namely $\mathcal{V}_z, \mathcal{P}_c, {q}_0$. These scales are not independent. For a pressure controlled case, the pressure scale $\mathcal{P}_c$ is derived from the pressure imposed at the boundary, and the remaining two scales are derived from the pressure scale. For the flow rate controlled case, the flow rate scale $\bar{q}_0$ is derived from the flow rate imposed at the boundary, while the remaining two scales are related to the flow rate scale. Details follow in Sec.\ref{sec:FluMech}. 

\section{Structural Mechanics}
\label{sec:StructMech}
We first deliberate upon the structural mechanical aspect of this FSI problem. To render the pursuit of the solution amenable to analytical treatment, we make some simplifying assumptions as given below:

\begin{enumerate}
    \item The hydrodynamic shear stress acting on the structure is negligible compared to the hydrodynamic pressure.
    
    This assumption leads to the structure being treated as a \textit{pressure vessel}.This assumption, in turn, is derived from the lubrication approximation for the flow field, which postulates that shear stress scales as $a/\ell$ times the hydrodynamic pressure.
    
    \item The pressure vessel is clamped at both its ends.
    
    Thus, there is normal stress in the axial direction in the structure, which is introduced due to the clamping.
    
    \item $t \ll a$
    
    \item $a \ll \ell$
    
\end{enumerate}

The last two assumptions ensure that there is no bending or twisting moments sustained by the structure, and thus the structure acts like a \textit{membrane}, or more precisely, like a \textit{thin walled pressure vessel}. These assumptions, even though they simplify the solution considerably, are not artificial; as shown by \cite{AC18b} , a FSI theory predicated upon these assumptions is valid in the real world in the pertinent parameter space.

\subsection{Kinematics}
\label{sec:kinematics}
In the undeformed, material coordinate system, the coordinate of point are denoted by:$\bar{r} =a $, $\theta \in [0,2\pi]$, and $\bar{z} \in [0,\ell]$.

As mentioned in Sec.~\ref{sec:Problem_Description}, in this problem we account for geometrical nonlinearity of the structural mechanical response by allowing the structure to exhibit large deformations and exhibit large strain. This consideration then necessitates the differentiation of the undeformed and deformed coordinates. To make analytical progress, we need to \textit{apriori}  map the deformed (spatial) coordinates in terms of the undeformed (material) coordinates.Such a mapping is contingent upon the assumptions that follow.

The structure is assumed to undergo \textit{axisymmetric} deformations, because the material comprising the structure is isotropic and the loading on the structure is also invariant along the $\bar{\theta}$ direction. We also assume that the deformation is homogeneous in the $\bar{z}$ direction.

Based on these assumptions, the coordinates of the material point in the deformed (spatial) coordinates system is given by:
\begin{equation}
     \bar{R} = \bar{R}(\bar{r}), \qquad \Theta = \theta, \qquad \bar{Z} = \varepsilon \bar{z},
     \label{Eq:Deformed_Coordinates_1}
\end{equation}
where $\varepsilon = L/\ell$. Now, since the tube is clamped at both its ends, its length does not change and $L = \ell$, which makes $\varepsilon =1$. Equation ~\eqref{Eq:Deformed_Coordinates_1} thus changes to:
\begin{equation}
    \bar{R} = \bar{R}(\bar{r}), \qquad \Theta = \theta, \qquad \bar{Z} =  \bar{z}
     \label{Eq:Deformed_Coordinates_2}
\end{equation}
The case of deformation mapping defined in Eq.~\eqref{Eq:Deformed_Coordinates_2} introduces the following expression for  the deformation gradient tensor $\doubleunderline{\mathcal{F}}$\cite{Mihai2017,Bonet_Book}:

\begin{equation}
\label{eq:deformation_tensor_reduced}
\doubleunderline{\mathcal{F}} = 
\begin{pmatrix}
{\partial \bar{R}}/{\partial \bar{r}} & 0 & 0 \\
0&  {\bar{R}}/{\bar{r}}& 0\\
0 & 0 & 1
\end{pmatrix}.
\end{equation}.

The diagonal nature of the $\doubleunderline{\mathcal{F}}$ suggests that
the rotation tensor is an indentity matrix $\doubleunderline{\mathcal{R}} = \doubleunderline{\mathcal{I}} $, and the deformation gradient tensor is same as the stretch tensor, expressed as:
\begin{equation}
\label{eq:stretch_tensor}
\doubleunderline{\mathcal{U}} = \doubleunderline{\mathcal{F}} = \begin{pmatrix}
\lambda_1 & 0 & 0 \\
0&  \lambda_2& 0\\
0 & 0 & \lambda_3 
\end{pmatrix}.
\end{equation},
where $\lambda_1, \lambda_2, \lambda_3$ are the principal stretches.

We also know from the incompressibility of the material, that det.[$\doubleunderline{\mathcal{F}}]=1$, which means:
\begin{equation}
    \lambda_1\lambda_2\lambda_3=1.
    \label{eq:incompressibility}
\end{equation}
Now, we employ the Eqs.~\eqref{eq:deformation_tensor_reduced},~\eqref{eq:stretch_tensor}, ~\eqref{eq:incompressibility} to determine explicit expressions for the principal stretches as:
\begin{equation}
    \lambda_1 = \bar{r}/\bar{R}, \qquad \lambda_2 = \bar{R}/\bar{r}, \qquad \lambda_3 = 1.
\label{eq:principal_stretches}
\end{equation}
Since, the deformation gradient tensor is symmetric, the left Cauchy-Green tensor $\doubleunderline{\mathcal{C}}$ and the right Cauchy-Green tensor  $\doubleunderline{\mathcal{B}}$ are same, and are given as:
\begin{equation}
\label{eq:Cauchy_Green}
    \doubleunderline{\mathcal{C}} =\doubleunderline{\mathcal{B}} =\doubleunderline{\mathcal{U}}^2 =\begin{pmatrix}
\lambda_1^2 & 0 & 0 \\
0&  \lambda_2^2& 0\\
0 & 0 & \lambda_3^2  
\end{pmatrix}.
\end{equation}
The pertinent strain measure for large deformation namely Green strain tensor ${\doubleunderline{\mathcal{E}}}$ given as:
\begin{equation}
\label{eq:strain_tensor}
{\doubleunderline{\mathcal{E}}} = \frac{\doubleunderline{\mathcal{C}}-\doubleunderline{I}}{2} = \begin{pmatrix}
\frac{\lambda_1^2-1}{2} & 0 & 0 \\
0&  \frac{\lambda_2^2-1}{2}& 0\\
0 & 0 & \frac{\lambda_3^2-1}{2} 
\end{pmatrix}.
\end{equation}
where $\doubleunderline{I}$ is the identity tensor. 

The deformation field $\underline{u}$ corresponding to the mapping of Eq.~\eqref{Eq:Deformed_Coordinates_2} is given as:
\begin{equation}
    \underline{u} =\bar{R}-\bar{r,} \quad\Theta -\theta, \quad \bar{Z}-\bar{z}
\end{equation}
and the gradient of the deformation field is therefore given as:
\begin{equation}
    \nabla\underline{u} =\frac{\doubleunderline{\mathcal{F}}-\doubleunderline{I}}{2}
\end{equation}

The small strain tensor then is given as:
\begin{equation}
\label{eq:small_strain}
    {\doubleunderline{{e}}} =\frac{\nabla\underline{u}+\nabla\underline{u}^T}{2} ={\doubleunderline{\mathcal{F}}-\doubleunderline{I}} = \begin{pmatrix}
{\lambda_1-1} & 0 & 0 \\
0&  {\lambda_2-1}& 0\\
0 & 0 & {\lambda_3-1} 
\end{pmatrix}.
\end{equation}

\subsection{Static Equilibrium}
\label{sec:statics}
The structural mechanical problem of the thin walled pressure vessel is a \textit{statically determinate} one. It is possibly to write the expressions for the stress distribution inside the cylinder without taking recourse to the constitutive equations, which are only required to find the strain or deformation field. To that end, we know from the stress distribution inside the pressure vessel is given as:
\begin{subequations}\begin{align}
    \sigma_{1} &= \sigma_{rr}= -\bar{p}, \\
    \sigma_{2} &= \sigma_{\theta\theta} = \frac{\bar{p}\bar{R}}{t},\\
    \sigma_{3} &= \sigma_{zz}= \frac{\bar{p}\bar{R}}{2t}.
\end{align}\label{eq:sigma_11_22_33}\end{subequations}

\subsection{Constitutive equation}
\label{sec:consti}

It is possible to define a material as  {"hyperelastic"} in two ways \cite{BW00,Bonet_Book}. First, in terms of \textit{mechanics}, a hyperelastic body is one for which the stress can be expressed as a function of strains only. This separates a hyperelastic material from a viscoelastic material, where stress is a function of strain as well as strain rate. A hyperelastic material, thus defined, has no "memory". On the other hand, a hyperelastic body may also be identified as such by its \textit{thermodynamics}, i.e. work-energy relations. A hyperelastic body is one for which the potential energy stored in the body during loading is recovered fully and reversibly during unloading, without any viscous dissipation. This means that the constitutive equation for a hyperelastic body, unlike a viscoelastic one, may also be expressed in terms of a strain energy functional. The strain energy functional $W$ of an isotropic hyperelastic solid can be expressed in terms of principal stretches, $W=W(\lambda_1,\lambda_2,\lambda_3)$. The exact form and expression of this functional is what separates the different models of hyperelasticity.

For an istropic hyperelastic material, the principal Cauchy stresses are co-axial with the principal stretches, and are expressed as:
\begin{subequations}\begin{align}
    \sigma_1 - \sigma_3 &= \lambda_1\frac{\partial W}{\partial \lambda_1}-\lambda_3\frac{\partial W}{\partial \lambda_3}, \\
    \sigma_2 - \sigma_3  &= \lambda_2\frac{\partial W}{\partial \lambda_2}-\lambda_3\frac{\partial W}{\partial \lambda_3}.
\end{align}\label{eq:StressEquation}\end{subequations}

The hyperelastic models considered in the present paper are namely, neo-Hookean, Mooney-Rivlin, Fung, Gent, and $\text{Ogden}_{3}$ models \cite{Ogden_main,F93,Bonet_Book,Gent_Original}. neo-Hookean and Mooney-Rivlin are generally used to model elastomers (rubber like materials), while Fung  model and Gent model are used to model body tissues \cite{Holzapfel2005SimilaritiesMaterials}, especially because these models replicate severe strain stiffening responses as observed in the body tissues ligaments and tendons, which is due to the presence of a protein named collagen \cite{Mihai2015,Holzapfel2005SimilaritiesMaterials}. On the other hand, Mooney- Rivlin model describes the shear deformation in elastomers better than the neo-Hookean model.

The (2 parameter) Gent model gives theoretical predictions very similar to that  furnished by more complicated Arruda and Boyce \cite{Arruda_Boyce_Main}(not considered in this paper), even though the Gent model, like the Mooney-Rivlin model, is a phenomenological model, while the Arruda-Boyce model is derived from microscopic response of polymer chains. 

There are many, many constitutive models for hyperelastic materials and it is neither possible nor desirable to include all of them, or even a large fraction of them here; the list of the hyderelastic models considered in this paper is clearly not exhaustive \cite{Selvadurai2006DeflectionsMembrane,Mihai2015, Shearer15}. Our choice of hyperelastic models here is then motivated by the following considerations:

\begin{enumerate}
    \item This cohort of the hyperelastic models includes models for rubber like materials as well as models for biological tissues.
    \item The models included are broad and foundational, and more specialised models which account for heterogeneity, anisotropy, hysteresis,difference in tissue behavior etc. can/have been derived using these foundational models \cite{Shearer15,Holzapfel_ogden}.
    \item Apart from the possible exclusion of Gent and Fung, the models here are built in commercial and popular engineering programs like ANSYS and Abaqus, and therefore the mathematical theory developed in this paper may be easily verified computationally \cite{ANSYS,Selvadurai2006DeflectionsMembrane}.
    \item The models included in this list are \textit{phenomenological}; they explain "how" a material behaves but not "why". In general, the material properties/fitting parameters of phenomenological models are easily obtained through experiments, and therefore the mathematical theory developed here can be easily verified by experimentalists \cite{Mihai2015}.
\end{enumerate}

\subsubsection{neo-Hookean model}
The neo-Hookean model can be derived from first principles from molecular statistical consideration and is also considered as a special case of the Mooney-Rivlin model.
The neo-Hookean model was developed for modelling of elastomers, but has been used extensively for modeling of modeling biological tissues as well. The strain energy density function $W$ for this material is expressed in terms of the first invariant $\mathcal{I}_1$ of the deformation tensor, where $\mathcal{I}_1 = (\lambda_{1}^2+\lambda_{2}^2+\lambda_{3}^2)$. For the neo-Hookean model, the strain energy density function $W(\lambda_1,\lambda_2,\lambda_3)$ takes the explicit form :

\begin{equation}
W^{nH} =\frac{C}{2}{(\lambda_{1}^2+\lambda_{2}^2+\lambda_{3}^2-3)},
\label{eq:constutive_neo_Hookean}
\end{equation}
where C is a material constant.
Here, for consistency with theory of linear elasticity,
\begin{equation}
    G=C,
    \label{eq:Neo_Hookean_Linear_Elastic_Equivalence}
\end{equation}
where $G$ refers to shear modulus of the linearly elastic solid. We also record here a well known equation for linearly elastic solid:
\begin{equation}
    2G(1+\nu)=E,
    \label{eq:Constitutive_Equation_Linear_Elastic}
\end{equation}
where $E$ is the elastic modulus of the linearly elastic solid, while $\nu$ is the Poisson ratio, which has a value of $\nu=1/2$ for incompressible material.
 We substitute the above expression for $W=W(\lambda_1,\lambda_2,\lambda_3)$ from Eq.~\eqref{eq:constutive_neo_Hookean} into Eq.~\eqref{eq:StressEquation} to obtain the following:
 \begin{subequations}\begin{align}
     \sigma_1-\sigma_3 =C[\lambda_1^2-\lambda_3^2]\\
      \sigma_2-\sigma_3 =C[\lambda_2^2-\lambda_3^2]
 \end{align}\end{subequations}

The expressions for $\sigma_1,\sigma_2,\sigma_3$ are then substituted from Eq.~\eqref{eq:sigma_11_22_33}, while the expressions for principal stretches are substituted from Eq.~\eqref{eq:principal_stretches} to finally obtain:

\begin{equation}
\label{eq:neoHookean_equi_large}
    -\bar{p}\bar{R}/t =C\left[\left(\frac{a}{\bar{R}}\right)^2-\left(\frac{\bar{R}}{a}\right)^2\right]
\end{equation}
Or,
\begin{equation}
    \frac{\bar{p}a}{tC} =\left[\left(\frac{\bar{R}}{a}\right)-\left(\frac{a}{\bar{R}}\right)^3\right]
\end{equation}
In non-dimensional terms:
\begin{equation}
    \gamma^{nH}{p} =\left[\left(1+{u}_r\right)-\left(\frac{1}{1+{u}_r}\right)^3\right] ;\quad \gamma^{nH} =\frac{\mathcal{P}_c}{C}\frac{a}{t}
    \label{eq:Pressure_Deformation_Dimless_Neo_Hookean}
\end{equation}

Here $\gamma^{nH}$ is the FSI parameter pertaining to the strength of FSI for a Neo-Hookean solid. 

\subsubsection{Mooney-Rivlin model}

The Mooney-Rivlin model is a \textit{phenomenological} model  and thus the material parameters (constants) for this model must be obtained from direct measurements. This model is more suited to describe the shear deformation in elastomers than the neo-Hookean model. Unlike the neo-Hookean model, the strain-energy function for the Mooney-Rivlin model is completely described in terms of the first, $\mathcal{I}_1 = \left(\lambda_{1}^2+\lambda_{2}^2+\lambda_{3}^2\right)$  and the second $\mathcal{I}_2 = \left(\lambda_1^2\lambda_2^2 + \lambda_2^2\lambda_3^2+\lambda_3^1\lambda_1^2\right)$ invariants of deformation tensor. Written explicitly in terms of the principal stretches, 
the Mooney-Rivlin model has the following strain-energy density function, namely:

\begin{equation}
    W^{MR} = \frac{\mathbb{C}_1}{2} \left(\lambda_1^2+\lambda_2^2+\lambda_3^2\right) + \frac{\mathbb{C}_2}{2} \left(\lambda_1^2\lambda_2^2 + \lambda_2^2\lambda_3^2+\lambda_3^1\lambda_1^2\right) \qquad(\lambda_1\lambda_2\lambda_3 =1).
    \label{eq:constitutive_equation_Mooney_Rivlin}
\end{equation}
Here, in equivalence of theory of linear elasticity:
\begin{equation}
    G = \mathbb{C}_1+\mathbb{C}_2
    \label{eq:MooneyRivlin_Linear_Elastic_Equivalence}
\end{equation}
Using the above constitutive law, the equations Eq.~\eqref{eq:StressEquation} are reduced to:
\begin{subequations}\label{eq:constitutive_sigmas_Mooney_Rivlin}\begin{align}
    \sigma_1 - \sigma_3 &= \mathbb{C}_1\left(\lambda_1^2-\lambda_3^2\right)-\mathbb{C}_2\left(\lambda_1^{-2}-\lambda_3^2\right), \\
    \sigma_2 -\sigma_3 &= \mathbb{C}_1\left(\lambda_2^2-\lambda_3^2\right)-\mathbb{C}_2\left(\lambda_1^{-2}-\lambda_3^2\right).
\end{align}\end{subequations}

Substitution of expressions for principal stretches from Eq.~\eqref{eq:principal_stretches} and the Cauchy stress from Eq.~\eqref{eq:sigma_11_22_33} into Eq.~\eqref{eq:constitutive_sigmas_Mooney_Rivlin} yields the following:
\begin{subequations}\begin{align}
    -\frac{\bar{p}\bar{R}}{2t} &= \mathbb{C}_1\left(\frac{a^2}{\bar{R}^2}-1\right)-\mathbb{C}_2\left(\frac{\bar{R}^2}{a^2}-1\right),\displaybreak[3]\\
    \frac{\bar{p}\bar{R}}{2t} &= \mathbb{C}_1\left(\frac{\bar{R}^2}{a^2}-1\right)-\mathbb{C}_2\left(\frac{a^2}{\bar{R}^2}-1\right).
\end{align}\end{subequations}

This yields, finally:
\begin{equation}
    \frac{\bar{p}a}{t(\mathbb{C}_1+\mathbb{C}_2)} = \frac{\bar{R}}{a}-\frac{a^3}{\bar{R}^3},
    \label{eq:Deformation_Dim_Final_Mooney_Rivlin}
\end{equation}

In non-dimensional terms, this can be written as:

\begin{equation}
    \gamma^{MR}{p} = (1+{u}_{{r}})-\frac{1}{(1+{u}_{{r}})^3},\qquad \gamma^{MR} := \frac{\mathcal{P}_c}{(\mathbb{C}_1+\mathbb{C}_2)}\frac{a}{t},
    \label{eq:pressure_deformation_dimless_MooneyRivlin}
\end{equation}

where $\gamma^{MR}$ is the FSI parameter pertaining to the coupling between the flow field and microtube composed of Mooney-Rivlin material.

Both neo- Hookean and Mooney-Rivlin models are used to describe constitutive behavior of the structure where the strain is moderate, and cannot be used to describe the experimental data for crystallized rubber with high strains. More importantly, both these models have weak strain-stiffening characteristics, unlike those observed in biological tissues.
\subsubsection{Fung Material}
Both the models considered till now do not have any strain-stiffening responses. On the other hand, soft biological tissues exhibit high strain stiffening even at moderate stretches. Fung postulated a constitutive law to capture such strain stiffening. The model proposed by Fung was meant to fit the experimental test data from uniaxial tension test on tissue from  rabbit mesentery as stress increases exponentially with strain. However, the model is able to approximate the behavior of other tissues as well (aorta, ligaments, tendons), but in a phenomenological sense- the model cannot be derived from first principles. For the Fung type material under the assumptions of isotropy, the expression for strain-energy density function takes the form:
\begin{equation}
    W^{F} =\frac{C}{2\alpha}\left[\alpha(\lambda_{1}^2+\lambda_{2}^2+\lambda_{3}^2-3)+e^{\alpha(\lambda_{1}^2+\lambda_{2}^2+\lambda_{3}^2-3)}-1\right].
    \label{eq:constitutive_equation_Fung}
\end{equation}
Here, for equivalence with the linearly elastic law:
\begin{equation}
G=C
\label{eq:equivalence_linear_elastic_Fung}
\end{equation}
and, $\alpha$ is the \textit{stiffening parameter}, with typical values ranging from $1<\alpha<5.5$ for soft tissues. For small strains, the term linear in $\alpha$ inside the brackets will dominate; on the other hand for large strains, the exponential term inside the brackets will dominate. As $\alpha \rightarrow 0$, the Fung's constitutive model reduces to the neo-Hookean model, namely Eq.~\eqref{eq:constutive_neo_Hookean}.
We also understand from Eq.~\eqref{eq:constitutive_equation_Fung} that like the neo-Hookean material, stress-energy density function for the Fung's material is only a function of the first invariant $\mathcal{I}_1 = \left(\lambda_{1}^2+\lambda_{2}^2+\lambda_{3}^2\right)$, and therefore the Fung's material can be classified as a \textit{generalized neo-Hookean material}.

We substitute the expression for $W^F$ from the above equation, Eq.~\eqref{eq:constitutive_equation_Fung} into the Eq.~\eqref{eq:StressEquation} to obtain:
\begin{subequations}\begin{align}
    \sigma_1-\sigma_2 = \left[C(\lambda_1^2-\lambda_2^2)\right]\left[1+e^{\alpha(\lambda_{1}^2+\lambda_{2}^2+\lambda_{3}^2-3)}\right] ;\\
    \sigma_1-\sigma_3 = \left[C(\lambda_1^2-\lambda_3^2)\right]\left[1+e^{\alpha(\lambda_{1}^2+\lambda_{2}^2+\lambda_{3}^2-3)}\right] 
    \end{align}
\label{eq:constitutive_sigmas_Fung}
\end{subequations}

Substitution of expressions for principal stretches from Eq.~\eqref{eq:principal_stretches} and the Cauchy stress from Eq.~\eqref{eq:sigma_11_22_33} into Eq.~\eqref{eq:constitutive_sigmas_Fung} yields us:
\begin{equation}
\label{eq:equilibrium_Fung_Large}
    -\frac{\bar{p}\bar{R}}{t} =C\left[\left(\frac{\bar{r}}{\bar{R}}\right)^2-\left(\frac{\bar{R}}{\bar{r}}\right)^2\right]\left[1+e^{\alpha(\frac{a}{\bar{R}}-\frac{\bar{R}}{a})^2}\right]
\end{equation}
Or, in final form:
\begin{equation}
    \frac{\bar{p}a}{tC} =\left[\left(\frac{\bar{R}}{a}\right)-\left(\frac{a}{\bar{R}}\right)^3\right]\left[1+e^{\alpha(\frac{a}{\bar{R}}-\frac{\bar{R}}{a})^2}\right]
\end{equation}
In non-dimensional terms, we obtain the following:

\begin{equation}
    \gamma^F{p} = \left[(1+{u}_{{r}})-\frac{1}{(1+{u}_{{r}})^3}\right]\left[1+e^{\alpha\{\frac{1}{1+{u}_{{r}}}-(1+{u}_{{r}})\}^2}\right],\qquad \gamma^F := \frac{\mathcal{P}_c}{\mathbb{C}}\frac{a}{t},
    \label{eq:pressure_deformation_dimless_Fung}
\end{equation}

where as before, $\gamma^F$ is the non-dimensional parameter pertaining to FSI coupling between the microtube composed of Fung's material and the lubrication flow field of a Newtonian fluid.

\subsubsection{Gent Model}
The model by Gent, like that by Fung, models strain stiffening. However, it also imposes a limit on the maximum amount of stretch the material can undergo.

For Gent's model, the expression for strain-energy density function, $W=W(\lambda_1,\lambda_2,\lambda_3)$ takes the form:
\begin{equation}
    W = -\frac{C}{2\eta}ln\left[1-\eta(\lambda_{1}^2+\lambda_{2}^2+\lambda_{3}^2-3)\right]
    \label{eq:strain_energy_Gent}
\end{equation}
For $\eta \rightarrow 0$, the neo-Hookean model is received. Note that there is a singularity when $\eta =1/(\lambda_{1}^2+\lambda_{2}^2+\lambda_{3}^2-3)$, which indicates a rapid strain-stiffening as well as an upper limit on the strain. Like Fung's model, the Gent model is also a generalized neo-Hookean model, as the constitutive equation is only expressed in terms of the first invariant, $\mathcal{I}_1 = \left(\lambda_{1}^2+\lambda_{2}^2+\lambda_{3}^2\right)$.
We then substitute the above expression into Eq.~\eqref{eq:StressEquation} to obtain :
\begin{subequations}\begin{align}
\sigma_1-\sigma_3 =\frac{C(\lambda_1^2-\lambda_3^2)}{1-\eta(\lambda_{1}^2+\lambda_{2}^2+\lambda_{3}^2-3)} \\
\sigma_2-\sigma_3 =\frac{C(\lambda_2^2-\lambda_3^2)}{1-\eta(\lambda_{1}^2+\lambda_{2}^2+\lambda_{3}^2-3)}
\end{align}\label{eq:Constitutive_Sigmas_Gent}
\end{subequations}

The equations for principal Cauchy's stress are obtained from the Eq.~\eqref{eq:sigma_11_22_33}, while those for the principal stretches are obtained from Eq.~\eqref{eq:principal_stretches}. When we insert these equations into Eq.~\eqref{eq:Constitutive_Sigmas_Gent}, we obtain after some algebraic manipulations:
\begin{equation}
\label{eq:Equilibrium_Large_Gent}
    \frac{\bar{p}\bar{R}}{t}=\frac{C\left[\left(\frac{\bar{R}}{a}\right)^2-\left(\frac{a}{\bar{R}}\right)^2\right]}{[1-\eta(\frac{a}{\bar{R}}-\frac{\bar{R}}{a})^2]},
\end{equation}
which after some minor algebraic manipulations reduces to:
\begin{equation}
\frac{\bar{p}a}{tC}=\frac{\left[\frac{\bar{R}}{a}-\left(\frac{a}{\bar{R}}\right)^3\right]}{[1-\eta(\frac{a}{\bar{R}}-\frac{\bar{R}}{a})^2]}
\end{equation}
or, in non-dimensional terms, it is expressed as:

\begin{equation}
    \gamma^G{p} = \frac{(1+{u}_{{r}})-\frac{1}{(1+{u}_{{r}})^3}}{[{1-\eta\{\frac{1}{1+{u}_{{r}}}-(1+{u}_{{r}})\}^2}]},\qquad \gamma^G := \frac{\mathcal{P}_c}{\mathbb{C}}\frac{a}{t},
    \label{eq:pressure_deformation_dimless_Gent}
\end{equation}

\subsubsection{$\text{Ogden}_{3}$ Model}
\label{sec:consti_Ogden}
Uptil now, we have considered models where the strain energy functional is expressed in terms of invariants of Green-Langrangian strain measure namely $\mathcal{I}_i,\mathcal{I}_2,\mathcal{I}_3$. The model by Ogden differs from these because here the strain energy functional depends on  a novel strain invariant $\phi$ defined as:
\begin{equation}
    \phi(2m_p) =\frac{(\lambda_1^{2m_p}+\lambda_2^{2mp}+\lambda_3^{2m_p}-3)}{2m_p},
\end{equation}
 and the corresponding strain energy functional $W$ is written as a linear combination of $\phi$:
\begin{equation}
\label{eq:strain_enegy_functional_Ogden}
W =\sum_{p=1}^NC_p\phi(2m_p)=\sum_{p=1}^N C_p\left(\lambda_1^{2 m_p}+\lambda_2^{2 m_p}+\lambda_3^{2 m_p}-3\right) /\left(2 m_p\right)
\end{equation}
Here $N$ can take values till up to $N =8$. For $N = 1 =m_p$, the Ogden model reduces to the neo Hookean model. For any value of $N$, the number of material parameters for the corresponding $\text{Ogden}_N$ model is $2N$. The large number of available parameters make the Ogden model astonishingly versatile. The efficacy of Ogden model is gauged from the fact that this model has been employed to fit experimental data across a bevy of materials such as vulcanised rubber \cite{Ogden_main}, aorta \cite{Breslavsky_plate1} and brain and fat tissue \cite{Mihai2015}. For this paper, we fix $N=3$ calling it $\text{Ogden}_3$ model.
The strain energy functional for $\text{Ogden}_3$ model is given in terms of the principal stretches as:
\begin{equation}
W =\sum_{p=1}^{3}\frac{C_p}{2m_p}(\lambda_1^{2m_p}+\lambda_2^{2mp}+\lambda_3^{2m_p}-3)
\end{equation}

Using the above expression for strain energy functional, and borrowing the expressions for principal Cauchy stresses from Eq.~\eqref{eq:sigma_11_22_33} and those for principal stretches from Eq.~\eqref{eq:principal_stretches}, we obtain the expressions for the pressure in terms of deformation as:
\begin{equation}
\label{eq:Equilibrium_Large_Ogden}
    -\frac{\bar{p}\bar{R}}{t}=C_1\left[\left(\bar{r}/\bar{R}\right)^{2m_1}-\left(\bar{R}/\bar{r}\right)^{2m_1}\right]+C_2\left[\left(\bar{r}/\bar{R}\right)^{2m_2}-\left(\bar{R}/\bar{r}\right)^{2m_2}\right]+C_3\left[\left(\bar{r}/\bar{R}\right)^{2m_3}-\left(\bar{R}/\bar{r}\right)^{2m_3}\right], 
\end{equation}
which yields after some algebraic manipulations as:
\begin{align}
    -\frac{\bar{p}a}{t}=C_1\left[\left(a/\bar{R}\right)^{2m_1+1}-\left(\bar{R}/a\right)^{2m_1-1}\right]+\\ C_2\left[\left(a/\bar{R}\right)^{2m_2+1}-\left(\bar{R}/a\right)^{2m_2-1}\right]+C_3\left[\left(a/\bar{R}\right)^{2m_3+1}-\left(\bar{R}/a\right)^{2m_3-1}\right], 
\end{align}

Or, in non-dimensional terms as:

\begin{multline}
    \gamma^{Og}{p}=\frac{pa}{tC_1}=\left[-\left(1/(1+{u}_{{r}})\right)^{2m_1+1}+\left(1+{u}_{{r}}\right)^{2m_1-1}\right]+ \bar{C}_{2}\left[-\left(1/(1+{u}_{{r}})\right)^{2m_2+1}+\left(1+{u}_{{r}}\right)^{2m_2-1}\right]\\+\bar{C}_{3}\left[-\left(1/(1+{u}_{\bar{r}})\right)^{2m_3+1}+\left(1+{u}_{{r}}\right)^{2m_3-1}\right], \\ \gamma^{Og} = \frac{\mathcal{P}_c}{C_1}\frac{a}{t}\quad \bar{C}_{2}=\frac{C_2}{C_1} \quad \bar{C}_{3} =\frac{C_3}{C_1}
    \label{eq:pressure_deformation_dimless_Ogden}
\end{multline}

Thus, we have defined the pressure versus the deformation field for the five different models of hyperelasticity. For each of these theories, an expression of FSI parameter has been proposed, which is defined self consistently from the equations.
\section{Fluid Mechanics}
\label{sec:FluMech}
In the previous section, we have delineated the expression relating the local pressure to the local (deformed) radius of the tube for various hyperelastic models, by considering only the structural mechanics (deformation, constitution and equilibrium) of the system. Any discussion of the pressure load acting on the system has escaped the ambit of our analysis so far. In this section, we discuss the origin of the pressure load acting on the tube by delving into the fluid mechanics of the system.

\subsection{Assumptions}
We start with some simplifying assumptions.

\begin{enumerate}
    \item The fluid being conveyed is Newtonian with constant viscosity.
    \item The flow is steady.
    \item The flow is axisymmetric and without any swirl.
    \item The slenderness of the tube allows us to invoke the lubrication approximation for the flow field.
\end{enumerate}

As per the lubrication approximation, the inertia of the flow is neglected. And the pressure gradient is only in the axial/flow wise direction \cite{panton}. From a mathematical perspective, lubrication approximation involves neglecting the terms of $\mathcal{O}(a/\ell)$ and higher in the analysis, after choosing the viscous stress as the pressure scale \cite{AC19a}.

\subsection{Governing equations}
Due to these assumptions, the ODE governing the conservation of fluid momentum reduces to:

\begin{equation}
\label{eq:mom_balance_flu_z}
0=\frac{1}{r} \frac{\partial}{\partial r}\left(r \frac{\partial v_z}{\partial r}\right)-\frac{\partial p}{\partial z}
\end{equation}
and
\begin{equation}
\label{eq:mom_balance_flu_r}
0=\frac{\partial p}{\partial r}
\end{equation}

The above differential equations Eqs.~\eqref{eq:mom_balance_flu_z} and \eqref{eq:mom_balance_flu_r} are predicated upon the following scaling relationship between pressure and velocity:
\begin{equation}
\label{eq:velocity_scale}
\mathcal{V}_c=\frac{\mathcal{P}_c a^2}{\mu \ell}
\end{equation}

\subsection{Boundary conditions}
 To solve the equations governing the momentum balance for fluid, we impose the following boundary conditions:
 
 \subsubsection{Wall}
 
 At the wall of tube, the following boundary conditions are imposed:
 
 \paragraph{Continuity of velocity}
 
 \begin{align}
 \label{eq:velocity_bc1}
 \bar{v}_{\bar{z}}|_{\bar{r}=\bar{R}} &= \frac{D\bar{u}_{\bar{z}}}{D \bar{t}} \\
 &=\frac{\partial \bar{u}_{\bar{z}}}{\partial \bar{t}}+\bar{v}_{\bar{z}}\frac{\partial \bar{u}_{\bar{z}}}{\partial \bar{z}}+\bar{v}_{\bar{r}}\frac{\partial \bar{u}_{\bar{z}}}{\partial \bar{r}} \\
 &= 0 
 \end{align}
Therefore, the continuation of velocity boundary condition imposed at $R = 1+u_r$, under the assumptions of steady, axial flow and neglecting the terms of $\mathcal{O}(a/\ell)$ or higher reduces to  the no slip boundary condition as:
\begin{equation}
\label{eq:velocity_noslip}
    v_z|_{r =R} =0
\end{equation}

The equation for velocity field of fluid obtained by solving the differential equation of Eq.~\eqref{eq:mom_balance_flu_z} subject to the boundary condition of Eq.~\eqref{eq:velocity_noslip} is given as

\begin{equation}
\label{eq:velocity_profile}
v_{z}(r)=-\frac{1}{2} \frac{\partial p}{\partial z}\left[\frac{\left(1+ u_{r}\right)^{2}-r^{2}}{2}\right]
\end{equation}

\subsubsection{Outlet}

At the outlet, the flow vents to zero gauge pressure:
\begin{equation}
\label{eq:outlet_bc}
    p(1) =0
\end{equation}
The above boundary condition for pressure $p$ at $z =1$, when inserted into any of the constitutive equations namely Eqs.~\eqref{eq:Pressure_Deformation_Dimless_Neo_Hookean},\eqref{eq:pressure_deformation_dimless_MooneyRivlin},\eqref{eq:pressure_deformation_dimless_Fung},\eqref{eq:pressure_deformation_dimless_Gent},\eqref{eq:pressure_deformation_dimless_Ogden} automatically gives us $u_r(1) =0$ thereby enforcing the clamped boundary condition for structure at $z=1$ naturally.

\subsubsection{Inlet}

At the inlet, one of the two conditions are possible. 

\paragraph{Pressure controlled}
The flow is pressure controlled and the pressure is imposed (fixed) at the inlet.
 \begin{equation}
 \label{eq:Inlet_pressure}
     \bar{p}(0) =\mathcal{P}_c
 \end{equation}
 
 In this case, the value of the pressure fixed at the inlet namely, $\mathcal{P}_c$ acts as the pressure scale. So the velocity scale can then be obtained from Eq.~\eqref{eq:velocity_scale}. And the dimensionless form of Eq.~\eqref{eq:Inlet_pressure} then becomes:
 \begin{equation}
 \label{eq:Inlet_pressure_dimless}
     {p}(0) =1
 \end{equation}
 
\paragraph{Flow rate controlled}
 
 The other case which is more relevant to our problem is when the flow rate is imposed at the inlet.
 \begin{equation}
     \label{eq:Inlet_flow_bc}
     \bar{q}(0) =q_0
 \end{equation}
The dimensionless form of Eq.~\eqref{eq:Inlet_flow_bc} is simply
\begin{equation}
        \label{eq:Inlet_flow_bc_dimless}
    {q}(0) =1
\end{equation}

For the flow rate controlled, the velocity scale is defined as:
\begin{equation}
\label{eq:v_scale_q}
    \mathcal{V}_c =\frac{{q}_o}{\pi a^2}
\end{equation}
Once the velocity scale is determined from Eq.~\eqref{eq:v_scale_q}, the pressure scale for the flow rate controlled case is then obtained from Eq.~\eqref{eq:velocity_scale}.

The flow rate is determined by the area integral of the velocity profile, and is given as:
 \begin{equation}
     \bar{q} = \int_{0}^{2\pi}\int_{0}^{\bar{R}(\bar{z})}\bar{v}_{\bar{z}}(\bar{r})\bar{r}d\bar{r}d\theta =2\pi\int_{0}^{\bar{R}(\bar{z})}\bar{v}_{\bar{z}}(\bar{r})\bar{r}d\bar{r},
 \end{equation}
 
 or in dimensionless terms:
\begin{equation}
\label{eq:flow_rate_defined}
q(z, t) \equiv \frac{\bar{q}(\bar{z}, \bar{t})}{\mathcal{V}_{z} \pi a^{2}}=\int_{0}^{R(z, t)} v_{z}(r, z, t) 2 r \mathrm{~d} r\stackrel{\text{Eq.~\eqref{eq:velocity_profile}}}{=}-\frac{\partial p}{\partial z}\left[\frac{1}{8}\left(1+ u_{r}\right)^{4}\right]
\end{equation}

For steady, incompressible flows, the flow rate at any cross-section is constant and invariant across the tube, $q(z, t) =q$ (constant).

For flow rate controlled cases, $q =1$ and therefore the above equation reduces to:
\begin{equation}
\label{eq:FluidMechanics}
    \frac{d p}{d z} =-\frac{8}{(1+u_r)^4}
\end{equation}

\section{Coupling and approach to solution}
\label{sec:coupling}

The mathematical statement of the FSI problem therefore consists of Eq.~\eqref{eq:FluidMechanics} from the fluid domain coupled with any of the equations (Eqs.~\eqref{eq:Pressure_Deformation_Dimless_Neo_Hookean},\eqref{eq:pressure_deformation_dimless_MooneyRivlin},\eqref{eq:pressure_deformation_dimless_Fung},\eqref{eq:pressure_deformation_dimless_Gent},\eqref{eq:pressure_deformation_dimless_Ogden}) from the structural mechanical domain. This set of algebraic and differential equations must be solved in tandem to obtain the deformation and pressure profile inside a hyperelastic tube conveying a Newtonian fluid at steady state.

To elaborate further, we first differentiate any of the structural mechanical equations (for different hyperelastic models) namely Eqs.~\eqref{eq:Pressure_Deformation_Dimless_Neo_Hookean},\eqref{eq:pressure_deformation_dimless_MooneyRivlin},\eqref{eq:pressure_deformation_dimless_Fung},\eqref{eq:pressure_deformation_dimless_Gent},\eqref{eq:pressure_deformation_dimless_Ogden} to obtain an expression for $\frac{dp}{dz}$. We use this expression  for $\frac{dp}{dz}$ thus obtained to eliminate $\frac{dp}{dz}$ from Eq.~\eqref{eq:FluidMechanics} and obtain an ODE for $u_r(z)$. This ODE is then solved subject to clamped end boundary condition at $z = 1$ to obtain an (implicit) expression for $u_r(z)$. Substitution of the expression for $u_r(z)$ back in the pertinent structural mechanical equation namely any of Eqs.~\eqref{eq:Pressure_Deformation_Dimless_Neo_Hookean},\eqref{eq:pressure_deformation_dimless_MooneyRivlin},\eqref{eq:pressure_deformation_dimless_Fung},\eqref{eq:pressure_deformation_dimless_Gent},\eqref{eq:pressure_deformation_dimless_Ogden} yields the expression for the pressure profile $p(z)$. Finally, the expressions for $u_r(z)$ and $p(z)$ thus obtained may be used to compute the circumferential stress in the structure using Eq.~\eqref{eq:sigma_11_22_33}(b).
This approach to solution has been summarised in Fig 2.

For the neo Hookean hyperelastic tube, differentiation of Eq.~\eqref{eq:Pressure_Deformation_Dimless_Neo_Hookean} yields the following:

\begin{equation}
\gamma^{NH} \frac{\mathrm{d} {p}}{\mathrm{~d} {z}}=\left[1+\frac{3}{\left(1+{u}_{{r}}\right)^{4}}\right] \frac{\mathrm{d} {u}_{{r}}}{\mathrm{~d} {z}},
\end{equation}
which when substituted into Eq.~\eqref{eq:FluidMechanics} yields the following ODE for $u_r(z)$:
\begin{equation}
\label{eq:NH_ODE_U}
   \left[1+\frac{3}{\left(1+{u}_{{r}}\right)^{4}}\right] \frac{\mathrm{d} {u}_{{r}}}{\mathrm{~d} {z}} =-\frac{8\gamma^{NH}}{(1+u_r)^4}
\end{equation}
The solution for the above ODE along with the clamping boundary condition at $z=1$, where $u_r(1) =1$, yield the following deformation (implicit) profile :

\begin{equation}
\label{eq:Deformation_NH}
    (1+u_r(z))^5+15u_r(z) =40\gamma^{NH}(1-z)+1
\end{equation}
The numerical solution of the above equation when inserted into the pressure deformation profile of Eq.~\eqref{eq:Pressure_Deformation_Dimless_Neo_Hookean} yields the pressure profile numerically. 

For Mooney-Rivlin model, whose pressure deformation equation as given by Eq.~\eqref{eq:pressure_deformation_dimless_MooneyRivlin} is same as that for neo Hookean model (Eq.~\eqref{eq:Pressure_Deformation_Dimless_Neo_Hookean} except for the definition of FSI parameter $\gamma^{MR}$, the implicit expression for deformation profile $u_r(z)$ is also given by Eq.~\eqref{eq:Deformation_NH}, with $\gamma^{NH}$ replaced by $\gamma^{MR}$.

For Fung model, the same procedure leads to the following ODE governing the deformation:
\begin{equation}
\label{eq:ODE_Fung}
  \frac{du_r}{dz} = \frac{\text{Nr}^F}{\text{Dr}^F} \end{equation}
  where
  \begin{equation}
    \text{Nr}^F =- 8 \gamma \left(u{\left(z \right)} + 1\right)^{2}
  \end{equation}
\begin{multline}  
     \text{Dr}^F =2 \alpha \left(u{\left(z \right)} + 1\right)^{6} u^{2}{\left(z \right)} e^{\frac{\alpha \left(\left(u{\left(z \right)} + 1\right)^{2} - 1\right)^{2}}{\left(u{\left(z \right)} + 1\right)^{2}}} 
     + 4 \alpha \left(u{\left(z \right)} + 1\right)^{6} u{\left(z \right)} e^{\frac{\alpha \left(\left(u{\left(z \right)} + 1\right)^{2} - 1\right)^{2}}{\left(u{\left(z \right)} + 1\right)^{2}}} \\ + 2 \alpha \left(u{\left(z \right)} + 1\right)^{4} u^{2}{\left(z \right)} e^{\frac{\alpha \left(\left(u{\left(z \right)} + 1\right)^{2} - 1\right)^{2}}{\left(u{\left(z \right)} + 1\right)^{2}}} + 4 \alpha \left(u{\left(z \right)} + 1\right)^{4} u{\left(z \right)} e^{\frac{\alpha \left(\left(u{\left(z \right)} + 1\right)^{2} - 1\right)^{2}}{\left(u{\left(z \right)} + 1\right)^{2}}} \\ - 2 \alpha \left(u{\left(z \right)} + 1\right)^{2} u^{2}{\left(z \right)} e^{\frac{\alpha \left(\left(u{\left(z \right)} + 1\right)^{2} - 1\right)^{2}}{\left(u{\left(z \right)} + 1\right)^{2}}} - 4 \alpha \left(u{\left(z \right)} + 1\right)^{2} u{\left(z \right)} e^{\frac{\alpha \left(\left(u{\left(z \right)} + 1\right)^{2} - 1\right)^{2}}{\left(u{\left(z \right)} + 1\right)^{2}}} \\ - 2 \alpha u^{2}{\left(z \right)} e^{\frac{\alpha \left(\left(u{\left(z \right)} + 1\right)^{2} - 1\right)^{2}}{\left(u{\left(z \right)} + 1\right)^{2}}} - 4 \alpha u{\left(z \right)} e^{\frac{\alpha \left(\left(u{\left(z \right)} + 1\right)^{2} - 1\right)^{2}}{\left(u{\left(z \right)} + 1\right)^{2}}} \\+ \left(u{\left(z \right)} + 1\right)^{6} e^{\frac{\alpha \left(\left(u{\left(z \right)} + 1\right)^{2} - 1\right)^{2}}{\left(u{\left(z \right)} + 1\right)^{2}}} + \left(u{\left(z \right)} + 1\right)^{6} + 3 \left(u{\left(z \right)} + 1\right)^{2} e^{\frac{\alpha \left(\left(u{\left(z \right)} + 1\right)^{2} - 1\right)^{2}}{\left(u{\left(z \right)} + 1\right)^{2}}} + 3 \left(u{\left(z \right)} + 1\right)^{2}
\end{multline}

The solution for Eq.~\eqref{eq:ODE_Fung} imposing the clamped boundary condition at $z =1$ gives us the following implicit expression for $u_r(z)$:

\begin{multline}
\label{eq:Deformation_Fung}
0=- 70 \alpha u^{9}{\left(z \right)} e^{\frac{\alpha \left(u^{2}{\left(z \right)} + 4 u{\left(z \right)} + 4\right) u^{2}{\left(z \right)}}{u^{2}{\left(z \right)} + 2 u{\left(z \right)} + 1}} + 2520 \gamma \left(1 - z\right) \left(u^{2}{\left(z \right)} + 2 u{\left(z \right)} + 1\right)\\ - 1260 \left(e^{\frac{\alpha \left(u^{2}{\left(z \right)} + 4 u{\left(z \right)} + 4\right) u^{2}{\left(z \right)}}{u^{2}{\left(z \right)} + 2 u{\left(z \right)} + 1}} + 1\right)  u{\left(z \right)} \\- 210 \left(16 \alpha e^{\frac{\alpha \left(u^{2}{\left(z \right)} + 4 u{\left(z \right)} + 4\right) u^{2}{\left(z \right)}}{u^{2}{\left(z \right)} + 2 u{\left(z \right)} + 1}} + 9 e^{\frac{\alpha \left(u^{2}{\left(z \right)} + 4 u{\left(z \right)} + 4\right) u^{2}{\left(z \right)}}{u^{2}{\left(z \right)}  + 2 u{\left(z \right)} + 1}} + 9\right) u^{3}{\left(z \right)} \\-  315 \left(24 \alpha e^{\frac{\alpha \left(u^{2}{\left(z \right)} + 4 u{\left(z \right)} + 4\right) u^{2}{\left(z \right)}}{u^{2}{\left(z \right)} + 2 u{\left(z \right)} + 1}} + 5 e^{\frac{\alpha \left(u^{2}{\left(z \right)} + 4 u{\left(z \right)} + 4\right) u^{2}{\left(z \right)}}{u^{2}{\left(z \right)} + 2 u{\left(z \right)} + 1}} + 5\right) u^{4}{\left(z \right)} \\ - 45 \left(56 \alpha e^{\frac{\alpha \left(u^{2}{\left(z \right)} + 4 u{\left(z \right)} + 4\right) u^{2}{\left(z \right)}}{u^{2}{\left(z \right)} + 2 u{\left(z \right)} + 1}} + e^{\frac{\alpha \left(u^{2}{\left(z \right)} + 4 u{\left(z \right)} + 4\right) u^{2}{\left(z \right)}}{u^{2}{\left(z \right)} + 2 u{\left(z \right)} + 1}} + 1\right) u^{7}{\left(z \right)} \\- 105 \left(56 \alpha e^{\frac{\alpha \left(u^{2}{\left(z \right)} + 4 u{\left(z \right)} + 4\right) u^{2}{\left(z \right)}}{u^{2}{\left(z \right)} + 2 u{\left(z \right)} + 1}} + 3 e^{\frac{\alpha \left(u^{2}{\left(z \right)} + 4 u{\left(z \right)} + 4\right) u^{2}{\left(z \right)}}{u^{2}{\left(z \right)} + 2 u{\left(z \right)} + 1}} + 3\right) u^{6}{\left(z \right)} \\- 63 \left(136 \alpha e^{\frac{\alpha \left(u^{2}{\left(z \right)} + 4 u{\left(z \right)} + 4\right) u^{2}{\left(z \right)}}{u^{2}{\left(z \right)} + 2 u{\left(z \right)} + 1}} + 15 e^{\frac{\alpha \left(u^{2}{\left(z \right)} + 4 u{\left(z \right)} + 4\right) u^{2}{\left(z \right)}}{u^{2}{\left(z \right)} + 2 u{\left(z \right)} + 1}} + 15\right) u^{5}{\left(z \right)} \\- 630 \left(\alpha u^{6}{\left(z \right)} e^{\frac{\alpha \left(u^{2}{\left(z \right)} + 4 u{\left(z \right)} + 4\right) u^{2}{\left(z \right)}}{u^{2}{\left(z \right)} + 2 u{\left(z \right)} + 1}} + 3 e^{\frac{\alpha \left(u^{2}{\left(z \right)} + 4 u{\left(z \right)} + 4\right) u^{2}{\left(z \right)}}{u^{2}{\left(z \right)} + 2 u{\left(z \right)} + 1}} + 3\right) u^{2}{\left(z \right)}
\end{multline}

 Similarly, the ODE governing the evolution of deformation for $\text{Ogden}_{3}$ model is given as:
 \begin{equation}
     \frac{du_r}{dz}=- \frac{8 \gamma}{\text{Dr}^{O}}
 \end{equation}
 where 
 \begin{multline}
     \text{Dr}^{O}=\left(u{\left(z \right)} + 1\right)^{3} \Bigg(2 C_{2} m_{2} \left(u{\left(z \right)} + 1\right)^{2 m_{2} - 1} + 2 C_{2} m_{2} \left(\frac{1}{u{\left(z \right)} + 1}\right)^{2 m_{2} + 1} - C_{2} \left(u{\left(z \right)} + 1\right)^{2 m_{2} - 1} \\ + C_{2} \left(\frac{1}{u{\left(z \right)} + 1}\right)^{2 m_{2} + 1} + 2 C_{3} m_{3} \left(u{\left(z \right)} + 1\right)^{2 m_{3} - 1}  + 2 C_{3} m_{3} \left(\frac{1}{u{\left(z \right)} + 1}\right)^{2 m_{3} + 1} \\- C_{3} \left(u{\left(z \right)} + 1\right)^{2 m_{3} - 1} + C_{3} \left(\frac{1}{u{\left(z \right)} + 1}\right)^{2 m_{3} + 1} + 2 m_{1} \left(u{\left(z \right)} + 1\right)^{2 m_{1} - 1} + 2 m_{1} \left(\frac{1}{u{\left(z \right)} + 1}\right)^{2 m_{1} + 1} \\- \left(u{\left(z \right)} + 1\right)^{2 m_{1} - 1} + \left(\frac{1}{u{\left(z \right)} + 1}\right)^{2 m_{1} + 1}\Bigg)
 \end{multline}
 
 The solution of the above equation with $u_r(1) =0$ is given as:
 \begin{multline}
 \label{eq:Deformation_Ogden}
   0=32 \gamma \left(1 - z\right) \left(u{\left(z \right)} + 1\right) \\ - \Bigg(2 C_{2} m_{2} \left(u{\left(z \right)} + 1\right)^{2 m_{2}} + 2 C_{2} m_{2} \left(\frac{1}{u{\left(z \right)} + 1}\right)^{2 m_{2}} - C_{2} \left(u{\left(z \right)} + 1\right)^{2 m_{2}} + C_{2} \left(\frac{1}{u{\left(z \right)} + 1}\right)^{2 m_{2}} \\ + 2 C_{3} m_{3} \left(u{\left(z \right)} + 1\right)^{2 m_{3}} + 2 C_{3} m_{3} \left(\frac{1}{u{\left(z \right)} + 1}\right)^{2 m_{3}} - C_{3} \left(u{\left(z \right)} + 1\right)^{2 m_{3}} + C_{3} \left(\frac{1}{u{\left(z \right)} + 1}\right)^{2 m_{3}} \\+ 2 m_{1} \left(u{\left(z \right)} + 1\right)^{2 m_{1}} \\ + 2 m_{1} \left(\frac{1}{u{\left(z \right)} + 1}\right)^{2 m_{1}} - \left(u{\left(z \right)} + 1\right)^{2 m_{1}} + \left(\frac{1}{u{\left(z \right)} + 1}\right)^{2 m_{1}}\Bigg) u^{4}{\left(z \right)} \\- 6 \Bigg(2 C_{2} m_{2} \left(u{\left(z \right)} + 1\right)^{2 m_{2}} + 2 C_{2} m_{2} \left(\frac{1}{u{\left(z \right)} + 1}\right)^{2 m_{2}} - C_{2} \left(u{\left(z \right)} + 1\right)^{2 m_{2}} + C_{2} \left(\frac{1}{u{\left(z \right)} + 1}\right)^{2 m_{2}}\\ + 2 C_{3} m_{3} \left(u{\left(z \right)} + 1\right)^{2 m_{3}} \\ + 2 C_{3} m_{3} \left(\frac{1}{u{\left(z \right)} + 1}\right)^{2 m_{3}} - C_{3} \left(u{\left(z \right)} + 1\right)^{2 m_{3}} + C_{3} \left(\frac{1}{u{\left(z \right)} + 1}\right)^{2 m_{3}} + 2 m_{1} \left(u{\left(z \right)} + 1\right)^{2 m_{1}} \\ + 2 m_{1} \left(\frac{1}{u{\left(z \right)} + 1}\right)^{2 m_{1}} - \left(u{\left(z \right)} + 1\right)^{2 m_{1}} + \left(\frac{1}{u{\left(z \right)} + 1}\right)^{2 m_{1}}\Bigg) u^{2}{\left(z \right)} \\+ 4 \Bigg(- 2 C_{2} m_{2} \left(u{\left(z \right)} + 1\right)^{2 m_{2}} - 2 C_{2} m_{2} \left(\frac{1}{u{\left(z \right)} + 1}\right)^{2 m_{2}} + C_{2} \left(u{\left(z \right)} + 1\right)^{2 m_{2}} - C_{2} \left(\frac{1}{u{\left(z \right)} + 1}\right)^{2 m_{2}} \\-  2 C_{3} m_{3} \left(u{\left(z \right)} + 1\right)^{2 m_{3}} \\ - 2 C_{3} m_{3} \left(\frac{1}{u{\left(z \right)} + 1}\right)^{2 m_{3}} + C_{3} \left(u{\left(z \right)} + 1\right)^{2 m_{3}} - C_{3} \left(\frac{1}{u{\left(z \right)} + 1}\right)^{2 m_{3}} - 2 m_{1} \left(u{\left(z \right)} + 1\right)^{2 m_{1}} \\- 2 m_{1} \left(\frac{1}{u{\left(z \right)} + 1}\right)^{2 m_{1}} + \left(u{\left(z \right)} + 1\right)^{2 m_{1}} \\+ \Bigg(- 2 C_{2} m_{2} \left(u{\left(z \right)} + 1\right)^{2 m_{2}} - 2 C_{2} m_{2} \left(\frac{1}{u{\left(z \right)} + 1}\right)^{2 m_{2}} + C_{2} \left(u{\left(z \right)} + 1\right)^{2 m_{2}} \\- C_{2} \left(\frac{1}{u{\left(z \right)} + 1}\right)^{2 m_{2}} - 2 C_{3} m_{3} \left(u{\left(z \right)} + 1\right)^{2 m_{3}} - 2 C_{3} m_{3} \left(\frac{1}{u{\left(z \right)} + 1}\right)^{2 m_{3}} + C_{3} \left(u{\left(z \right)} + 1\right)^{2 m_{3}} - C_{3} \left(\frac{1}{u{\left(z \right)} + 1}\right)^{2 m_{3}}\\ - 2 m_{1} \left(u{\left(z \right)} + 1\right)^{2 m_{1}}\\ - 2 m_{1} \left(\frac{1}{u{\left(z \right)} + 1}\right)^{2 m_{1}} + \left(u{\left(z \right)} + 1\right)^{2 m_{1}} - \left(\frac{1}{u{\left(z \right)} + 1}\right)^{2 m_{1}}\Bigg) u^{2}{\left(z \right)} - \left(\frac{1}{u{\left(z \right)} + 1}\right)^{2 m_{1}}\Bigg) u{\left(z \right)}
 \end{multline}
 
 Finally, the ODE governing the evolution of deformation for Gent's model is:
 \begin{equation}
     \frac{du_r}{dz} = \frac{\text{Nr}^G}{\text{Dr}^G}
 \end{equation}
 
 \begin{multline}
     \text{Nr}^G =8 \gamma \Bigg(- \alpha^{2} u^{8}{\left(z \right)} - 8 \alpha^{2} u^{7}{\left(z \right)} - 24 \alpha^{2} u^{6}{\left(z \right)} - 32 \alpha^{2} u^{5}{\left(z \right)} - 16 \alpha^{2} u^{4}{\left(z \right)} + 2 \alpha u^{6}{\left(z \right)} + 12 \alpha u^{5}{\left(z \right)} \\ + 26 \alpha u^{4}{\left(z \right)} + 24 \alpha u^{3}{\left(z \right)} + 8 \alpha u^{2}{\left(z \right)} - u^{4}{\left(z \right)} - 4 u^{3}{\left(z \right)} - 6 u^{2}{\left(z \right)} - 4 u{\left(z \right)} - 1\Bigg)
 \end{multline}
 and 
 \begin{multline}
     \text{Dr}^G=\alpha u^{10}{\left(z \right)} + 10 \alpha u^{9}{\left(z \right)} + 47 \alpha u^{8}{\left(z \right)} + 136 \alpha u^{7}{\left(z \right)} + 258 \alpha u^{6}{\left(z \right)} + 316 \alpha u^{5}{\left(z \right)} + 236 \alpha u^{4}{\left(z \right)} \\+ 96 \alpha u^{3}{\left(z \right)} + 16 \alpha u^{2}{\left(z \right)} + u^{8}{\left(z \right)} + 8 u^{7}{\left(z \right)} + 28 u^{6}{\left(z \right)} + 56 u^{5}{\left(z \right)} + 73 u^{4}{\left(z \right)} \\ + 68 u^{3}{\left(z \right)} + 46 u^{2}{\left(z \right)} + 20 u{\left(z \right)} + 4
 \end{multline}
, the solution of which with the boundary condition at $u_r(1) = 0$, gives us the following implicit solution for $u_r(z)$:

\begin{multline}
\label{eq:Deformation_Gent}
    0=- 315 \alpha u^{11}{\left(z \right)} + 27720 \gamma \left(1 - z\right) \Bigg(\alpha^{2} u^{8}{\left(z \right)} + 8 \alpha^{2} u^{7}{\left(z \right)} + 24 \alpha^{2} u^{6}{\left(z \right)} + 32 \alpha^{2} u^{5}{\left(z \right)}\\ + 16 \alpha^{2} u^{4}{\left(z \right)} - 2 \alpha u^{6}{\left(z \right)} - 12 \alpha u^{5}{\left(z \right)} - 26 \alpha u^{4}{\left(z \right)} - 24 \alpha u^{3}{\left(z \right)} - 8 \alpha u^{2}{\left(z \right)}\\ + u^{4}{\left(z \right)} + 4 u^{3}{\left(z \right)} + 6 u^{2}{\left(z \right)} + 4 u{\left(z \right)} + 1\Bigg)\\ - 385 \left(47 \alpha + 1\right) u^{9}{\left(z \right)} - 990 \left(129 \alpha + 14\right) u^{7}{\left(z \right)} - 693 \left(236 \alpha + 73\right) u^{5}{\left(z \right)}\\ - 2310 \left(8 \alpha + \left(79 \alpha + 14\right) u^{3}{\left(z \right)} + 23\right) u^{3}{\left(z \right)} - 3465 \left(\alpha u^{6}{\left(z \right)} + 24 \alpha + \left(17 \alpha + 1\right) u^{4}{\left(z \right)} + 17\right) u^{4}{\left(z \right)} \\- 34650 u^{2}{\left(z \right)} - 13860 u{\left(z \right)}
\end{multline}

\section{Results}
\label{sec:Results}
 Next, we discuss the results of our analysis in this section. The results are segregated into three categories. First, in Sec.~\ref{sec:Results}\ref{sec:tube_law}, we discuss the pressure deflection relationships derived in Sec.~\ref{sec:StructMech} for different hyperelastic models. Next, in Sec.~\ref{sec:Results}\ref{sec:FSI_Results}, we deliberate upon the FSI solution and delineate the quantitative and qualitative characteristics of the pressure, deformation and stress profiles. Finally, we discuss the contribution of geometric vs material nonlinearity and isolate the relative importance of each in the five models.
 Since the Mooney-Rivlin model furnishes the same expression as the neo Hookean model, we are not going to discuss the results pertaining to Mooney-Rivlin model from now on.

 \subsection{Pressure -deflection relationship}
 \label{sec:tube_law}
 
\begin{figure}[t]
\centering
\subfloat[]{\includegraphics[height =0.35\linewidth]{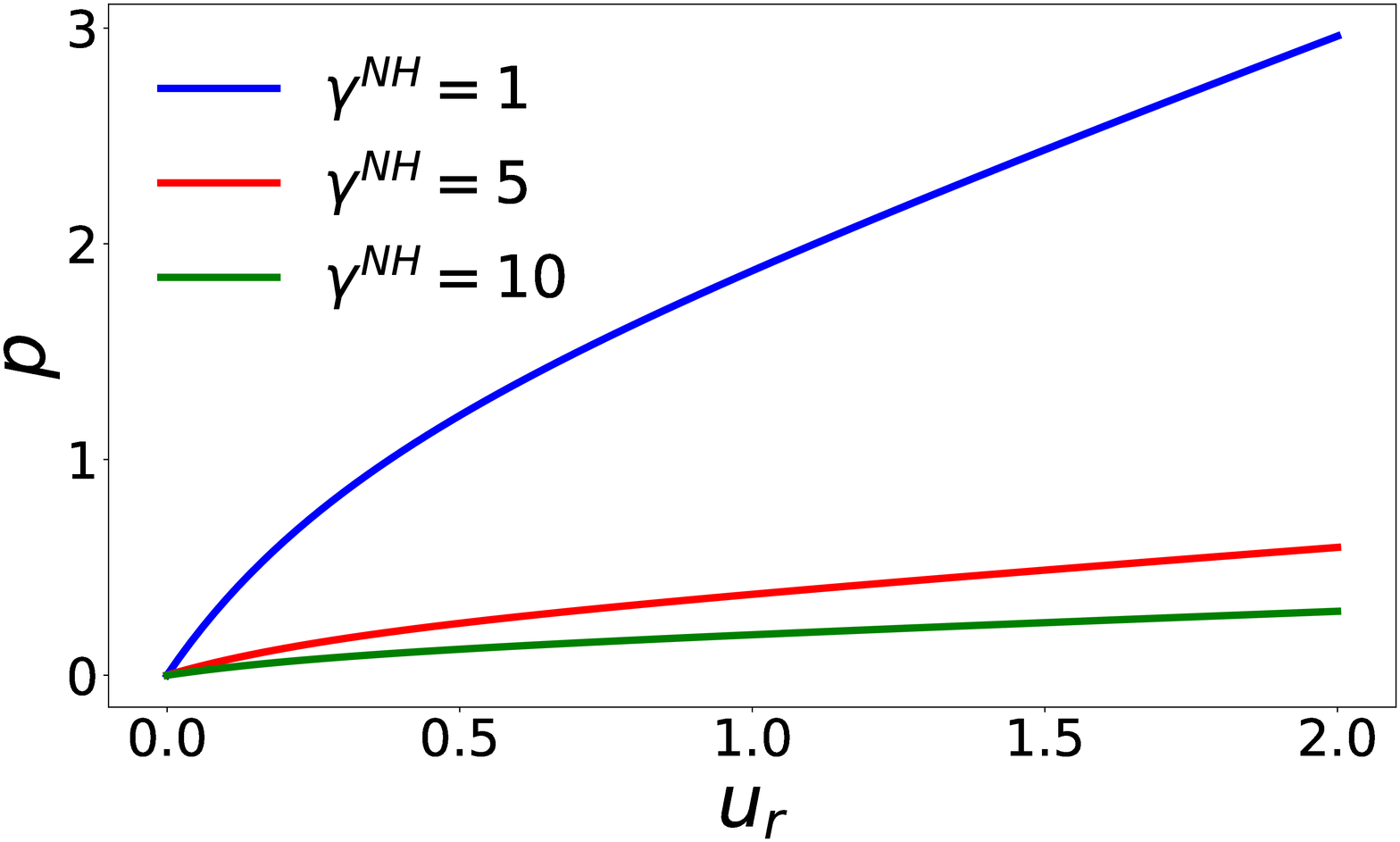}} \\
\subfloat[]{\includegraphics[height=0.35\linewidth]{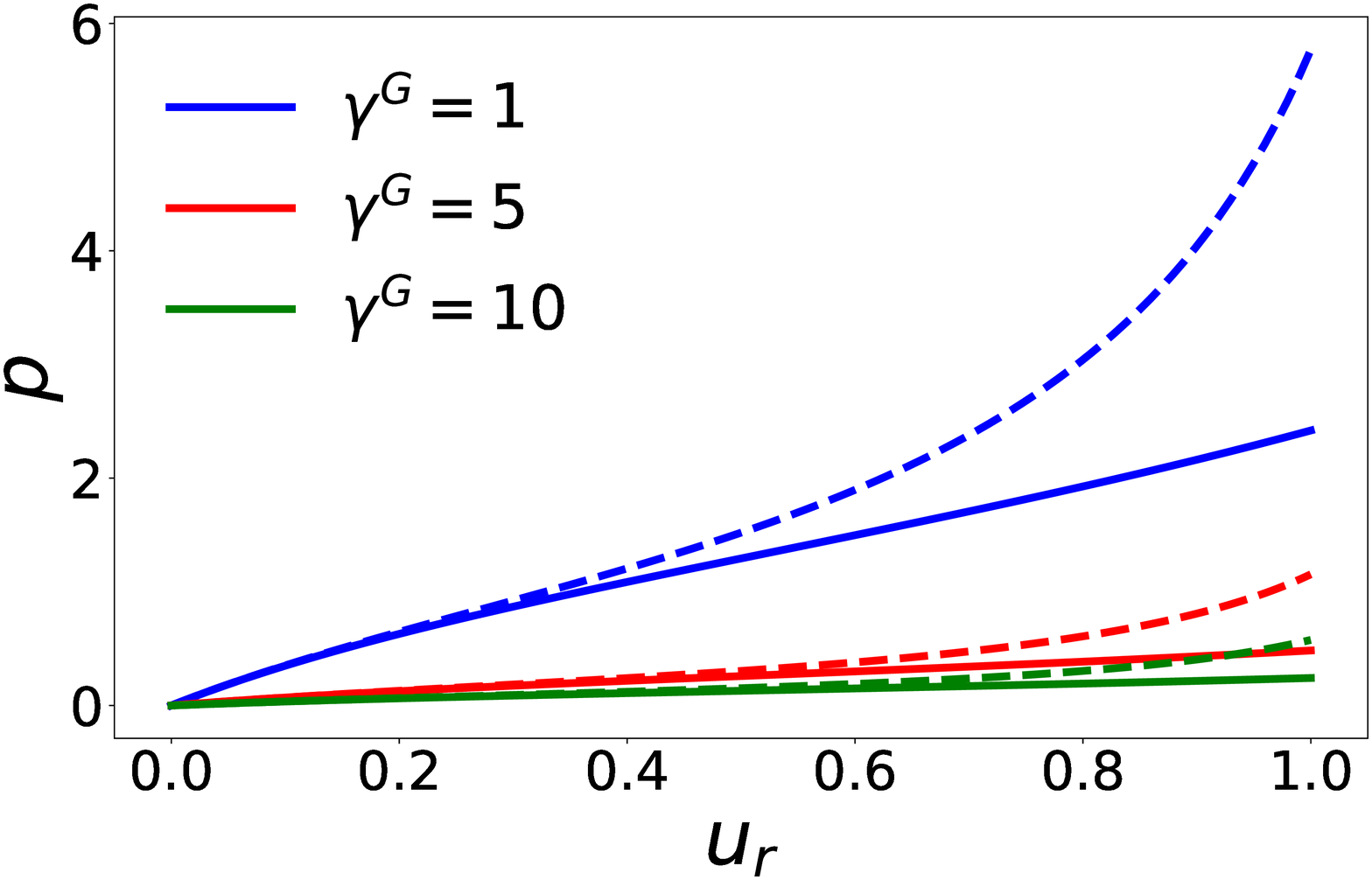}}\\
\subfloat[]{\includegraphics[height=0.35\linewidth]{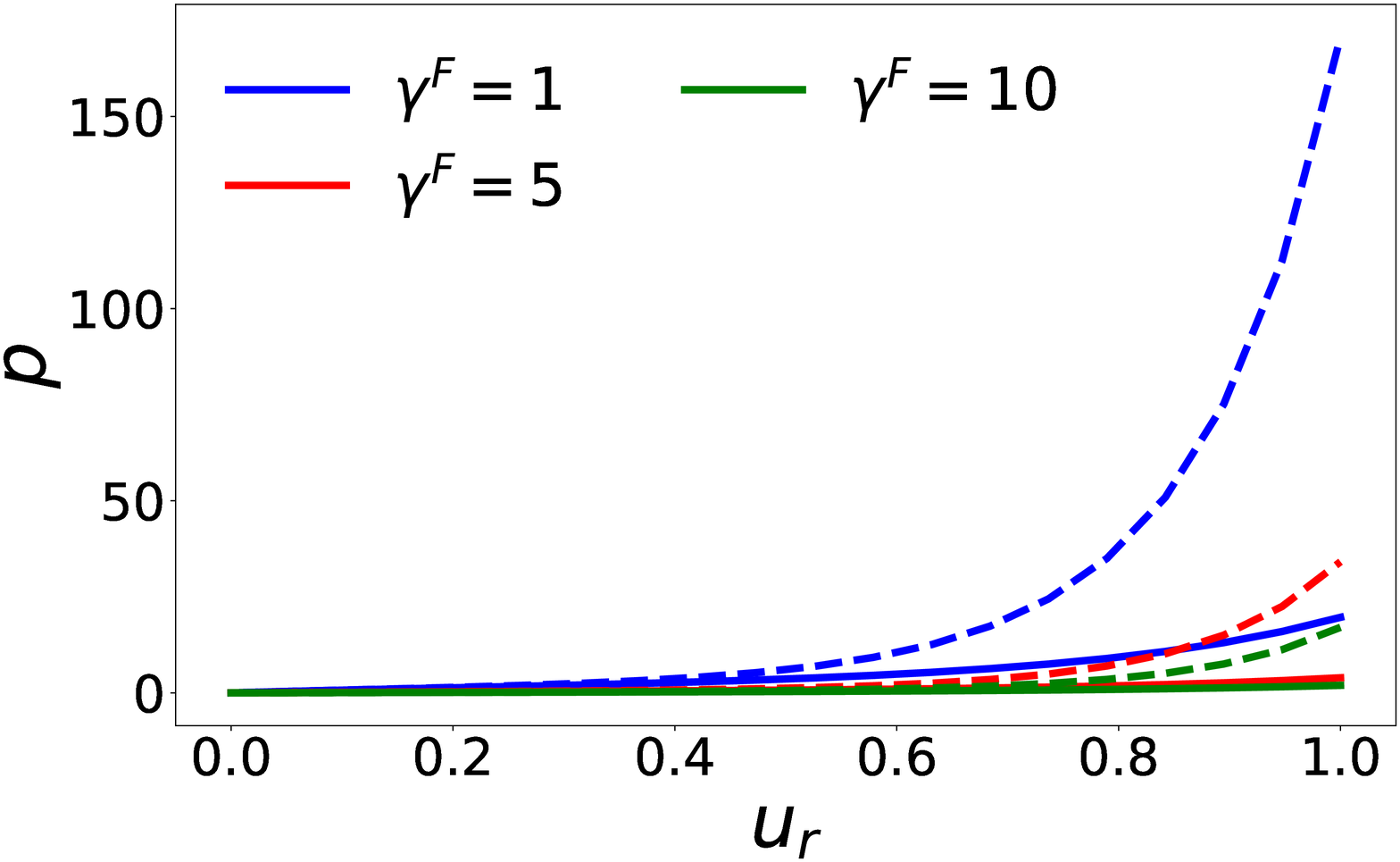}} \\
\subfloat[]{\includegraphics[height=0.35\linewidth]{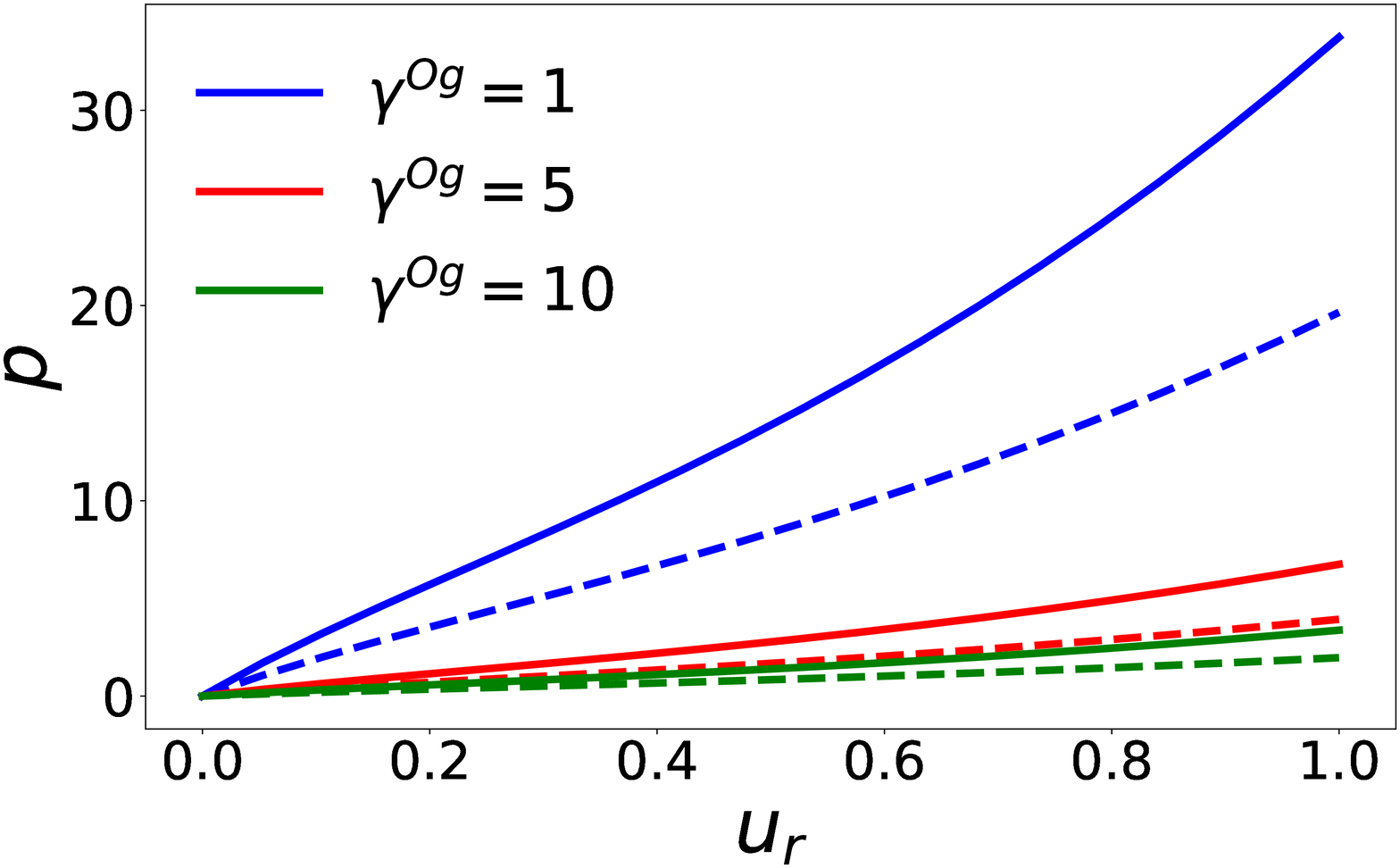}}
\caption{Pressure deflection curves for different hyperelastic tubes (a) neo Hookean (b) Gent's model: solid curves represent $\eta =0.1 $, dashed curves for $\eta =0.3$. (c) Fung's model: solid curves represent $\alpha =1$,dashed curves for $\alpha =5 $ (d) $\text{Ogden}_{3}$ model: solid curves represent $C_1= -3543,C_2 =-2723,C_3 =654,m_1 =1,m_2 =-1,m_3 =2$ , dashed curves represent $C_1= -3882,C_2 =-2113,C_3 =931,m_1 =1,m_2 =-1,m_3 =2$}
\label{fig:TubeLaw}
\end{figure}

 The pressure deflection relationships derived in Sec.\ref{sec:StructMech}\ref{sec:consti} for different hyperelastic models are shown in FIg.\ref{fig:TubeLaw}. We observe straightforward that the increase in respective $\gamma$'s leads to a corresponding decrease in pressure for all models. This trend is attributed to the definition of $\gamma$ itself, a lower $\gamma$ denotes a higher value of elastic modulus and therefore a stiffer tube for all models. And since it takes higher pressure to deform stiffer tubes, we see an increase in pressure for smaller $\gamma$.
 
For neo Hookean tube as shown in Fig.\ref{fig:TubeLaw}(a), we observe that at high $\gamma$ (stiffer tubes), the pressure curve tends to zero slope; the deformation continues to increase even when the pressure barely increases. This characteristic is symptomatic of rubber and has been observed experimentally in tubes made of latex rubber (compare Fig.1(b) of \cite{Holzapfel2005SimilaritiesMaterials}, also see Fig 1 in \cite{Osborne1909} which was later digitised and reprinted as Fig 1(a) in \cite{Destrade_NeoHookean_Instability} ). Ultimately this behavior of neo Hookean tube leads to \textit{snap} instability (or "limit point" instability, to use the term by \cite{Destrade_NeoHookean_Instability}) wherein the deformation continues to increase even at constant pressure and the tube eventually ruptures. This is also one of the reasons neo Hookean model is not popular at large strains.
 
 Quite an opposite trend is observed in the case of Fung's model in Fig.~\ref{fig:TubeLaw}(b). Here, there is a marked increase in pressure for incremental deformation, as the material paramter $\alpha$ increases, the increase in pressure is even higher. This trend is attributed to the famous exponential strain hardening response built in the Fung's model, symptomatic of the stress response of stiff biological tissues like tendons and ligaments, owing to the presence of copious amounts of collagen in them.

 Gent's model on the other hand, shown in Fig\ref{fig:TubeLaw}(c), exhibits strain hardening but not to the same extent as the Fung's model. The strain hardening of Gent's model becomes more severe with an increase in $\eta$, but still not comparable to that of Fung. However, the model by Gent employs a different strategy to curtail the deformation of the tube. Refer to appendix to see Fig.~\ref{fig:TubeLaw2_Gent}, where the pressure deformation curve for Gent's model has been plotted for higher values of $u_r$, we notice that the pressure increases for up to a particular value of $u_r$ say $u_{max}$ and after that the pressure nosedives and becomes negative - clearly an aphysical situation. This trend means that the Gent's model establishes a maximum value of $u_r$, deformation beyond which are not allowed. On the other hand, for $u_r < u_{max}$, the strain stiffening response of Gent's model is milder than that of Fung.
 
 Finally, the pressure deformation relationship for  $\text{Ogden}_{3}$ model is shown in Fig.\ref{fig:TubeLaw}(d). Here, we observe that the strain hardening response is weaker than that of Fung. Therefore, this model is mostly used to model soft tissues like fat and brain tissues.
 
 In fluid mechanics community and especially in FSI literature, it is common to express the pressure deformation expressions as pressure-area relationships known as "tube laws". Dating from tube law of Laplace \cite{Canic} to Fung's in ring model\cite{F97_ch3}, the tube laws are used as constitutive equations for tubes in a cross-sectional averaged sense. For instance, the tube law version of neo Hookean model given in Eq.~\eqref{eq:Pressure_Deformation_Dimless_Neo_Hookean} is given as:
 \begin{equation}
 \label{eq:tube_Law}
     p(A) =\left[A-\frac{1}{A}\right]\left(\frac{1}{\sqrt{A}}\right)\frac{1}{\gamma^{NH}}; \quad A=\frac{\pi\bar{R}^2}{\pi a^2}
 \end{equation}
 where $A$ is the dimensionless area of the deformed tube. 
 Similarly, we have derived the tube laws for tubes of other materials as well.
 Summary of the tube laws for different hyperelastic models is given in table\ref{tb:1}.
 \pagebreak

\begin{sidewaystable}
\caption[]{Summary of dimensionless characteristics of inflation problem and FSI problem for hyperelastic tube conveying Newtonian flow at steady state.}
\label{tb:1}
  \centering
  \begin{tabular}{@{\extracolsep{\fill}}lllll}
    \toprule
      Model & Strain energy functional & Tube Law & Deformation  & Pressure profile \\  
     \addlinespace \addlinespace
     \midrule
     neo Hookean & $\frac{C}{2}{(\lambda_{1}^2+\lambda_{2}^2+\lambda_{3}^2-3)}$ & $p(A) =\left[\sqrt{A}-\frac{1}{A^{3/2}}\right]\frac{1}{\gamma^{NH}}$ & Eq.\eqref{eq:Deformation_NH} & $
    \gamma^{nH}{p} =\left[\left(1+{u}_r\right)-\left(\frac{1}{1+{u}_r}\right)^3\right]$  
     \\ && $\quad \gamma^{nH} =\frac{\mathcal{P}_c}{C}\frac{a}{t} $ && \\
    \addlinespace
    Mooney-Rivlin & $\frac{\mathbb{C}_1}{2} \left(\lambda_1^2+\lambda_2^2+\lambda_3^2\right) + $ & $p(A) =\left[\sqrt{A}-\frac{1}{A^{3/2}}\right]\frac{1}{\gamma^{MR}}$ & Eq.~\eqref{eq:Deformation_NH} & $ \gamma^{MR}{p} =\left[\left(1+{u}_r\right)-\left(\frac{1}{1+{u}_r}\right)^3\right]$ 
      \\ 
    &$+\frac{\mathbb{C}_2}{2} \left(\lambda_1^2\lambda_2^2 + \lambda_2^2\lambda_3^2+\lambda_3^1\lambda_1^2\right)$&$\gamma^{MR} := \frac{\mathcal{P}_c}{(\mathbb{C}_1+\mathbb{C}_2)}\frac{a}{t}$ &&
   \\
    \addlinespace
   Fung & $\frac{C}{2\alpha}\Big[\alpha(\lambda_{1}^2+\lambda_{2}^2+\lambda_{3}^2-3)$ & $p(A) =\frac{\left[\sqrt{A}-\frac{1}{A^{3/2}}\right]}{\left[1+e^{\frac{\alpha}{A}\left(1-A\right)^2}\right]}\frac{1}{\gamma^{MR}}$ & Eq.~\eqref{eq:Deformation_Fung} &  $\gamma^F{p} = \frac{\left[(1+{u}_{{r}})-\frac{1}{(1+{u}_{{r}})^3}\right]}{\left[1+e^{\alpha\{\frac{1}{1+{u}_{{r}}}-(1+{u}_{{r}})\}^2}\right]}$ \\ & $+e^{\alpha(\lambda_{1}^2+\lambda_{2}^2+\lambda_{3}^2-3)}-1\Big]$ &$\gamma^F := \frac{\mathcal{P}_c}{\mathbb{C}}\frac{a}{t}$ &&\\ \\\addlinespace
   Gent &  $-\frac{C}{2\eta}ln\left[1-\eta(\lambda_{1}^2+\lambda_{2}^2+\lambda_{3}^2-3)\right] $ & $p(A) =\frac{\left[\sqrt{A}-\frac{1}{A^{3/2}}\right]}{\left[1-{\frac{\eta}{A}\left(1-A\right)^2}\right]}\frac{1}{\gamma^{G}}$ & Eq.~\eqref{eq:Deformation_Gent} &  $\gamma^{G}{p} = \frac{(1+{u}_{{r}})-\frac{1}{(1+{u}_{{r}})^3}}{[{1-\eta\{\frac{1}{1+{u}_{{r}}}-(1+{u}_{{r}})\}^2}]}$ \\ & &$\gamma^G := \frac{\mathcal{P}_c}{\mathbb{C}}\frac{a}{t}$ & & \\  \\
    \addlinespace
    $\text{Ogden}_{3}$ &  $\sum_{p=1}^{3}\frac{C_p}{2m_p}(\lambda_1^{2m_p}+\lambda_2^{2mp}+\lambda_3^{2m_p}-3) $ & $
    \gamma^{Og}{p}(A)=\left[-\left(\frac{1}{A}\right)^\frac{2m_1+1}{2}+A^\frac{2m_1-1}{2}\right]
    
$ & Eq.~\eqref{eq:Deformation_Ogden} &  $\gamma^{Og}{p}=\left[-\left(\frac{1}{(1+{u}_{{r}}}\right)^{2m_1+1}+\left(1+{u}_{{r}}\right)^{2m_1-1}\right]$  \\
& & $+  \bar{C}_{2}\left[-\left(\frac{1}{A}\right)^\frac{2m_2+1}{2}+A^\frac{2m_2-1}{2}\right]$  & & $+\bar{C}_{2}\left[-\left(\frac{1}{1+{u}_{r}}\right)^{2m_2+1}+\left(1+{u}_{{r}}\right)^{2m_2-1}\right]$ \\ & & $+\bar{C}_{3}\left[-\left(\frac{1}{A}\right)^\frac{2m_3+1}{2}+A^\frac{2m_3-1}{2}\right] $ & & $+\bar{C}_{3}\left[-\left(\frac{1}{1+{u}_{\bar{r}}}\right)^{2m_3+1}+\left(1+{u}_{{r}}\right)^{2m_3-1}\right]$ \\
\end{tabular}
\label{tbl:Dimless_Parameters}
\end{sidewaystable}

\subsection{Pressure,deformation, and stress profile}
\label{sec:FSI_Results}
In this section we move beyond the pure inflation problem and discuss the bfull fledged FSI between the hyperelastic tube and the Newtonian fluid flow conveyed therein.

\subsubsection{neo Hookean}
\begin{figure}[t]
\centering
\subfloat[]{\includegraphics[height=0.4\linewidth]{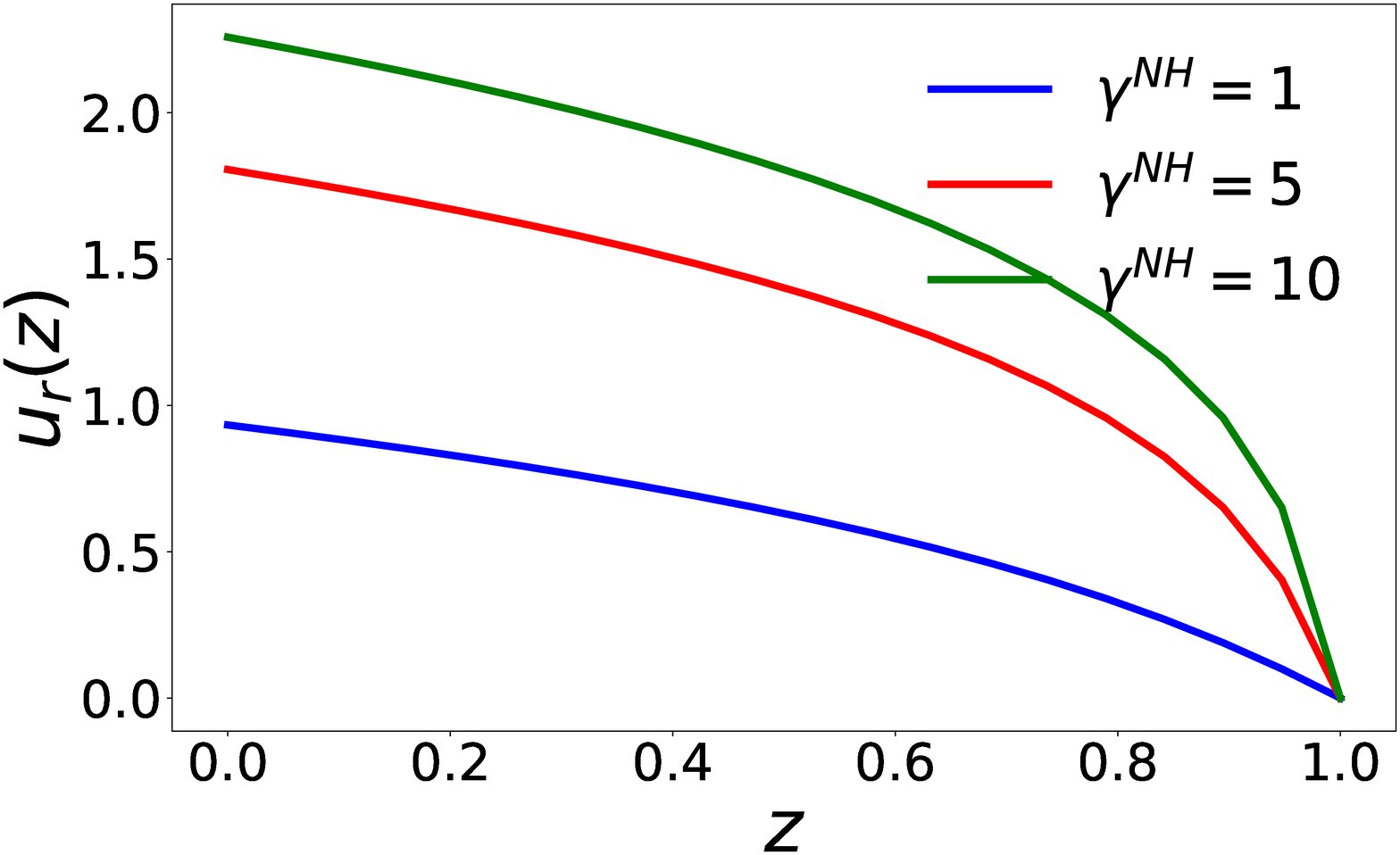}} \\
\subfloat[]{\includegraphics[height=0.4\linewidth]{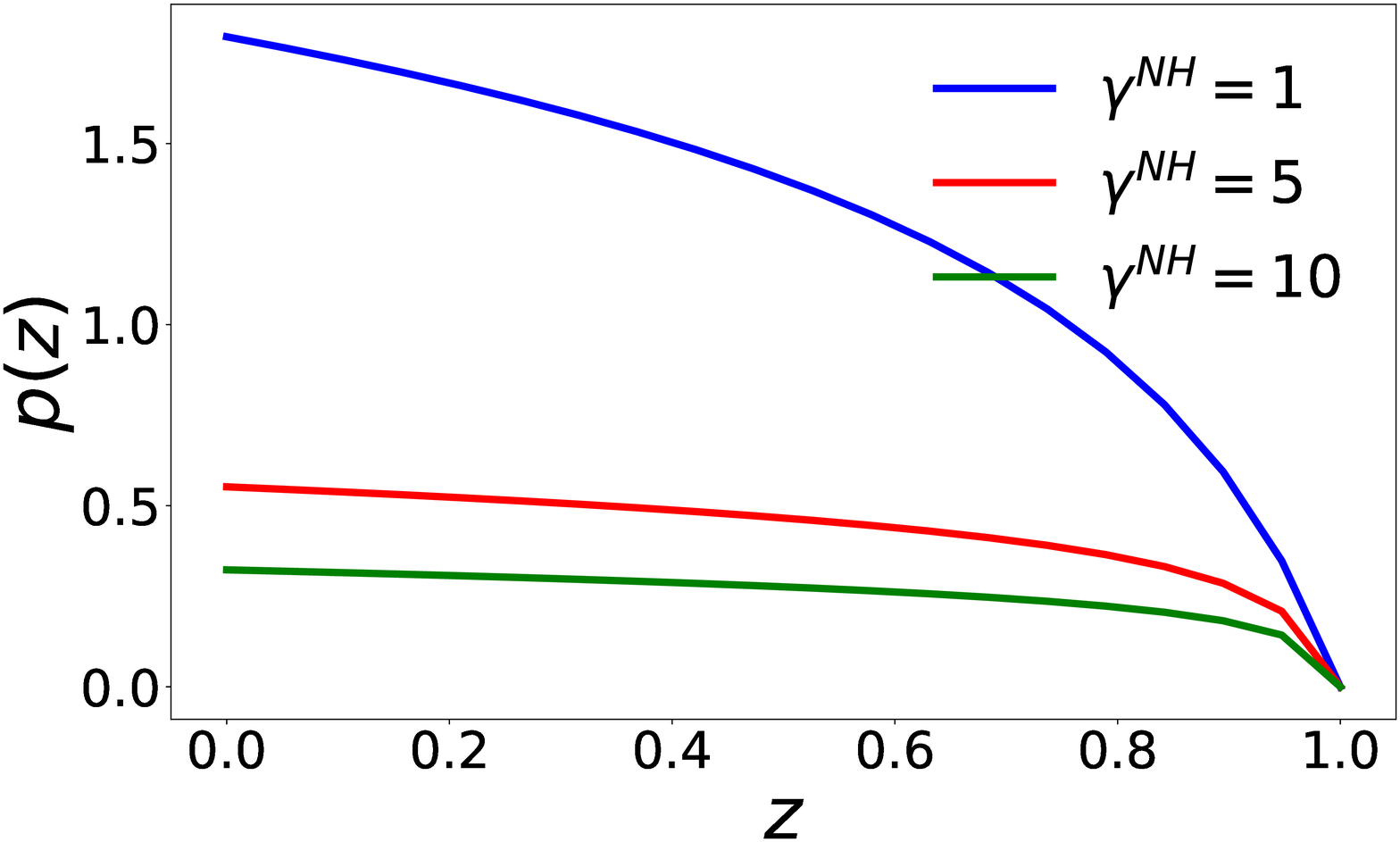}} \\
\subfloat[]{\includegraphics[height=0.4\linewidth]{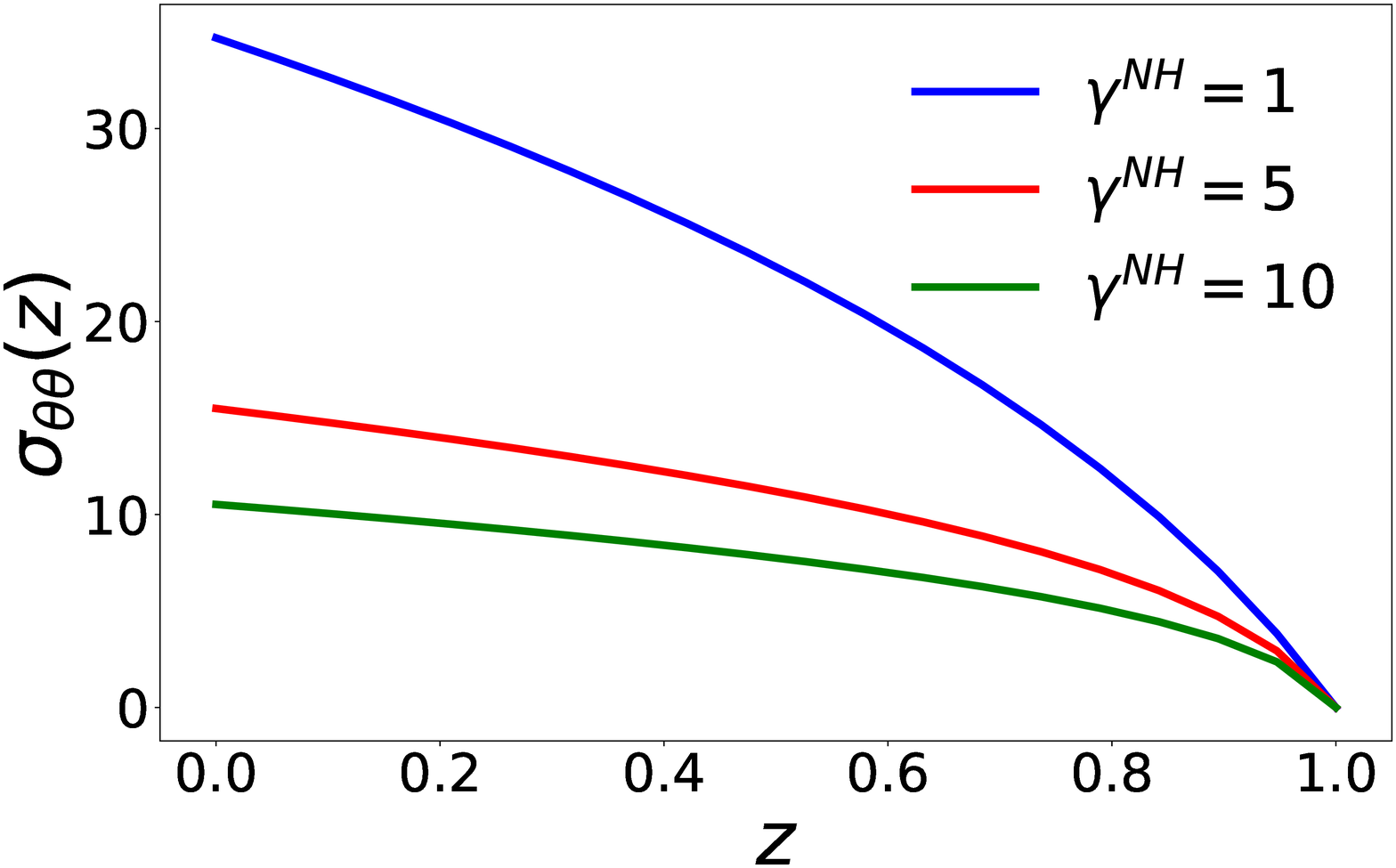}}
\caption{(a) Deformation profile, (b) pressure profile (c) circumferential structural stress profile for a hyperelastic tube constituted of neo Hookean material conveying Newtonian flow at steady state. }
\label{fig:neoHookean}
\end{figure}

The FSI characteristics of a hyperelastic tube constituted of neo Hookean material and carrying Newtonian fluid flow is shown in Fig .~\ref{fig:neoHookean} . The first observation we make for the deformation profile of the hyperelastic tube is that the deformation goes to zero at $z =1$, but not at $z =0$, where the deformation reaches a maximum value. This result seems conspicuous because the tube is clamped at both its ends $z =0$ and $z =1$. Our mathematical theory has not accounted for clamping at $z =0$ and therefore the same is reflected in the figure (Figs.~\ref{fig:neoHookean}(a),\ref{fig:Fung}(a),\ref{fig:Gent}(a) and \ref{fig:Ogden}(a)). In reality, due to clamping at $z =0$, there will be structural mechanical boundary layers near $z =0$ \cite{AC18b}, where $u_r$ will vary rapidly from $u_r =0$ to a maximum value in a short length of tube. In this boundary layer, bending becomes important and therefore the extent of boundary layer depends on tube thickness. However, for a thin tube like ourse (where $t/a \ll 1$) quantitatively the boundary layer has insignificant influence on the deformation and pressure profiles and has since been ignored.

Between the pressure and deformation profiles, the curves of different colors flip; obviously a smaller value of $\gamma^{NH}$ denotes a stiffer tube which deforms less and supports lower pressure. More importantly though, we observe at higher softness ($\gamma^{NH} =5,10$), the deformation profile shows a sharp change from its maximum value at the inlet, while the pressure profile is practically invariant. For a more precise understanding, observe the green curves in Fig.\ref{fig:neoHookean}(a) and (b). For the pressure profile the green curve remains almost parallel from $z =0$ to $z =0.8$, while in the same length of the tube, the deformation profile almost gets diminished by half. This behavior is clearly symptomatic of the limit  point /snap instability of the neo Hookean material model as discussed in Sec.\ref{sec:Results}\ref{sec:tube_law}. In fact, this behavior of the neo Hookean tube may be interpreted as a form of $\textit{strain softening}$, where the strain continues to rise even at near constant loads.

The profile for hoop stress follows the similar trend as that for pressure.

\subsubsection{Fung's model}

\begin{figure}[t]
\centering
\subfloat[]{\includegraphics[height=0.4\linewidth]{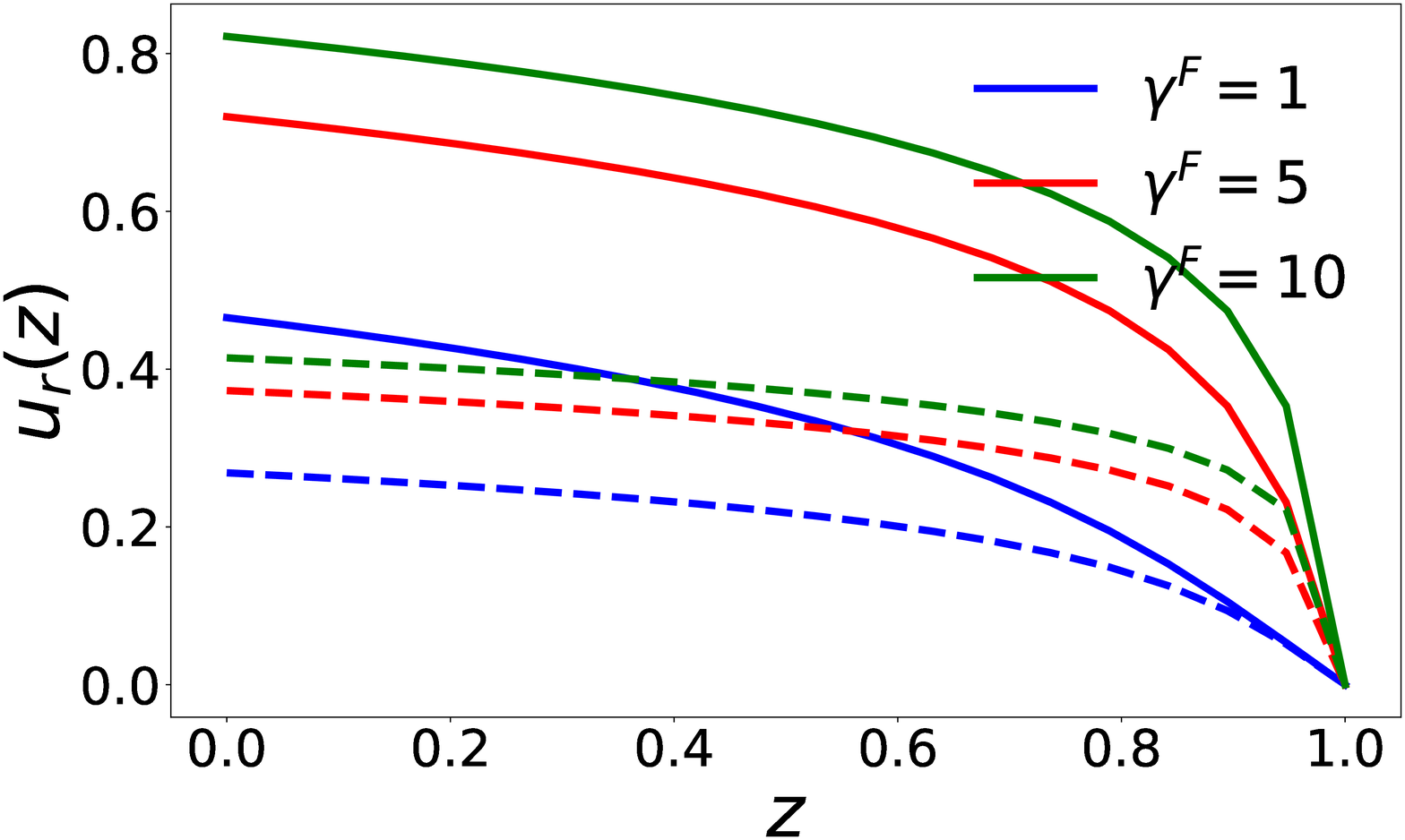}} \\
\subfloat[]{\includegraphics[height=0.4\linewidth]{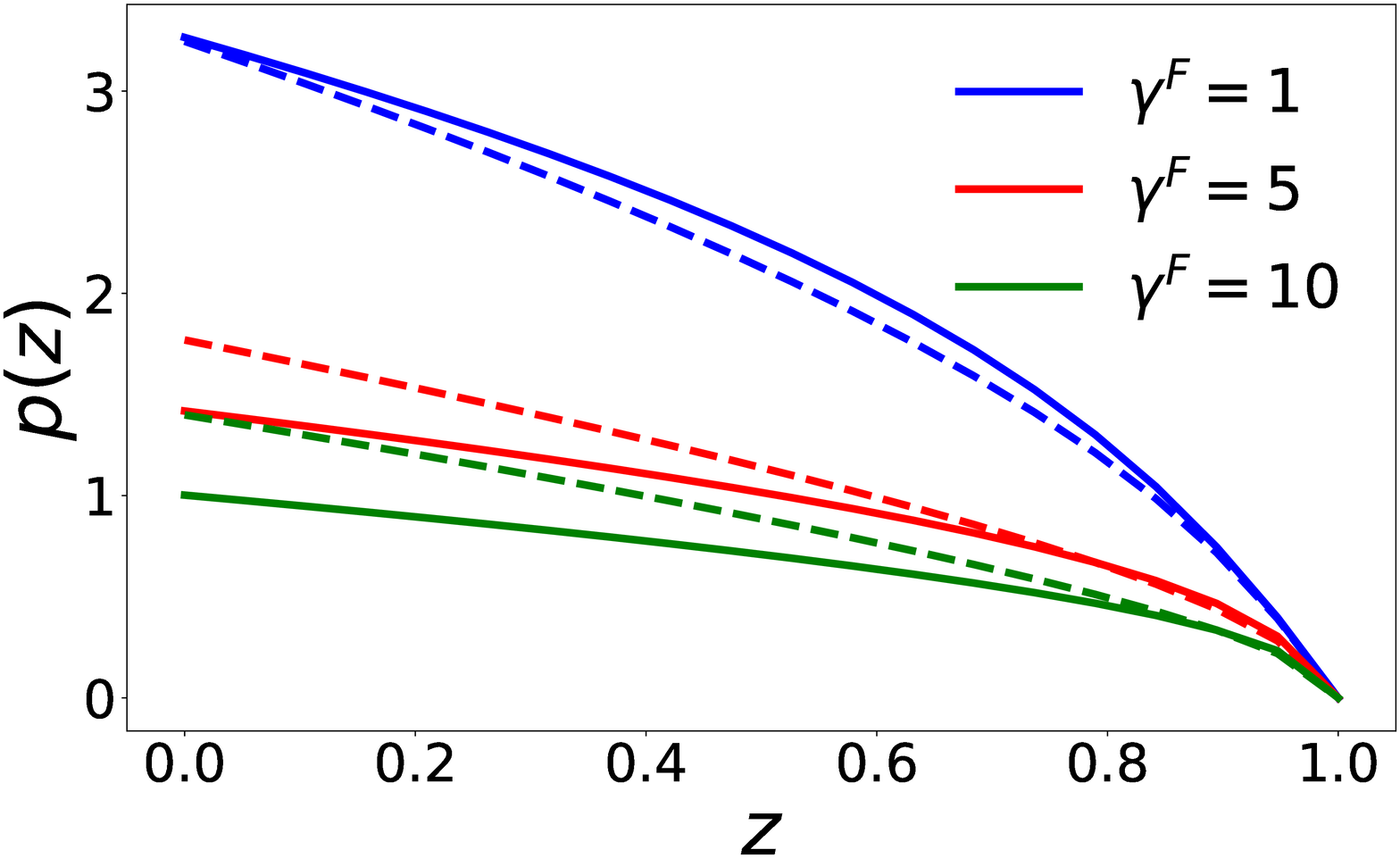}}\\
\subfloat[]{\includegraphics[height=0.4\linewidth]{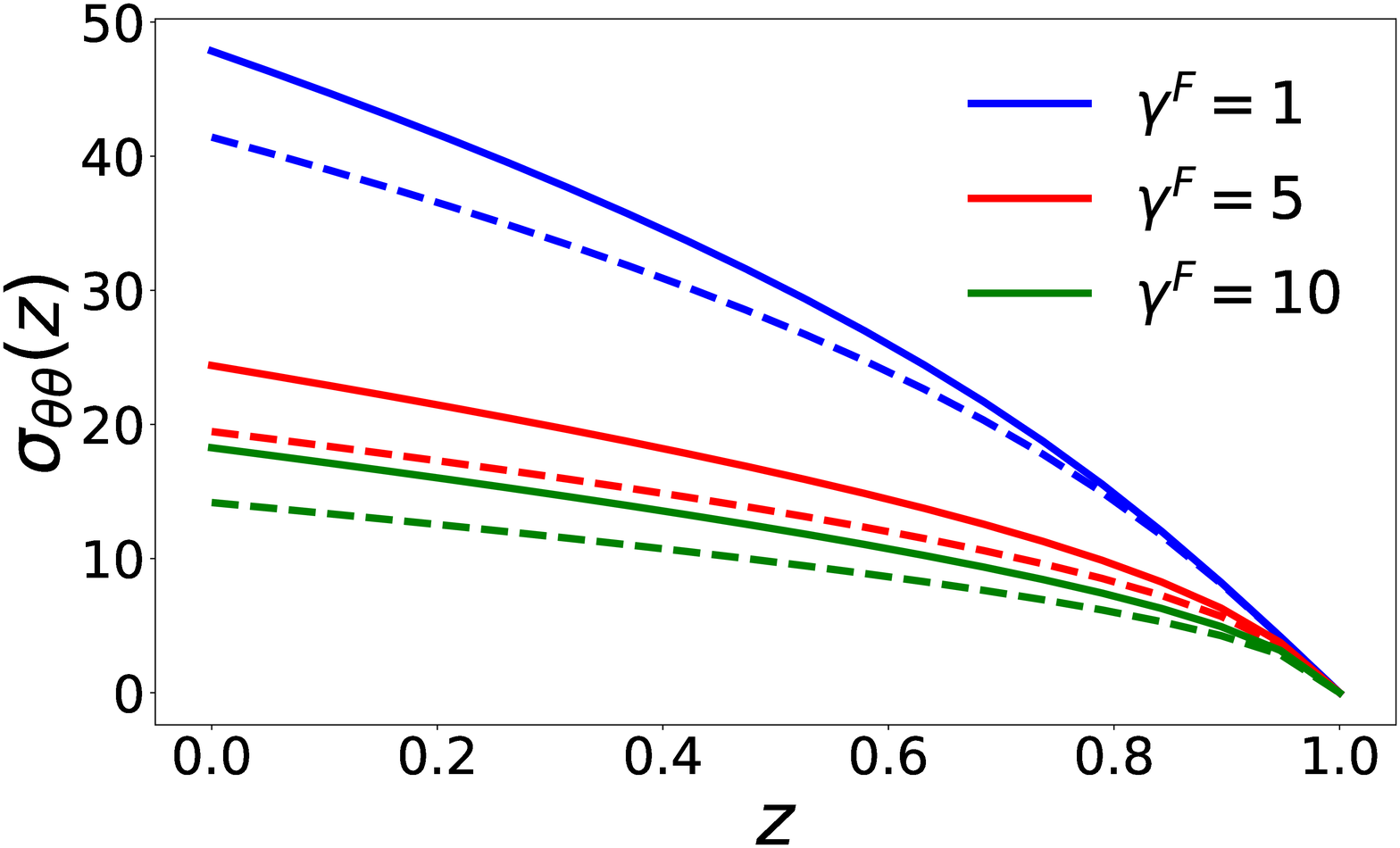}}
\caption{(a) Deformation profile, (b) pressure profile (c) circumferential structural stress profile for a hyperelastic tube constituted of Fung material conveying Newtonian flow at steady state. Solid curve represents $\alpha =1$, while dahsed curves are for $\alpha =5$.}
\label{fig:Fung}
\end{figure}

We discussed in Sec.~\ref{sec:Results}\ref{sec:tube_law}, which marked our foray into the pressure -deflection relationship for hyperelastic models, the model by Fung demonstrates the strongest strain hardening response among all the hyperelastic models studied. This quality of Fung's model has profound effect on FSI characteristics of the tube, both the deformation profile and the pressure/hoop stress profile (Fig.~\ref{fig:Fung}). The maximum deformation of the Fung's tube is much smaller than that of neo Hookean tube, an unmistakable consequence of the strain hardening response of the tube. A more vivid vivisection of the deformation profile in Fig.~\ref{fig:Fung}(a) additionally shows that deformation curve has a flatter slope  in Fung's tube than in neo Hookean tube; the Fung tube's deformation diminishes quite slowly from its maximum value at the inlet. To summarise, strain hardening not only curtails the maximum deformation of the tube but resists the change in the deformation across the tube length (see Fig.\ref{fig:Fung}(b)).

The pressure drop profile of Fung's tube exhibits trends reverse of that of deformation profile. Strain hardening precipitates much higher pressure drops in the Fung's tube compared to neo Hookean tube. Additionally, since all tubes must vent to the zero gauge pressure at the outlet, a strain hardened tube exhibits steeper pressure gradient across its length.

We already are aware that the parameter $\alpha$ controls the strain hardening response of Fung's model and higher $\alpha$ denotes higher strain hardening. However, the effect of $\alpha$ (strain hardening) diminishes as the tube becomes more rigid (at lower $\gamma^{NH}$) and has smaller deformations (near $z=1$). The gap between the solid curve ($\alpha =1$) and dashed curve ($\alpha =5$) diminishes as $\gamma^{NH}$ diminishes from $10$ to $1$. To understand this, we rewrite the strain energy functional for Fung's model as:
\begin{equation}
\label{eq:strain_energy_Fung}
    W =\frac{C}{2}\left(\mathcal{I}_1-3\right)+\left[\frac{C\left(e^{\alpha(\mathcal{I}_1-3)}-1\right)}{2\alpha}\right]
\end{equation}
Here, the term outside $[]$ is same as neo Hookean while the term inside $[]$ is the exponential term accounting for strain hardening. As the tube become stiff and the strain tends to zero:
\begin{equation}
    \lambda_i \to 1 ;\qquad \mathcal{I}_1 \to 3 ;
\end{equation}
 Therefore both, the exponential and neo Hookean, terms tend to zero but the exponential term goes to zero faster. It means that the tube's behavior becomes independent of $\alpha$, akin to a neo Hookean tube. This property of Fung's model helps explain the trend in Fig.\ref{fig:Fung}, especially the variation with respect to $\alpha$,i.e. the difference between dashed and solid curves.
 
 The variation of hoop stress across the tube length is shown in Fig.~\ref{fig:Fung}(c), and is observed to follow a trend similar to that of pressure.
 
\subsubsection{Gent's model}
\begin{figure}[t]
\centering
\subfloat[]{\includegraphics[height =0.4\linewidth]{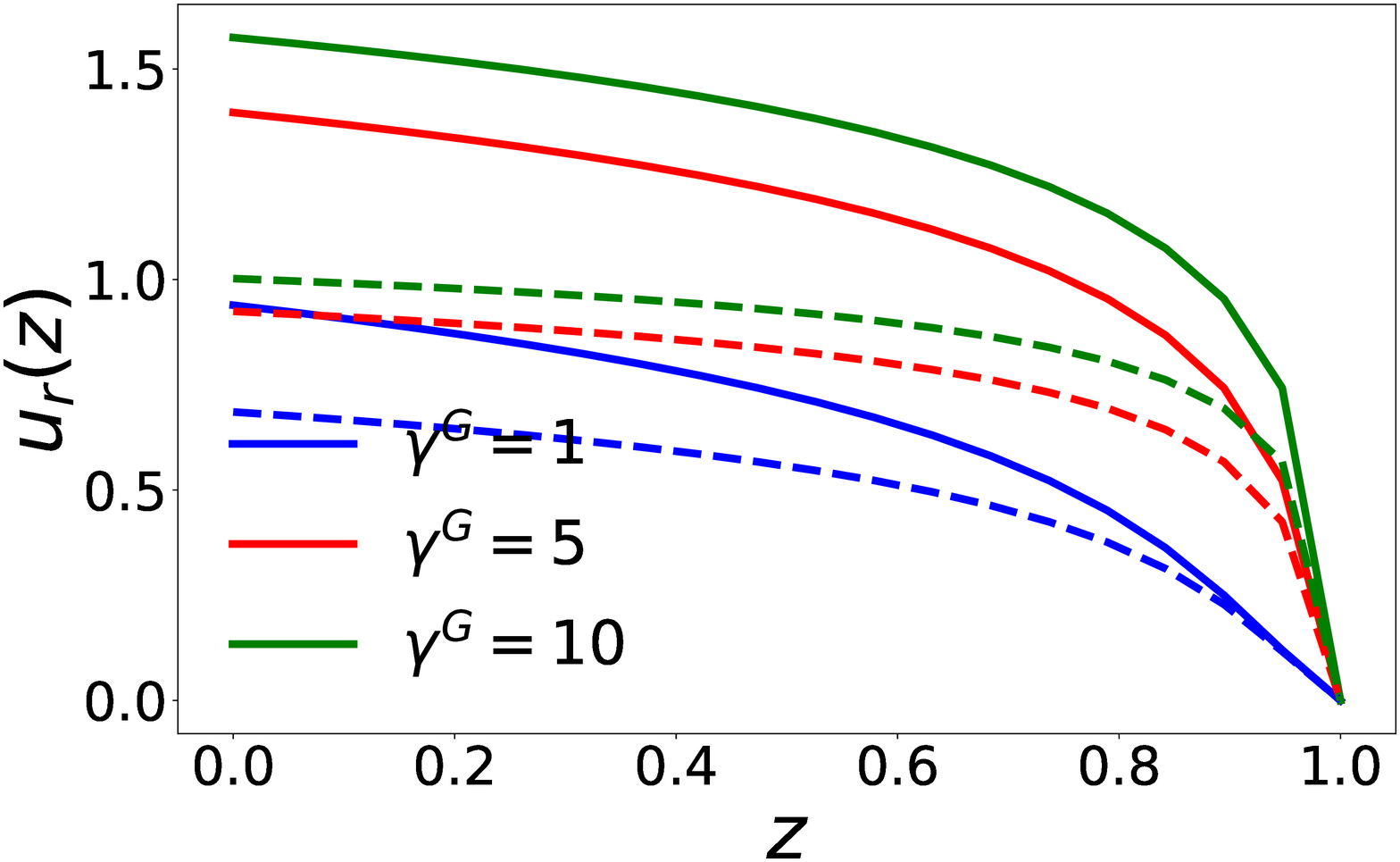}} \\
\subfloat[]{\includegraphics[height =0.4\linewidth]{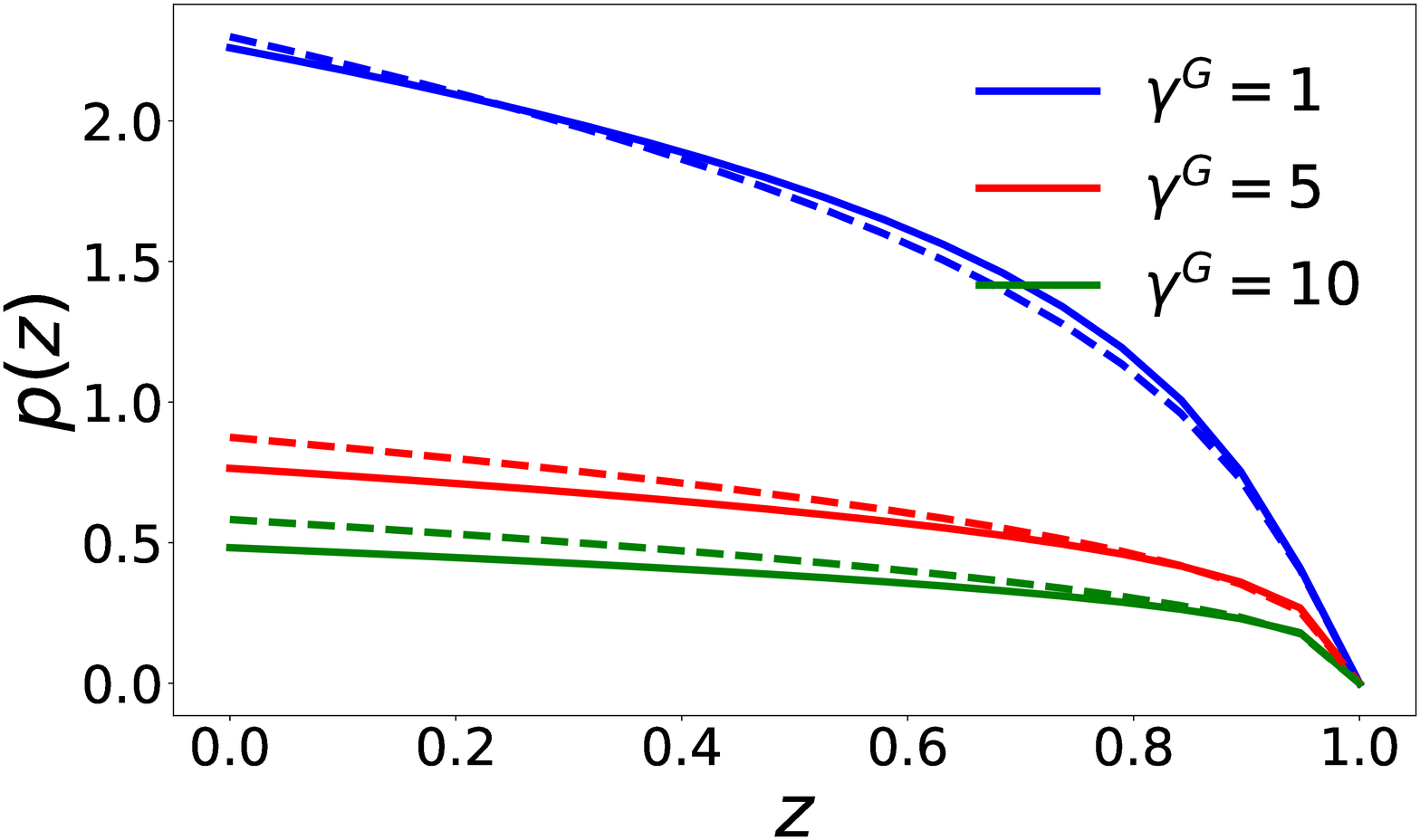}} \\
\subfloat[]{\includegraphics[height =0.4\linewidth]{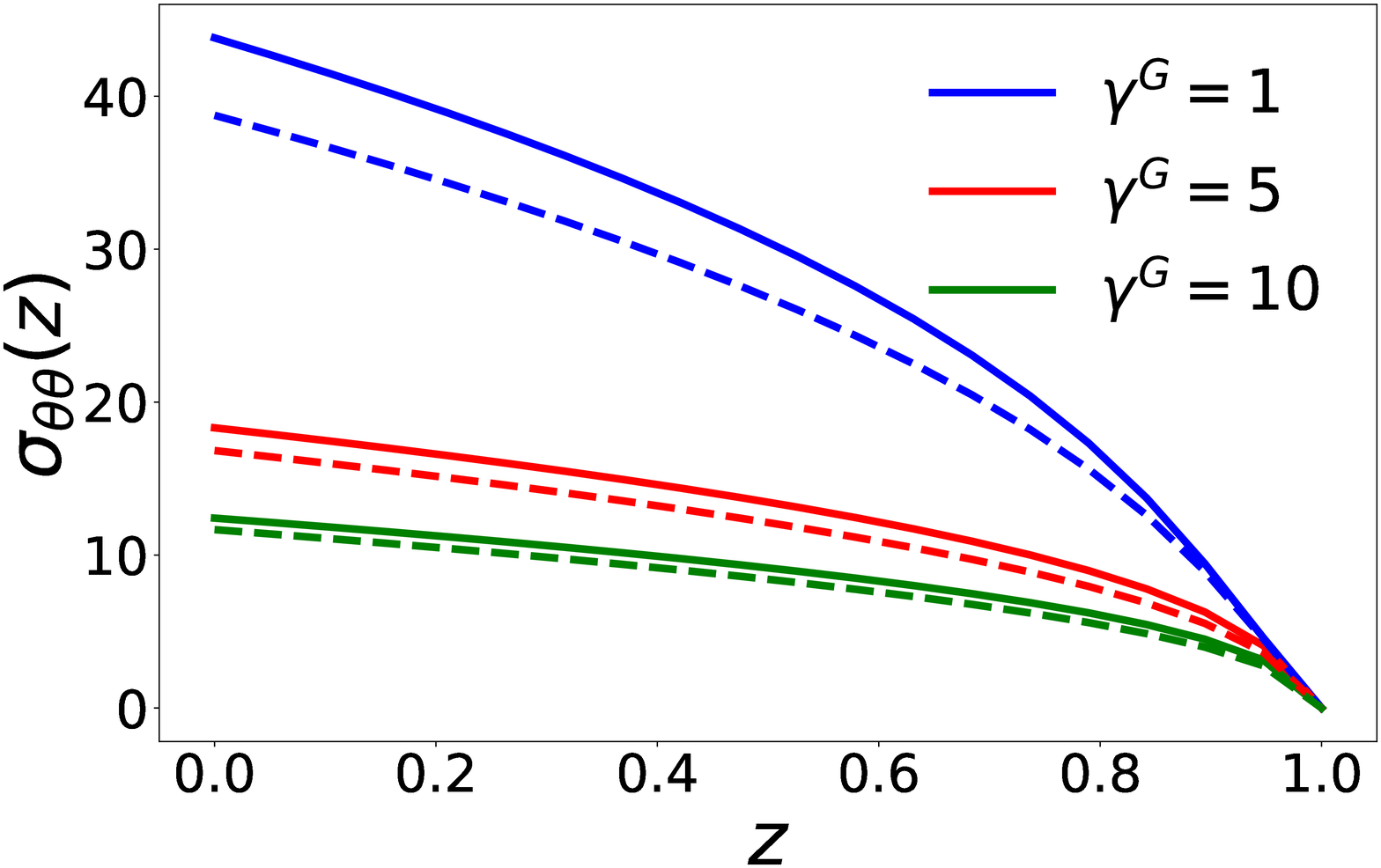}}

\caption{(a) Deformation profile, (b) pressure profile (c) circumferential structural stress profile for a hyperelastic tube constituted of Gent material conveying Newtonian flow at steady state. Solid curves represent $\eta =0.1$, while dashed curves represent $\eta =0.3$.}
\label{fig:Gent}
\end{figure}
Next, in Fig.~\ref{fig:Gent}, we discuss the FSI characteristics of a tube constituted of Gent's material. 

Overall, we observe that the trends of Gent's tube model are similar to that of Fung's tube. However, the Gent's tube does not show as strong as strain hardening as does the Fung's tube. 

The deformation profile of Gent's tube in Fig.~\ref{fig:Gent}(a) ,shows a smaller maximum deformation at the inlet than the Fung's tube but has flatter slope across the tube length similar to the Fung's tube model. 
In the pressure profile shown in in Fig.~\ref{fig:Gent}(b), we observe that the maximum pressure drop is smaller than that of Fung, but the pressure curve stills shows a steep decline compared to the neo Hookean tube in Fig.~\ref{fig:neoHookean}(b). These trends are tell-tale signs of the  strain hardening property entrenched inside the Gent's model.

In the Gent's model, the parameter $\eta$ is responsible for strain hardening response. From the Fig~\ref{fig:Gent} we observe that for small deformations, when the tube becomes stiff ($\gamma^{G} =1$) or near the outlet $z =1$, the influence of $\eta$ diminishes and the dashed and the solid curves begin to come together for both pressure and deformation profiles. To understand this trend, we rewrite the strain energy functional for Gent's model as below:

\begin{equation}
\label{eq:strain_energy_Gent}
    W =-\frac{C}{2\eta}\text{ln}\left[1-\eta\left(\mathcal{I}_1-3\right)\right]
\end{equation}
A Taylor's series expansion of the logarithmic term and subsequent simplification yields the following:
\begin{equation}
  \label{eq:strain_energy_Gent2}
    W =\frac{C}{2}\left[\left(\mathcal{I}_1-3\right)+\eta\left(\mathcal{I}_1-3\right)^2+\eta^2\left(\mathcal{I}_1-3\right)^3 +\dots\right]
\end{equation}  
Now, as before, for small strains $\lambda_i \to 1$, and $\mathcal{I}_1 \to 3$,and therefore all the terms inside $[]$ go to zero. However, the higher degrees of $\left(\mathcal{I}_1-3\right)$ go to zero faster and therefore we are left with the following limiting expression:
\begin{equation}
  \label{eq:strain_energy_Gent3}
    W =\frac{C}{2}\left[\left(\mathcal{I}_1-3\right)\right]
\end{equation}  
which as we see is an expression independent of $\eta$. This result explains that as the tube becomes stiffer for lower values of $\gamma^{G}$ (or near $z=1$ where the deformation is smaller), the curves pertaining to different values of $\eta$ (dashed and solid curves in Fig~\ref{fig:Gent}) tend to come closer.

\subsubsection{$\text{Ogden}_3$ model}

As discussed in Sec.\ref{sec:StructMech}\ref{sec:consti_Ogden}, Ogden's model is versatile where the astonishingly high number of material parameter ($6$ for $N=3$) can be fine tuned to suit an array of material response ranging from vulcanised rubber \cite{Ogden_main} to aorta \cite{Breslavsky_plate1} to soft tissues \cite{Mihai2015}.

On the flip side, the utilitarian allure of the $\text{Ogden}_{3}$ model also robs it of an originality. Since this model unequivocally and unabashedly mimics the characteristics of the material, any conclusions regarding the mechanical response of the model $\textit{per se}$ without taking into account the material it models (which in turn enforces the value of parameters), is whimsical and wishful. Therefore, to isolate and analyse the characteristics of $\text{Ogden}_{3}$ model, we must also specify what material it is modeling.

Our narrative of material modeling in this paper till now has wavered between two extremes. On one end of the spectrum, we have the rubber like materials modeled by neo Hookean and Mooney Rivlin models, and on the other end of the spectrum we have fibrous biological tissues like aorta, ligaments and tendons, modeled by Fung's and Gent's strain hardening response.

The versatility of $\text{Ogden}_{3}$ model then allows us to execute an exciting trade off. For $\text{Ogden}_{3}$ model, we choose a material which exhibits a milder version of strain hardening in comparison to ligaments and tendons, but is also soft like rubber. An ideal candidate is the brain and fat tissues \cite{Mihai2015}.

\begin{figure}[t]
\centering
\subfloat[]{\includegraphics[height =0.4\linewidth]{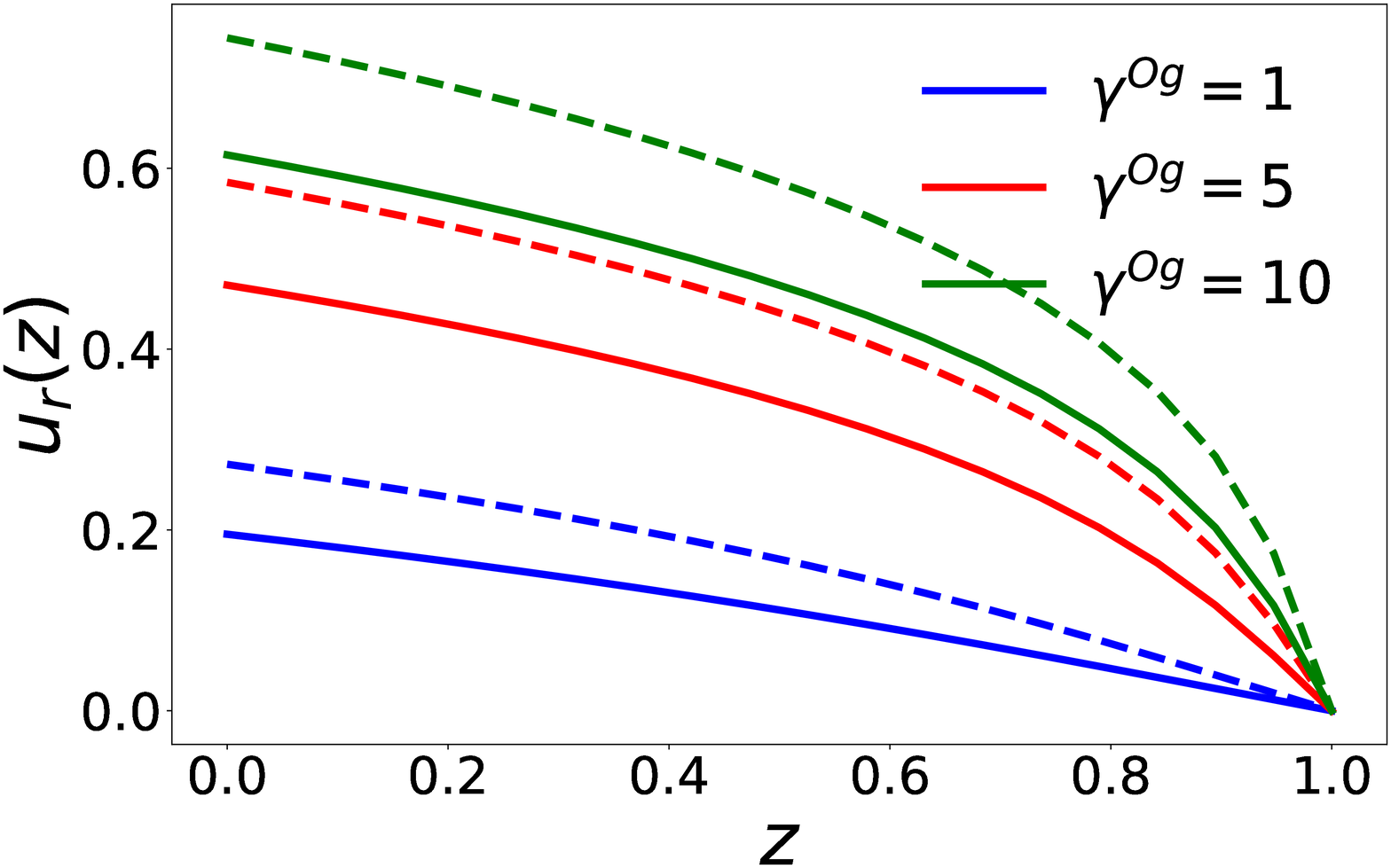}}\\
\subfloat[]{\includegraphics[height =0.4\linewidth]{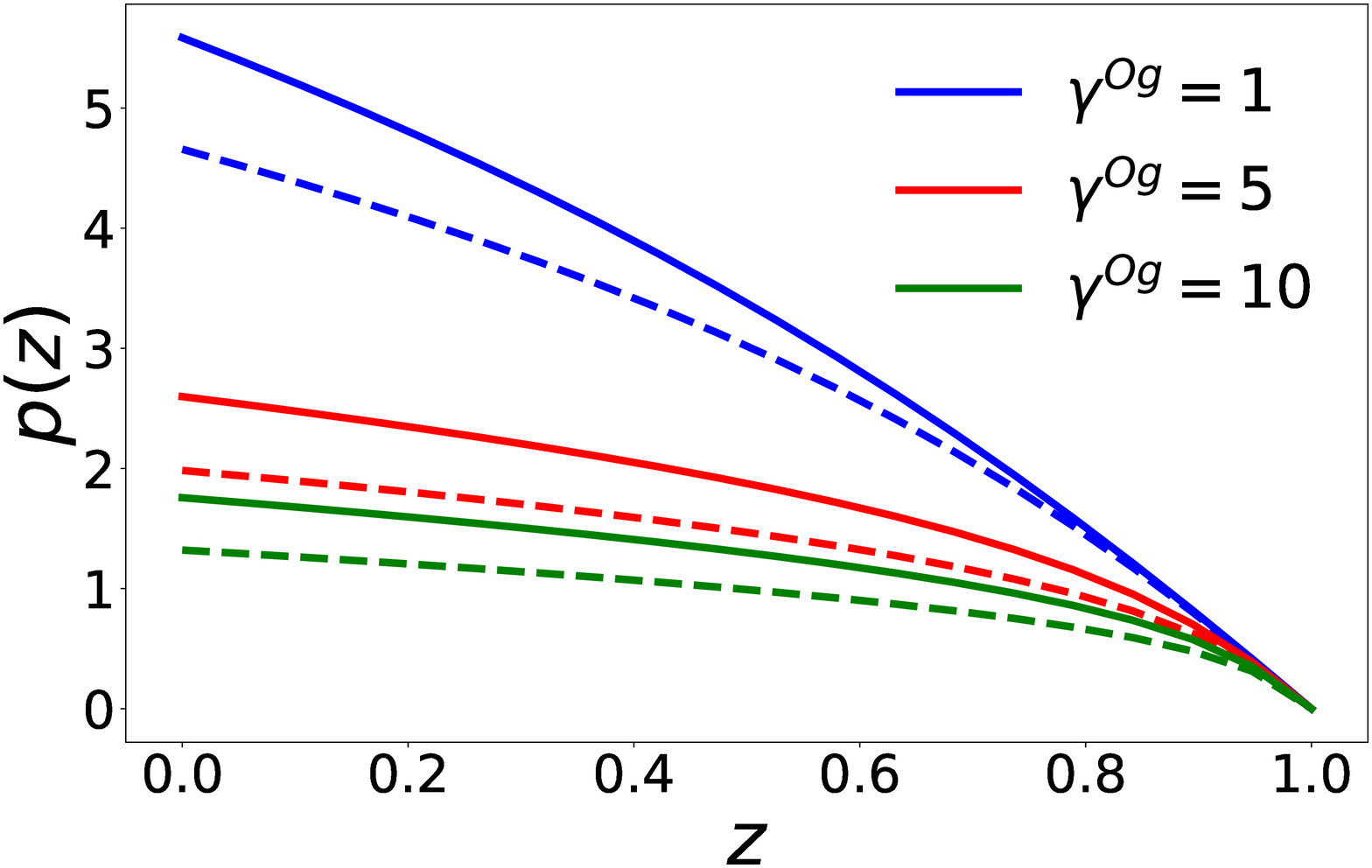}} \\
\subfloat[]{\includegraphics[height =0.4\linewidth]{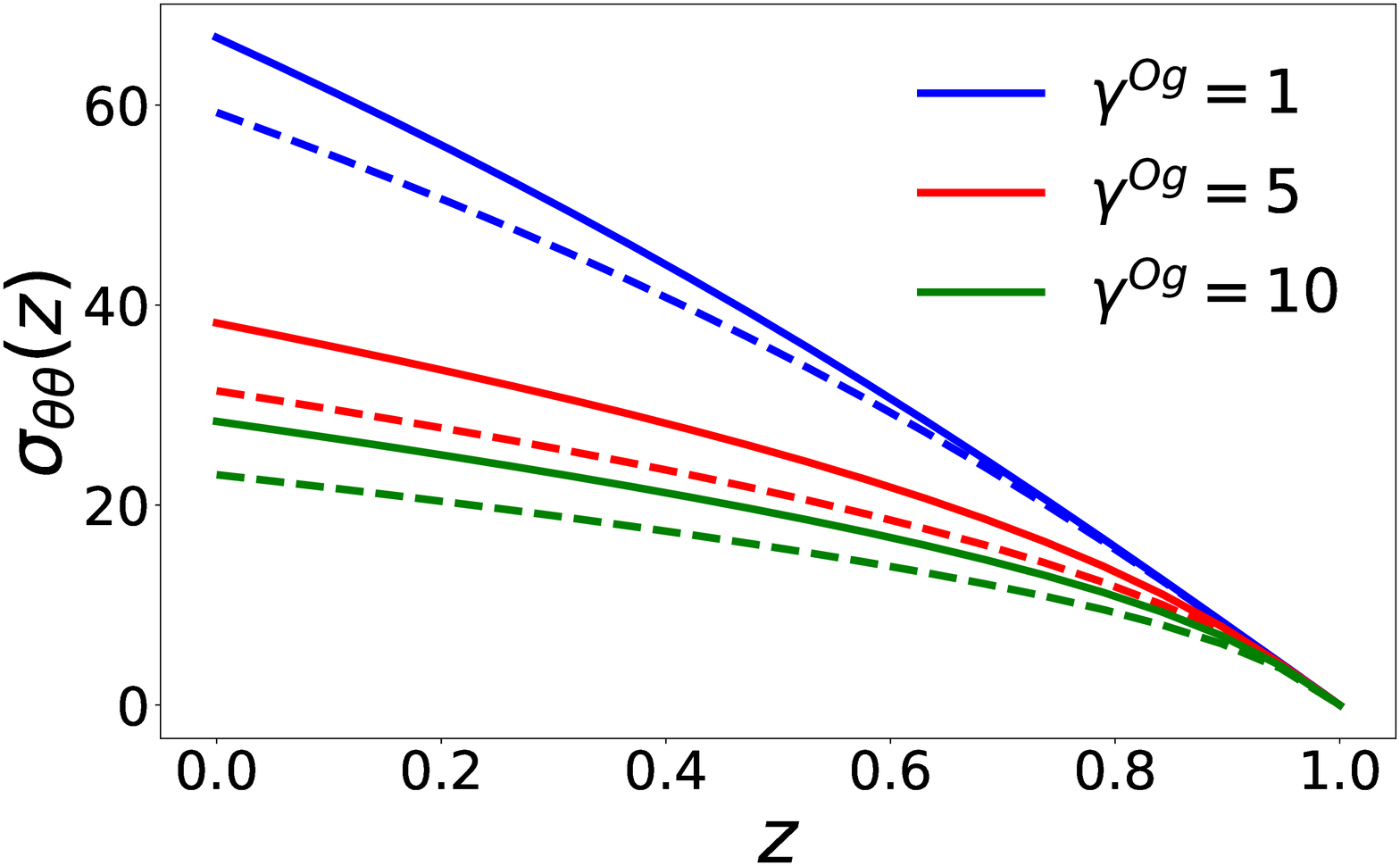}}
\caption{(a) Deformation profile, (b) pressure profile (c) circumferential structural stress profile for a hyperelastic tube constituted of $\text{Ogden}_{3}$  material conveying Newtonian flow at steady state. Solid curves represent $C_1= -3543,C_2 =-2723,C_3 =654,m_1 =1,m_2 =-1,m_3 =2$ , dashed curves represent $C_1= -3882,C_2 =-2113,C_3 =931,m_1 =1,m_2 =-1,m_3 =2$}
\label{fig:Ogden}
\end{figure}

For these tissues, an $\text{Ogden}_{3}$ model has already been fit and necessary parameters derived \cite{Mihai2015}, and we use these parameters as they are. The results are shown in Fig.~\ref{fig:Ogden}. The solid lines correspond to brain tissue as given in Table $3$ of \cite{Mihai2015}, while the dashed lines correspond to fat tissue as given in Table $4$ of \cite{Mihai2015}.

It is easy to conclude that the $\text{Ogden}_{3}$ model shows strain stiffening but to a far lesser extent than the Fung and even the Gent's model. The deformation and pressure profile for $\gamma^{Og} =1$ is almost linear like that of a linearly elastic tube. The profiles tend to become flatter and strain hardening more severe as the flow rate increases ($\gamma^{Og}$ becomes higher). For the choice of parameters, the fat tissue (as marked by dashed lines) are much softer and therefore undergo larger deformation and support smaller deformation than the brain tissues (shown by solid lines).

\subsection{Sensitivity Analysis: Geometric vs. material nonlinearity}

The nonlinearity in the theory of finite elasticity as employed in this paper has three attributes:
\begin{enumerate}
\item	The material law governing the relationship between stress and strain is nonlinear.
\item	The strain measure used in the material law is the  Green Langrangian strain ($\doubleunderline{\mathcal{E}}$) and not the small strain \doubleunderline{e}.
\item	The equations of equilibrium are written in deformed coordinates in terms of Cauchy stress, and not in the material coordinates in terms of first Piola Kirchoff stress tensor.
\end{enumerate}
Out of the above three (3) points, point 1 is related to material nonlinearity while point 2 and point 3 are borne out of geometric nonlinearity of the problem. Material nonlinearity is built in the material model itself and independent of the strains being small or the choice of strain measure. The only hyperelastic model which has linear constitutive equation is St Venant  model \cite{Bonet_Book}, which has limited practical use and is outside the purview of our analysis.

We intend to derive a FSI theory of finite elasticity which is geometrically linear. The reason behind this thought experiment are two fold. First, the practical reason is that a linear theory is easy to use and apply and makes the job of predicting pressure drop and deformation profile easier for the engineer/researcher. Second, the more philosophical reason is that we want to understand the how dependent each material model is on geometric nonlinearity to capture the true physics of hyperelastic tube undergoing steady state FSI. Such knowledge will help us to prepare guidelines whether or not to account for geometric nonlinearity in FSI modeling of hyperelastic tubes.

To derive a geometrically linear theory, we need to relax the constraints imposed by point(2) and point(3).  Replacement of $\doubleunderline{\mathcal{E}}$ by $\doubleunderline{e}$ will entail re-derivation of all the hyperelastic laws from scratch, and then using the new hyperelastic laws to establish new pressure deformation relationships - clearly a laborious and perhaps tedious enterprise. On the other hand, the constraint of point (2) above is automatically satisfied if:
\begin{equation}
    \doubleunderline{\mathcal{E}} \approx \doubleunderline{{e}} \qquad \Longleftrightarrow \qquad \lambda_i \approx  1 \qquad \Longleftrightarrow \qquad u_r \ll 1,
\end{equation}
i.e the deformation/strains is small.
Therefore, we will derive the geometrically linear theory, only by relaxing the constraint 3 explicitly.

For \textbf{neo Hookean}, Eq.~\eqref{eq:neoHookean_equi_large} in original coordinates gives us:
\begin{equation}
\label{eq:neoHookean_equi_small}
    -\bar{p}\bar{a}/t =C\left[\left(\frac{a}{\bar{R}}\right)^2-\left(\frac{\bar{R}}{a}\right)^2\right],
\end{equation}
which after some manipulation yields:
\begin{equation}
    \gamma^{nH}{p} =\left[\left(1+{u}_r\right)^2-\left(\frac{1}{1+{u}_r}\right)^2\right] ;\quad \gamma^{nH} =\frac{\mathcal{P}_c}{C}\frac{a}{t},
    \label{eq:Pressure_Deformation_Dimless_Neo_Hookean_small}
\end{equation}
which may be termed as a geometrically linear counterpart to Eq.~\eqref{eq:Pressure_Deformation_Dimless_Neo_Hookean}. 

For \textbf{Fung's} model, the equation of equilibrium Eq.~\eqref{eq:equilibrium_Fung_Large} when written in terms of undeformed coordinates gives us:
\begin{equation}
    -\frac{\bar{p}a}{t} =C\left[\left(\frac{\bar{r}}{\bar{R}}\right)^2-\left(\frac{\bar{R}}{\bar{r}}\right)^2\right]\left[1+e^{\alpha(\frac{a}{\bar{R}}-\frac{\bar{R}}{a})^2}\right],
\end{equation}
which after some minor algebraic manipulations gives us:
\begin{equation}
\gamma^{F}{p} =\left[\left(1+{u}_r\right)^2-\left(\frac{1}{1+{u}_r}\right)^2\right]\left[1+e^{\alpha(\frac{1}{1+u_r}-(1+u_r))^2}\right] ;\quad \gamma^{F} =\frac{\mathcal{P}_c}{C}\frac{a}{t},
    \label{eq:Pressure_Deformation_Dimless_Fung_small}
\end{equation}

For \textbf{Gent}'s model, the equation of equilibrium Eq.~\eqref{eq:Equilibrium_Large_Gent} when expressed in terms of original, undeformed coordinates gives us:
\begin{equation}
\label{eq:Equilibrium_Small_Gent}
    \frac{\bar{p}a}{t}=\frac{C\left[\left(\frac{\bar{R}}{a}\right)^2-\left(\frac{a}{\bar{R}}\right)^2\right]}{[1-\eta(\frac{a}{\bar{R}}-\frac{\bar{R}}{a})^2]},
\end{equation}
which when simplified and rendered dimensionless gives us:
\begin{equation}
\gamma^{G}{p} =\frac{\left[\left(1+{u}_r\right)^2-\left(\frac{1}{1+{u}_r}\right)^2\right]}{\left[1-\eta{(\frac{1}{1+u_r}-(1+u_r))^2}\right]} ;\quad \gamma^{G} =\frac{\mathcal{P}_c}{C}\frac{a}{t},
    \label{eq:Pressure_Deformation_Dimless_Gent_small}
\end{equation}

Finally, the equation of equilibrium for \textbf{$\text{Ogden}_{3}$} model, when referred to in original coordinates give us, from Eq.~\eqref{eq:Equilibrium_Large_Ogden}:
\begin{equation}
\label{eq:Equilibrium_Small_Ogden}
    -\frac{\bar{p}\bar{a}}{t}=C_1\left[\left(\bar{r}/\bar{R}\right)^{2m_1}-\left(\bar{R}/\bar{r}\right)^{2m_1}\right]+C_2\left[\left(\bar{r}/\bar{R}\right)^{2m_2}-\left(\bar{R}/\bar{r}\right)^{2m_2}\right]+C_3\left[\left(\bar{r}/\bar{R}\right)^{2m_3}-\left(\bar{R}/\bar{r}\right)^{2m_3}\right], 
\end{equation}
which when simplified and expressed in dimensionless variables yields:
\begin{multline}
    \gamma{p}=\frac{pa}{tC_1}=\left[-\left(1/(1+{u}_{{r}})\right)^{2m_1}+\left(1+{u}_{{r}}\right)^{2m_1}\right]+ \bar{C}_{2}\left[-\left(1/(1+{u}_{{r}})\right)^{2m_2}+\left(1+{u}_{{r}}\right)^{2m_2}\right]\\+\bar{C}_{3}\left[-\left(1/(1+{u}_{\bar{r}})\right)^{2m_3}+\left(1+{u}_{{r}}\right)^{2m_3}\right], \\ \gamma = \frac{\mathcal{P}_c}{C_1}\frac{a}{t}\quad \bar{C}_{2}=\frac{C_2}{C_1} \quad \bar{C}_{3} =\frac{C_3}{C_1}
    \label{eq:pressure_deformation_dimless_Ogden_small}
\end{multline}

To summarise, the expression relating the pressure to deformation for geometrically linear hyperelastic tube models are given by Eqs.~\eqref{eq:Pressure_Deformation_Dimless_Neo_Hookean_small},\eqref{eq:Pressure_Deformation_Dimless_Gent_small},\eqref{eq:Pressure_Deformation_Dimless_Fung_small},\eqref{eq:pressure_deformation_dimless_Ogden_small} for neo Hookean, Gent, Fung and $\text{Ogden}_{3}$ models respectively. These equations individually need to be solved in tandem with the flow equation Eq.~\eqref{eq:FluidMechanics} to obtain the solutions for pressure, deformation and hoop stress profile for the hyperelastic tube in geometrically linear FSI with Newtonian fluid flow at steady state. The results of this exercise are discussed now.

\begin{figure}[t]
\centering
\subfloat[]{\includegraphics[width =0.4\linewidth]{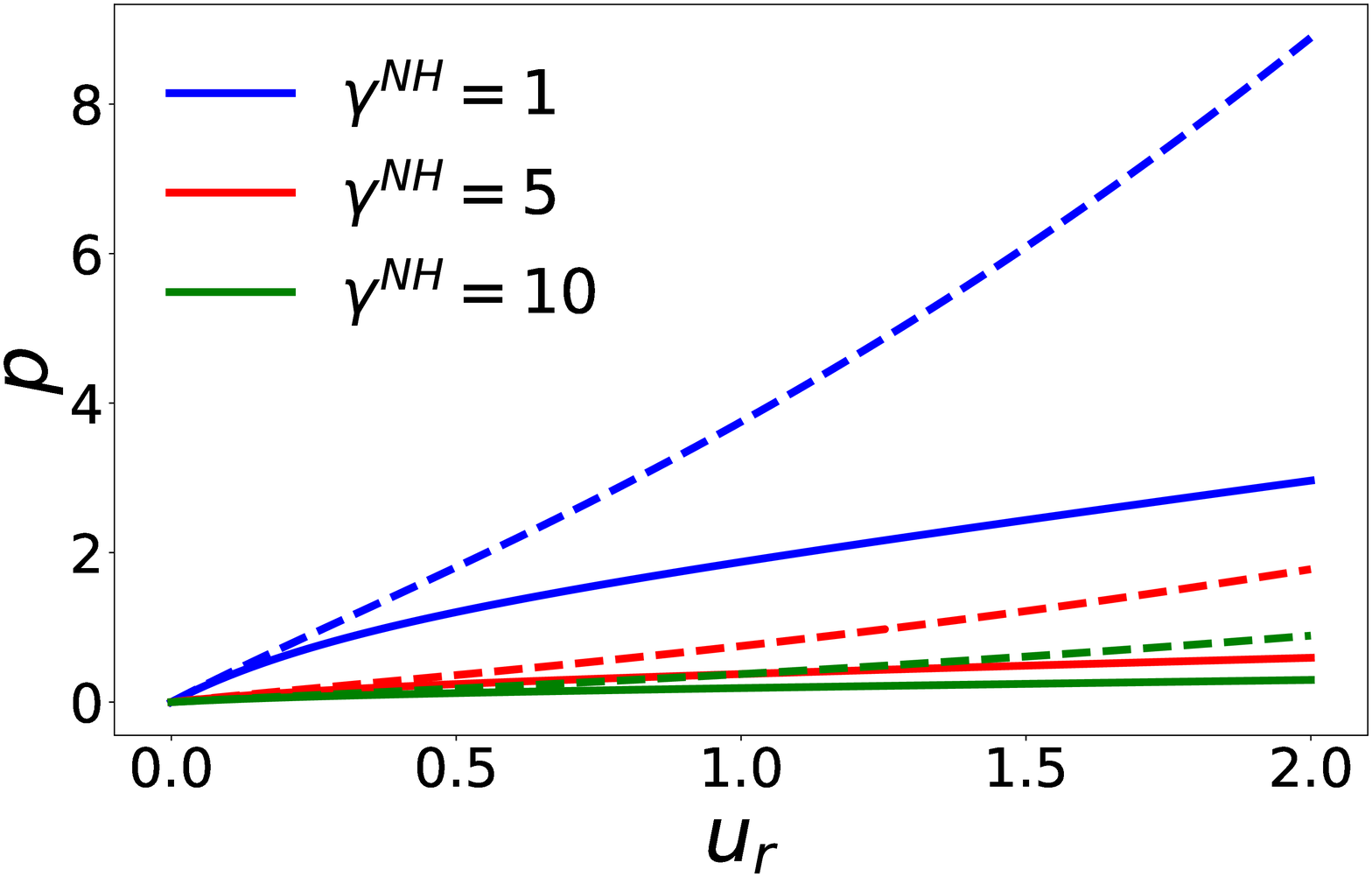}}
\subfloat[]{\includegraphics[width =0.4\linewidth]{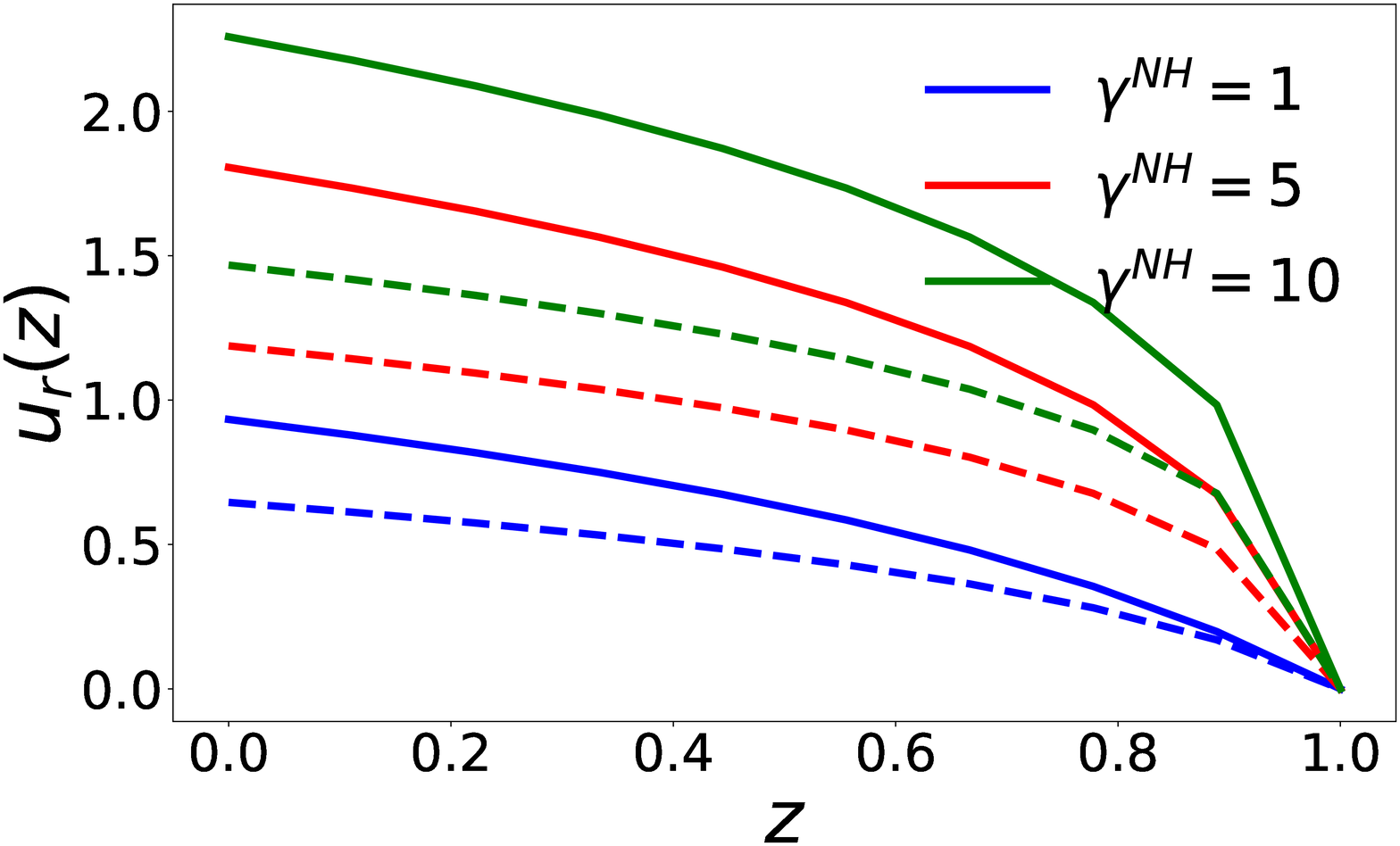}} \\
\subfloat[]{\includegraphics[width =0.4\linewidth]{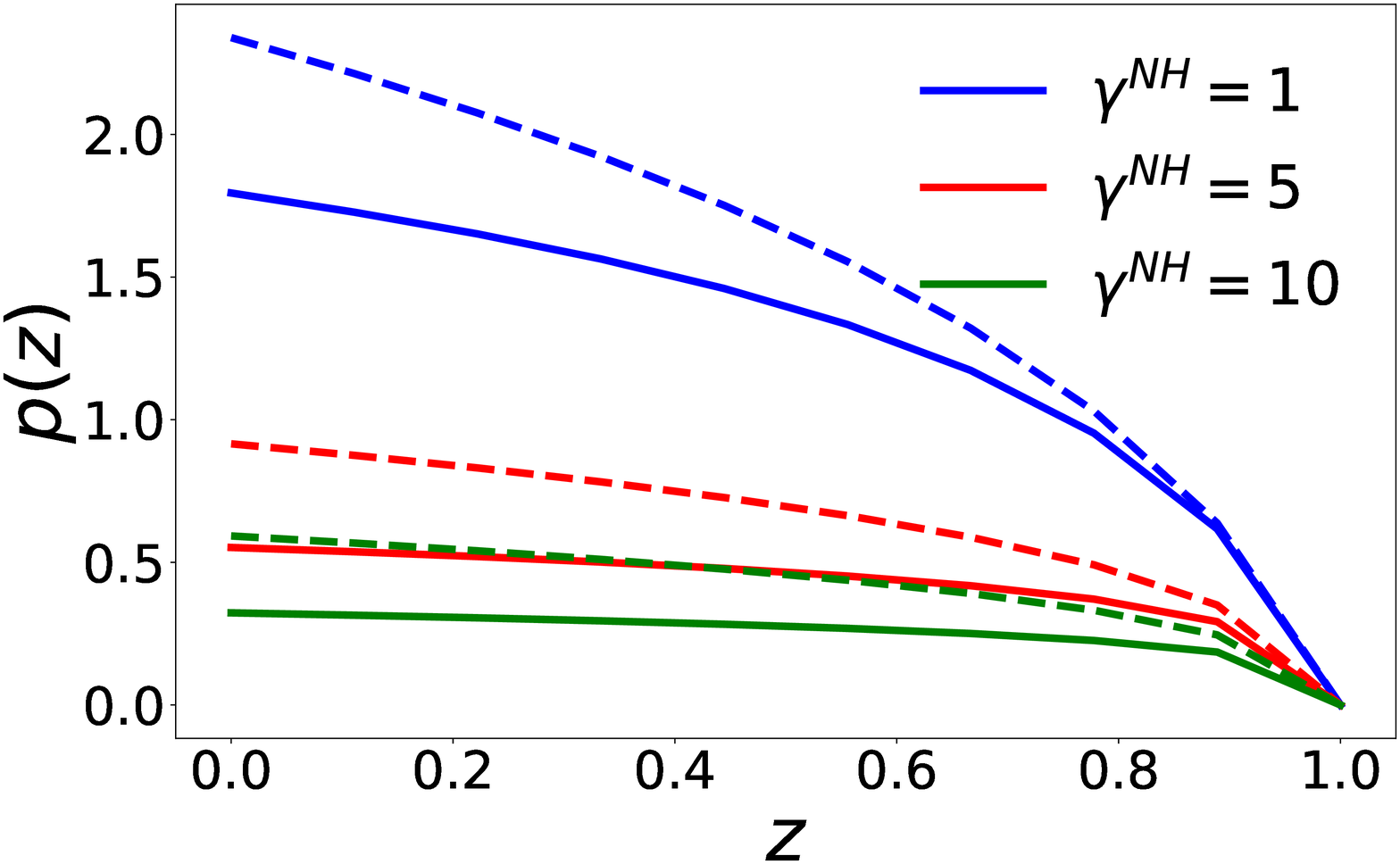}} 
\subfloat[]{\includegraphics[width =0.4\linewidth]{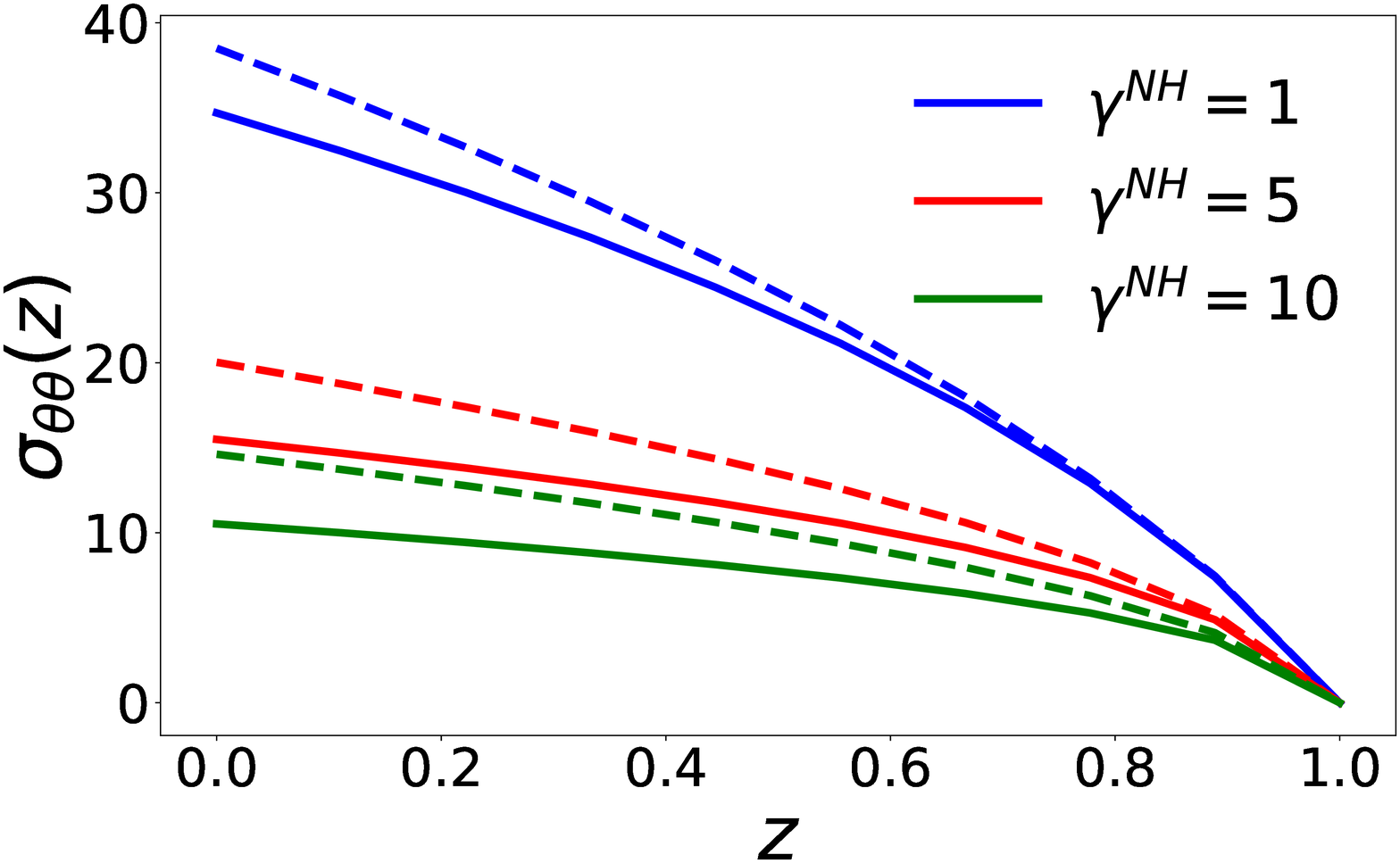}} 
\caption{(a) Pressure deflection curves (b) deformation profile, (b) pressure profile (c) circumferential structural stress profile for a hyperelastic tube constituted of neo Hookean material conveying Newtonian flow at steady state. Solid curves account for large strains and deformations (both material and geometrical nonlinearity) while dashed curves account for small strains and deformations (only material nonlinearity)}
\label{fig:neoHookean_Small}
\end{figure}

\begin{figure}[t]
\centering
\subfloat[]{\includegraphics[width =0.4\linewidth]{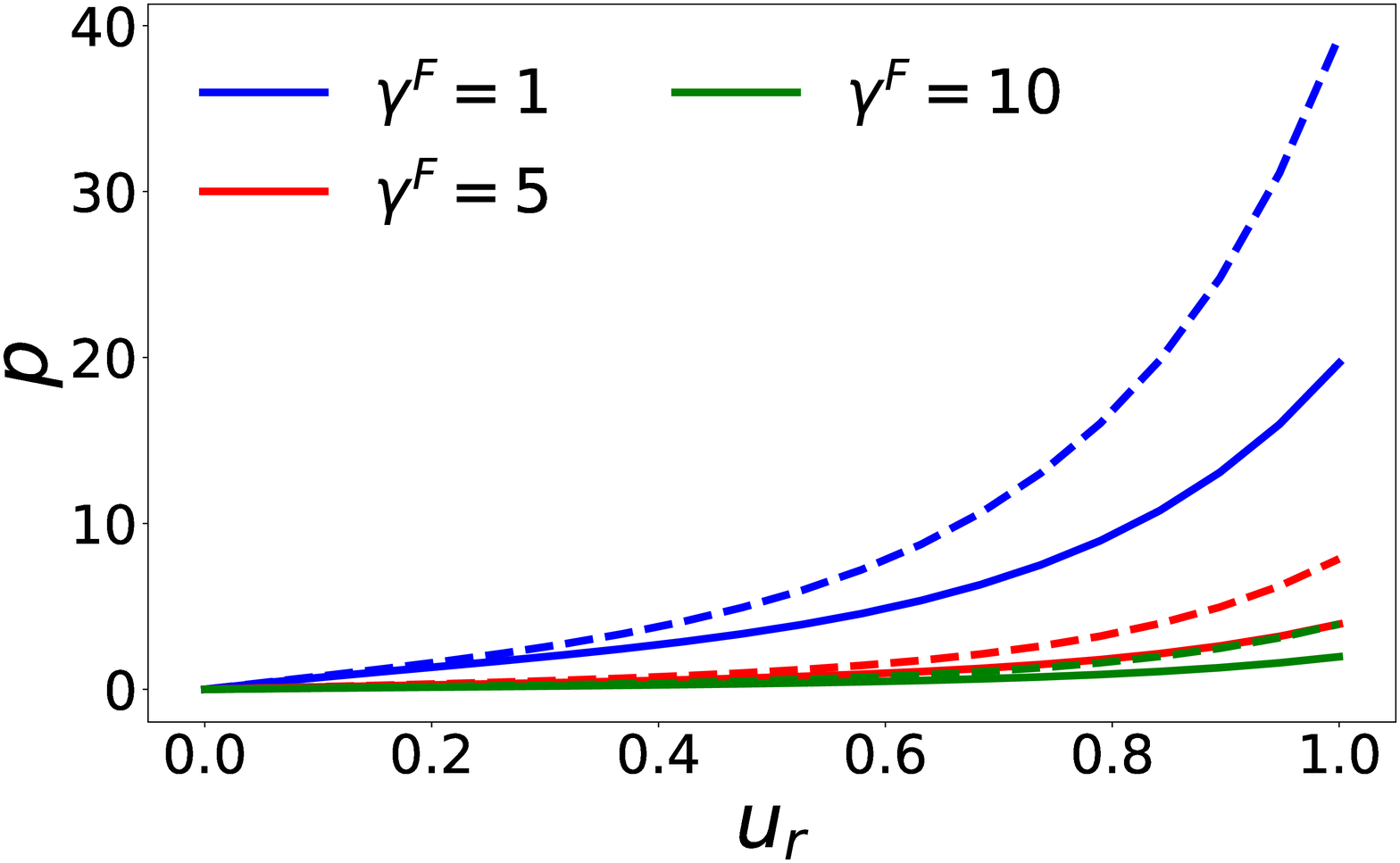}}
\subfloat[]{\includegraphics[width =0.4\linewidth]{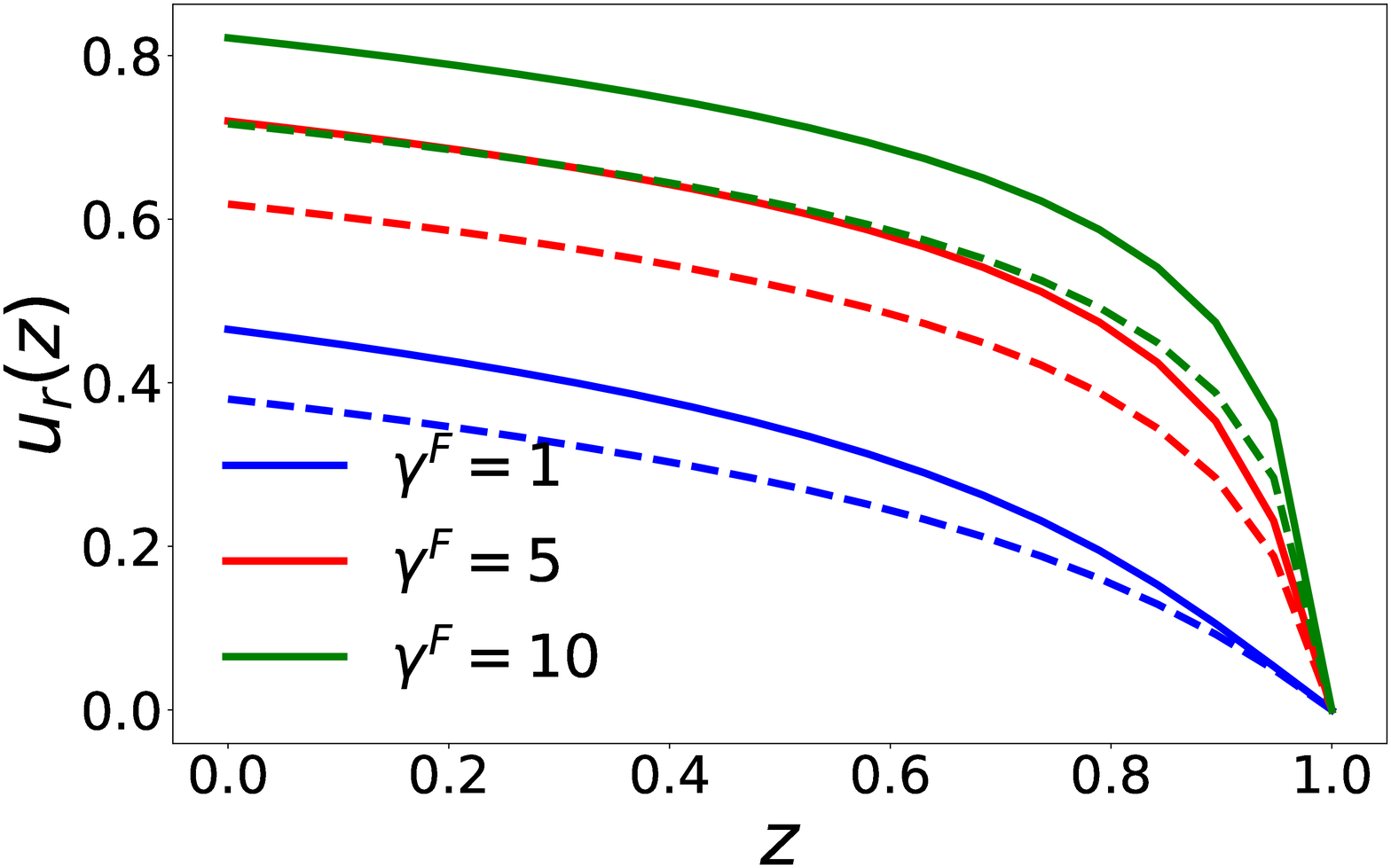}} \\
\subfloat[]{\includegraphics[width =0.4\linewidth]{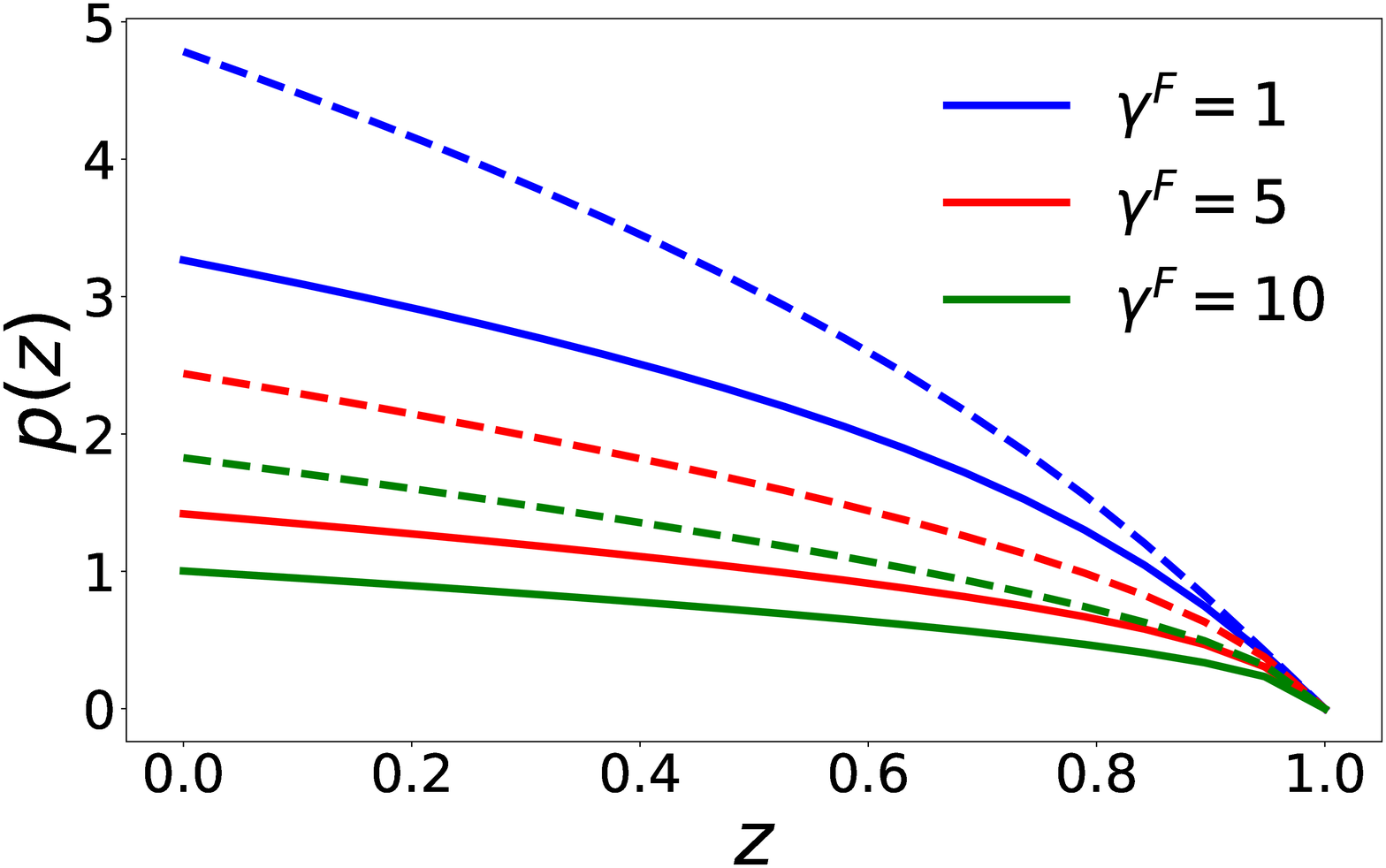}} 
\subfloat[]{\includegraphics[width =0.4\linewidth]{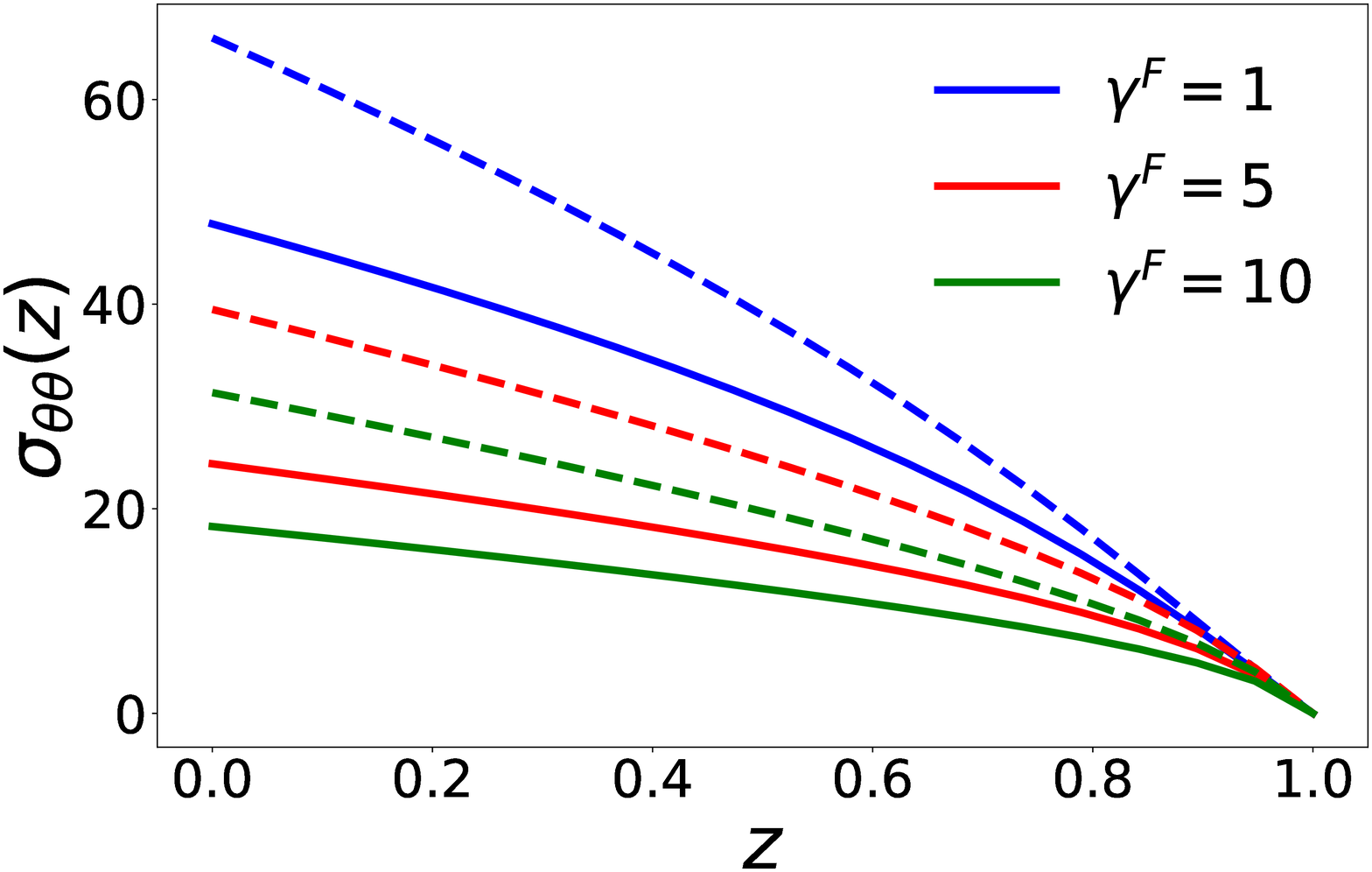}} 
\caption{(a) Pressure deflection curves (b) deformation profile, (b) pressure profile (c) circumferential structural stress profile for a hyperelastic tube constituted of Fung material conveying Newtonian flow at steady state. Solid curves account for large strains and deformations (both material and geometrical nonlinearity) while dashed curves account for small strains and deformations (only material nonlinearity). $\alpha =1$ for all the curves}
\label{fig:Fung_Small}
\end{figure}

\begin{figure}[t]
\centering
\subfloat[]{\includegraphics[width =0.4\linewidth]{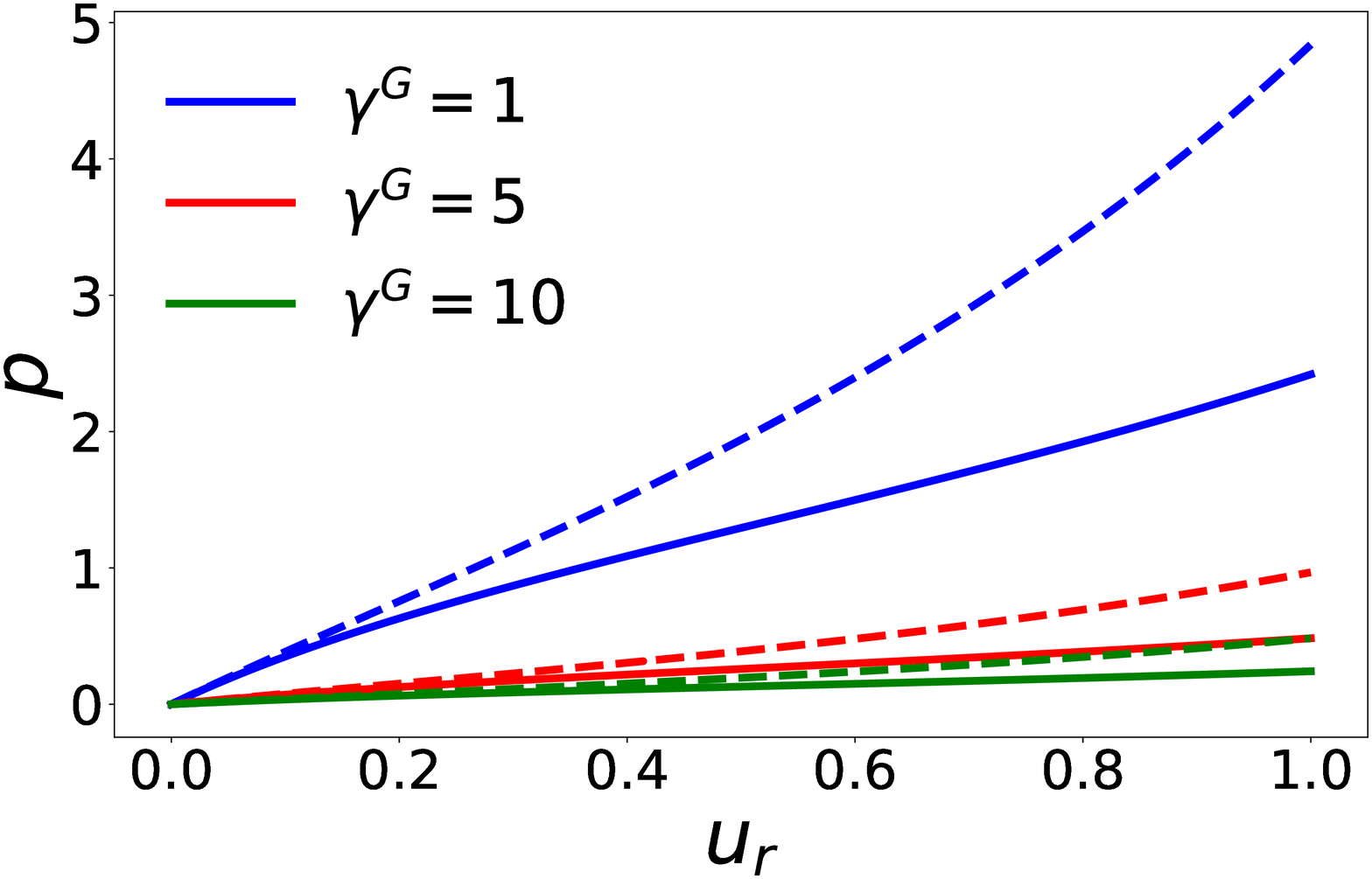}}
\subfloat[]{\includegraphics[width =0.4\linewidth]{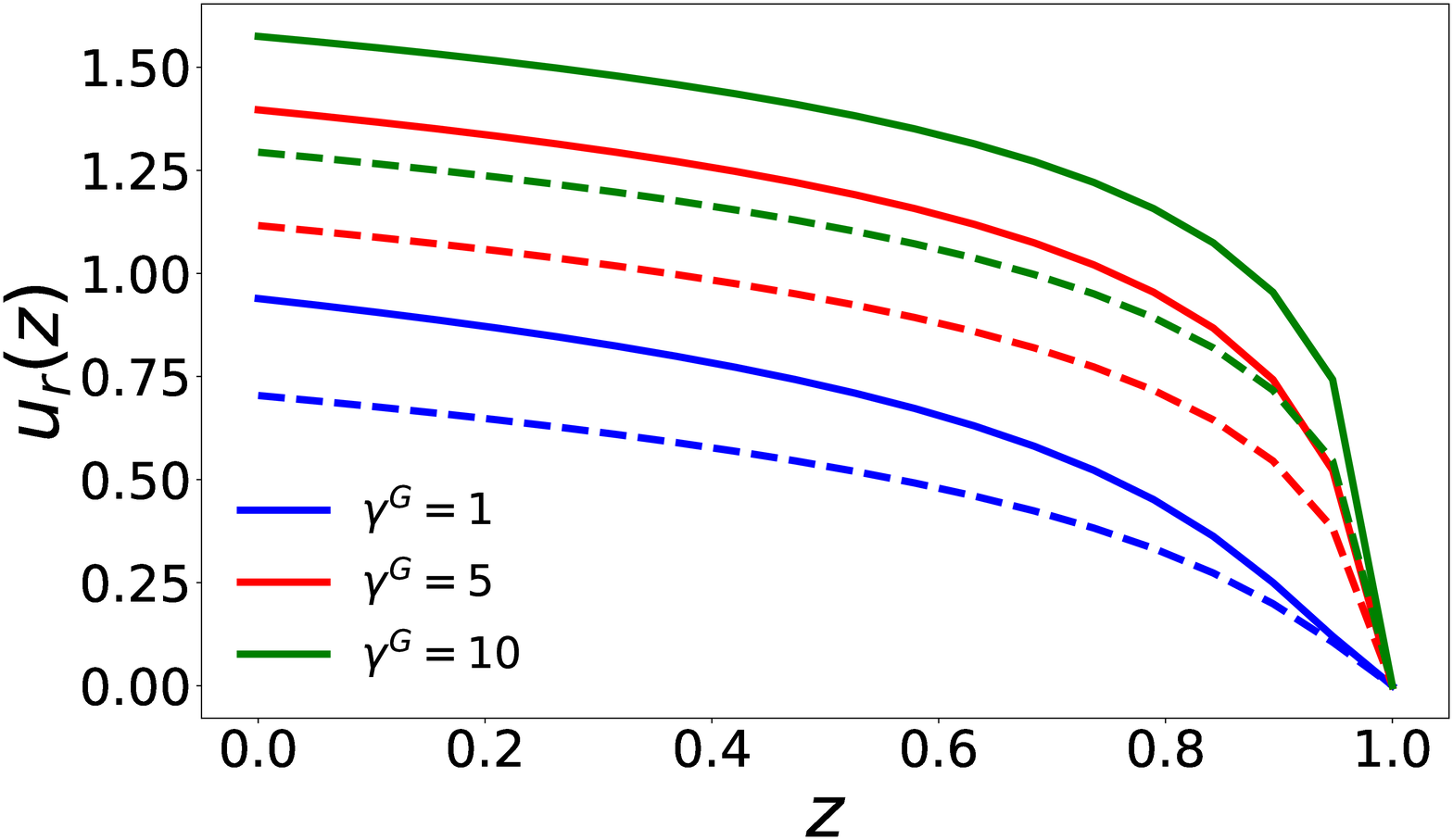}} \\
\subfloat[]{\includegraphics[width =0.4\linewidth]{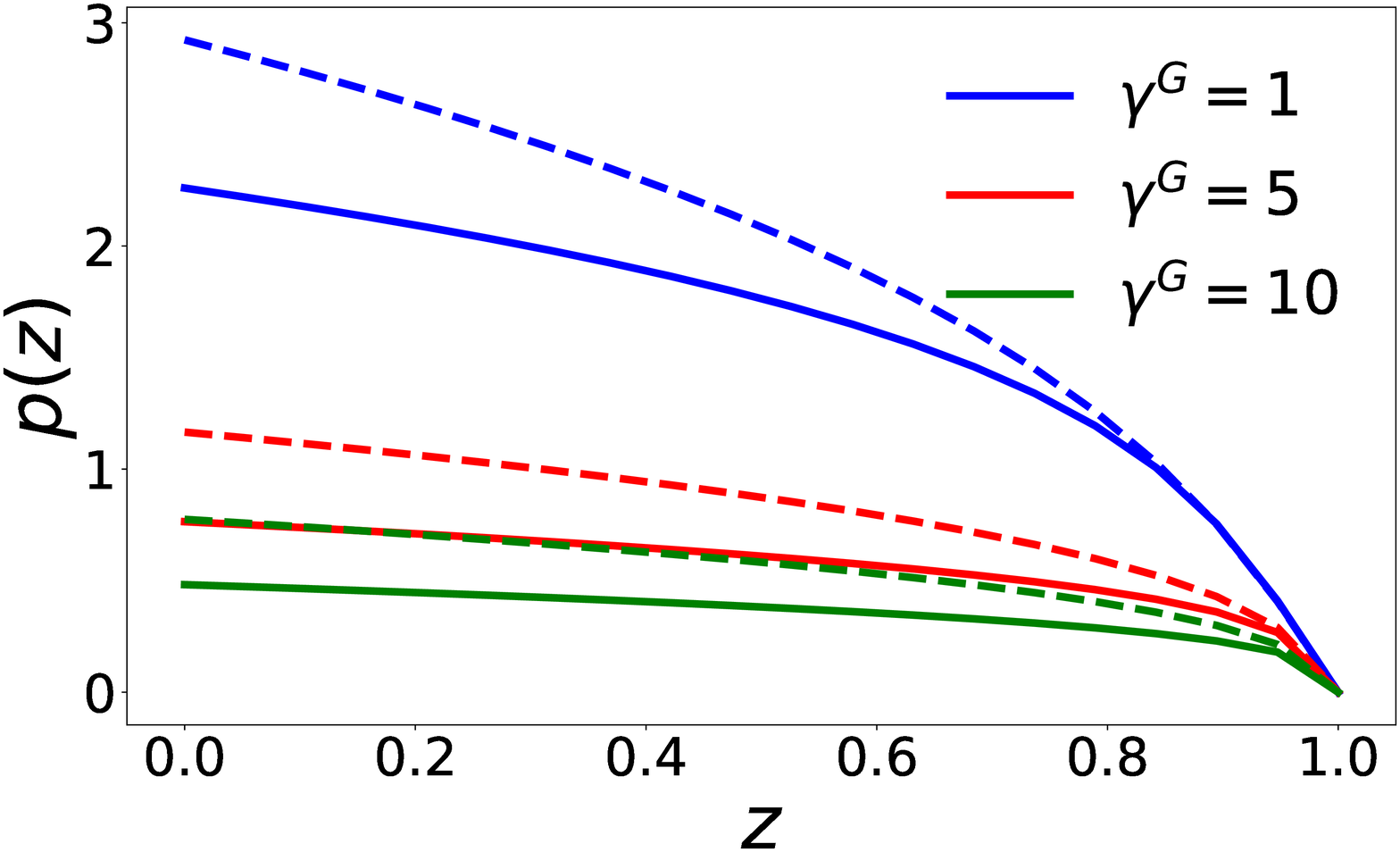}} 
\subfloat[]{\includegraphics[width =0.4\linewidth]{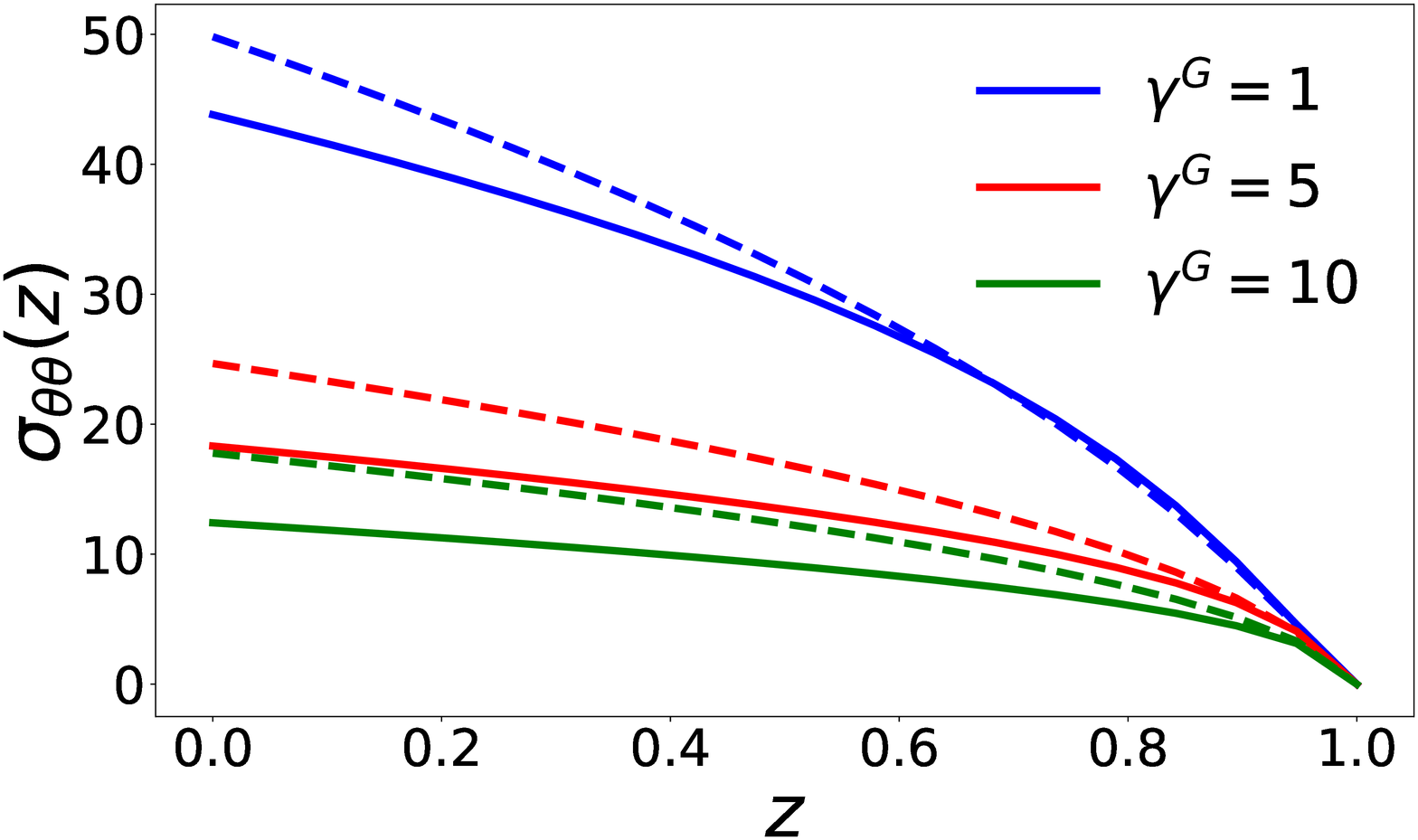}} 
\caption{(a) Pressure deflection curves (b) deformation profile, (b) pressure profile (c) circumferential structural stress profile for a hyperelastic tube constituted of Gent material conveying Newtonian flow at steady state. Solid curves account for large strains and deformations (both material and geometrical nonlinearity) while dashed curves account for small strains and deformations (only material nonlinearity). $\eta =0.1$ for all the curves.}
\label{fig:Gent_Small}
\end{figure}

\begin{figure}[t]
\centering
\subfloat[]{\includegraphics[width =0.4\linewidth]{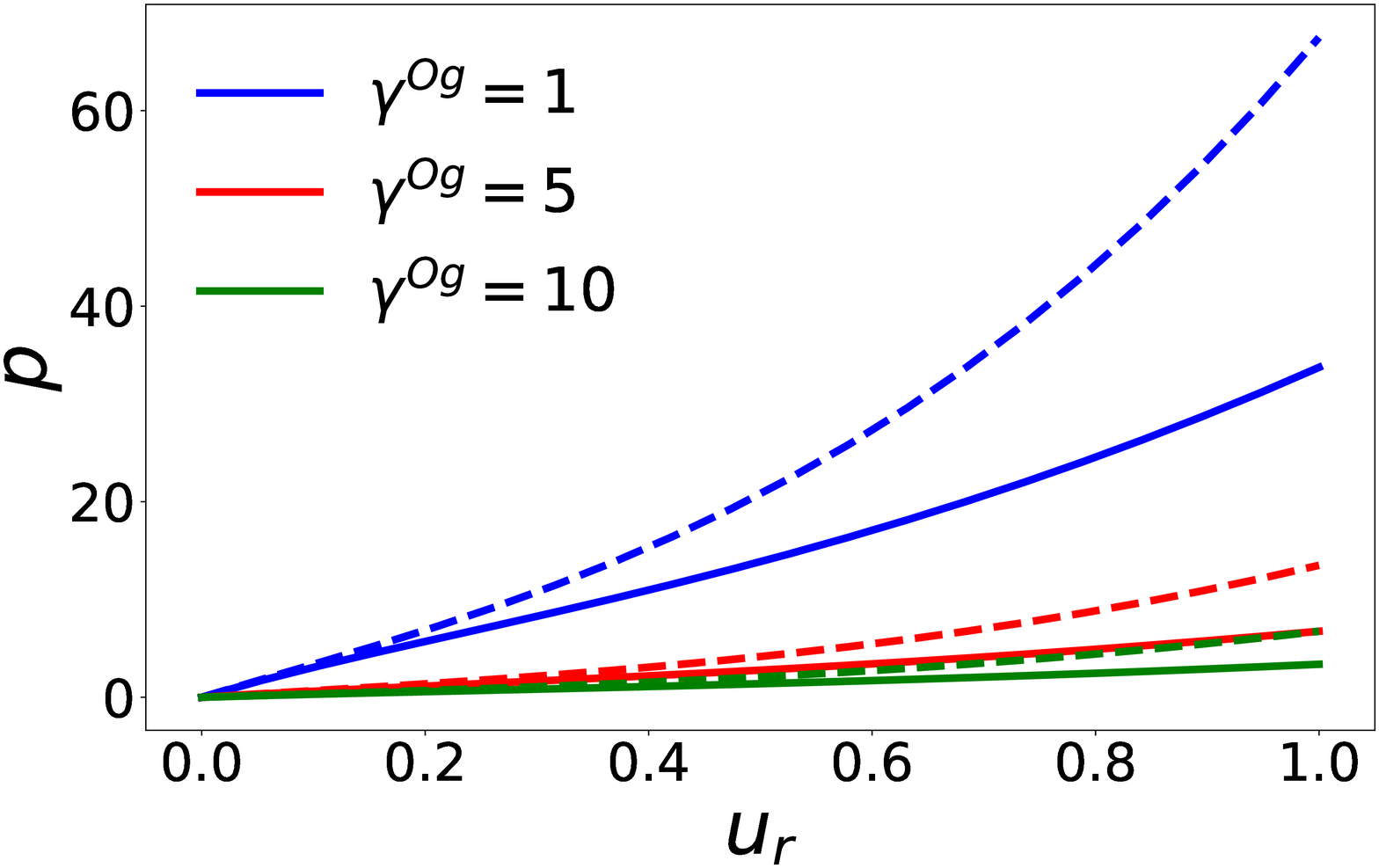}}
\subfloat[]{\includegraphics[width =0.4\linewidth]{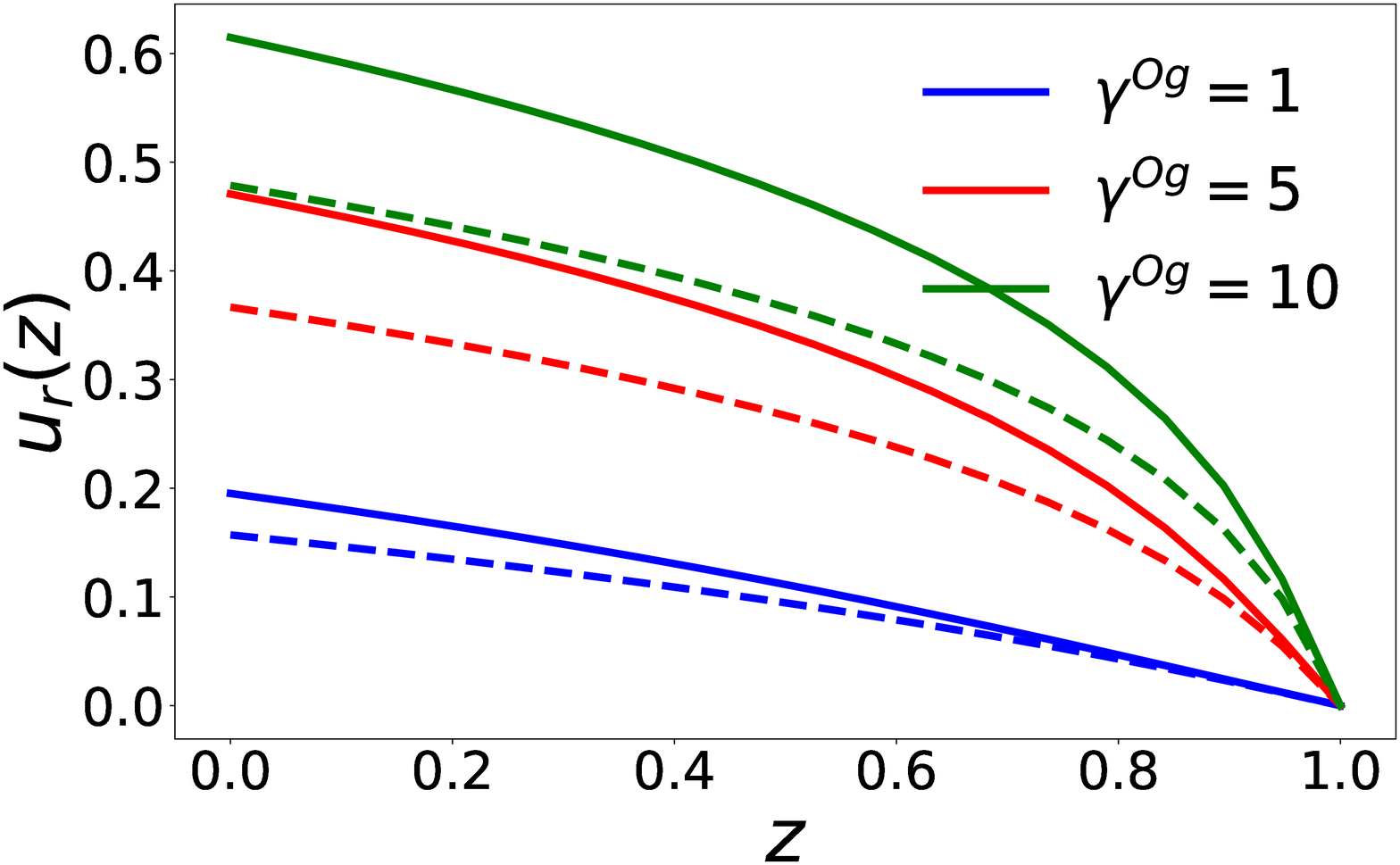}} \\
\subfloat[]{\includegraphics[width =0.4\linewidth]{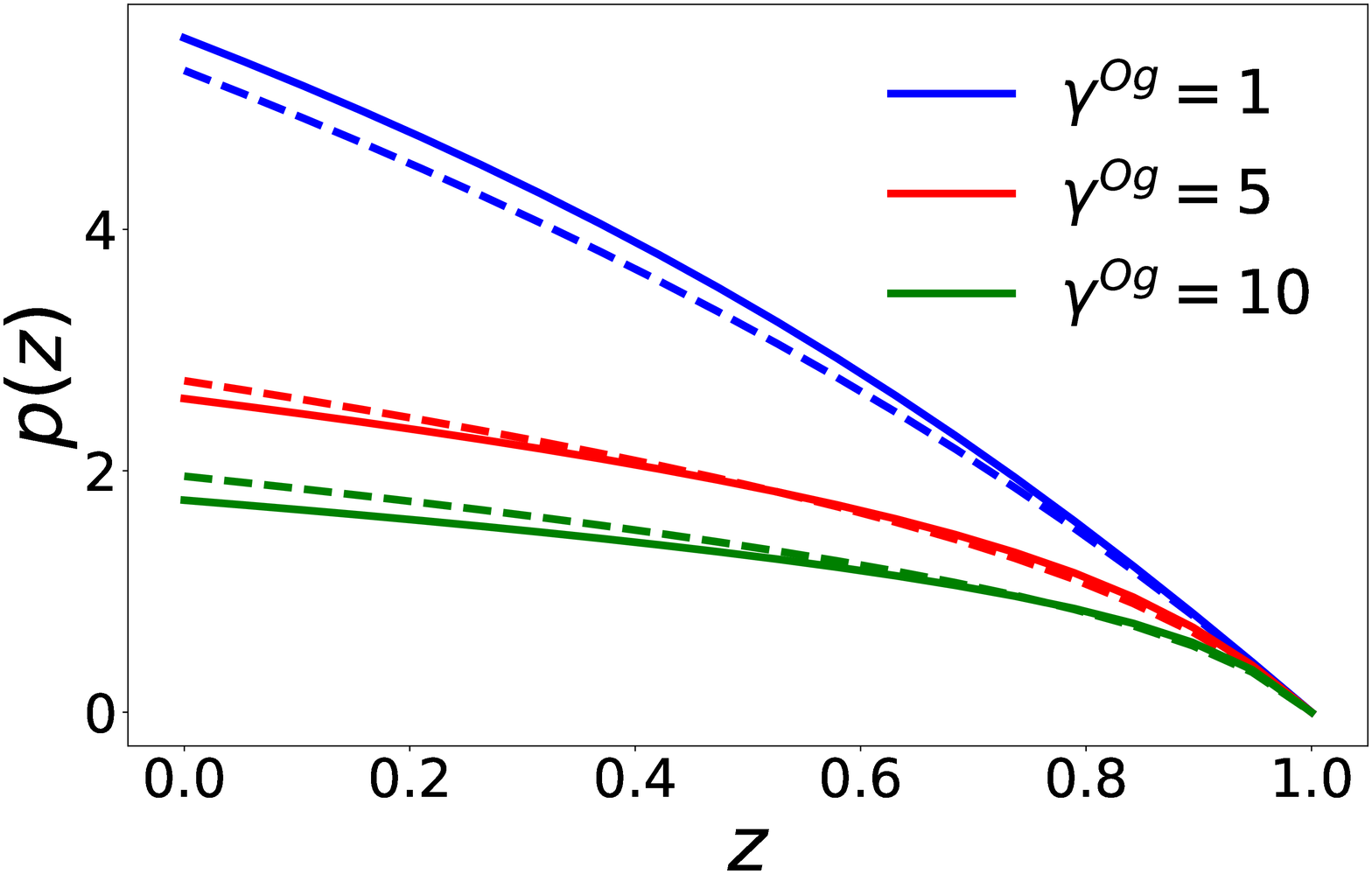}} 
\subfloat[]{\includegraphics[width =0.4\linewidth]{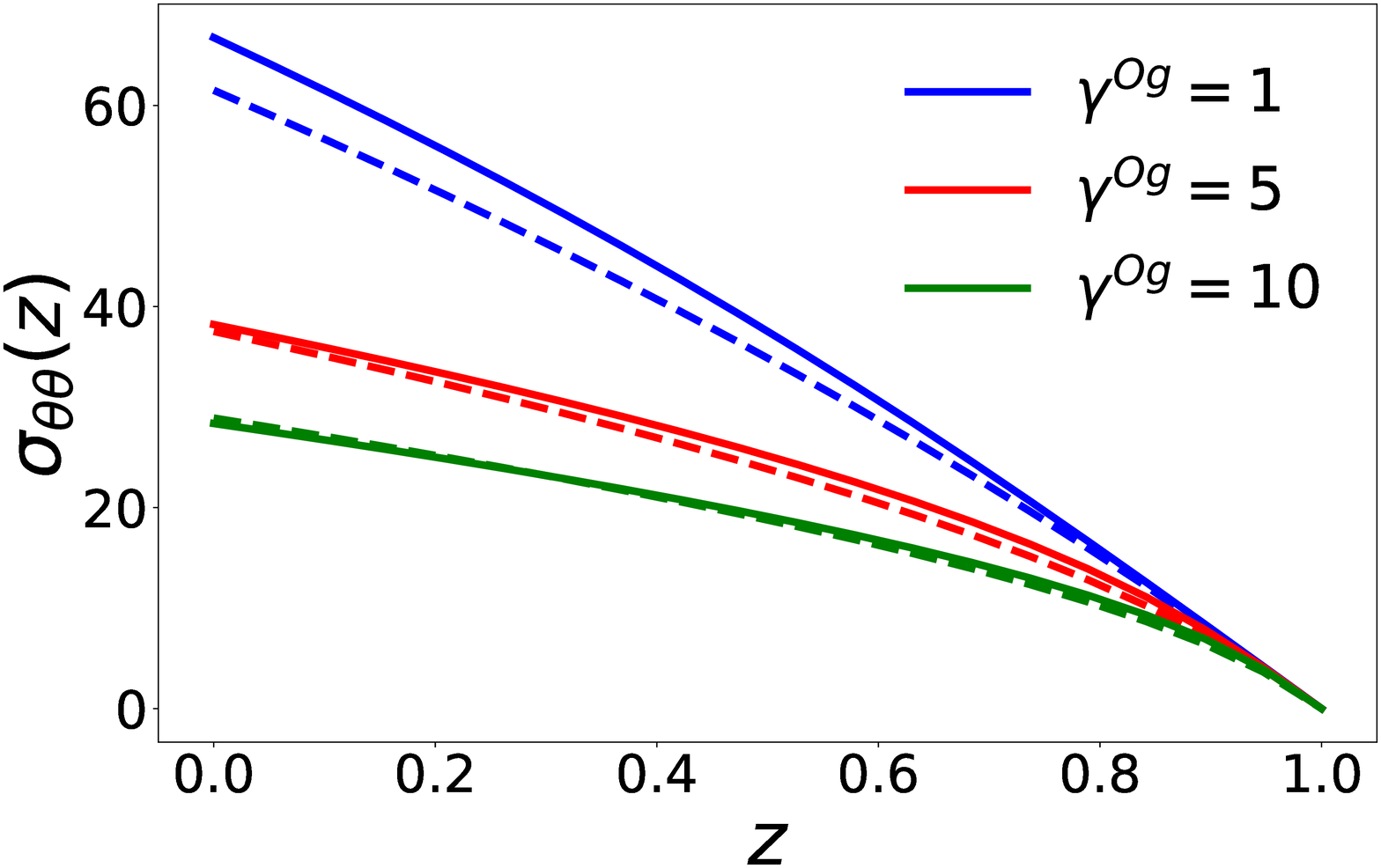}} 
\caption{(a) Pressure deflection curves (b) deformation profile, (b) pressure profile (c) circumferential structural stress profile for a hyperelastic tube constituted of Gent material conveying Newtonian flow at steady state. Solid curves account for large strains and deformations (both material and geometrical nonlinearity) while dashed curves account for small strains and deformations (only material nonlinearity).  $C_1= -3543,C_2 =-2723,C_3 =654,m_1 =1,m_2 =-1,m_3 =2$ for all the curves}
\label{fig:Ogden_Small}
\end{figure}

The plots corresponding to sensitivity analysis of the neo Hookean, Fung, Gent and $\text{Ogden}_{3}$ models are shown in Figs.~\ref{fig:neoHookean_Small},\ref{fig:Fung_Small},\ref{fig:Gent_Small},\ref{fig:Ogden_Small} respectively. Both inflation problem and FSI problem have been subjected to this sensitivity analysis; part (a) of the corresponding figures shows the pressure deflection curve for the inflation problem, part (b),(c) and (d) portray the deflection, pressure and hoop stress response respectively.

The underlying trend seen in the plots is that the neglection of geometrical nonlinearity  underpredicts the deflection and overpredicts the pressure for all models in both inflation and FSI problem; with the possible exception of FSI problem of $\text{Ogden}_{3}$. Clearly, geometrical linearity renders the tube softer and approximations induced by neglecting geometrical nonlinearity are far from inconsequential.

Geometrical nonlinearity, or lack thereof, has a stronger effect on the inflation problem than on the FSI problem. Comparison of Fig.\ref{fig:neoHookean_Small}(a) with \ref{fig:neoHookean_Small}(c) drives home the message that the pressure profile in the inflation problem deviates much more strongly than that in the FSI problem, as a response to geometrical linearity. The same trend is also replicated in Fung's model (Fig.~\ref{fig:Fung_Small}(a) and (c)), Gent's model (Fig.~\ref{fig:Gent_Small}(a) and \ref{fig:Gent_Small}(c)) and $\text{Ogden}_{3}$' model (Fig.~\ref{fig:Ogden_Small}(a) and (c)).

Across the different hyperelastic models, the role of geometrical nonlinearity in the inflation problem diminishes in proportion to strain hardening response of the tube; Fung, Gent, $\text{Ogden}_{3}$ (in that order) exhibit weaker influence of geometrical linearity than the neo Hookean model.

\section{Summary}
\label{sec:summ}
In this paper, we have analysed the inflation problem  and the FSI problem of a hyperelastic tube conveying Newtonian flow at steady state. Five ($5$) different hyperelastic models have been explored namely, neo Hookean, Mooney-Rivlin, Gent's, Fung's and $\text{Ogden}_{3}$. The local pressure and deformation relationship, pressure profile across the tube, deformation profile across the tube have been derived for each of the models and cataloged in Table \ref{tb:1} . The main scientific message wrung out from our analysis is as follows:
\begin{enumerate}
    \item The FSI characteristics of neo Hookean and Mooney Rivlin tubes are quite  the converse of Fung's and Gent's tubes. For neo Hookean and Mooney Rivlin models,  the pressure profiles  deviates only slightly across the tube from its maximum value at the inlet; the deformation profile registers a steep decline across the tube. . Conversely, Fung and Gent's tubes have their deformation profile practically invariant across the length of the tube, while the pressure profile descends sharply.

     The above observations are explained in the light of the fact that both Mooney Rivlin and neo Hookean models are used for modeling elastomers (rubber like materials) and exhibit the limit point instability commonly documented to be found in latex tubes \cite{Holzapfel2005SimilaritiesMaterials,Destrade_NeoHookean_Instability,Osborne1909}. Therefore, neo Hookean and Mooney Rivlin tubes exhibit \textit{strain softening} - wherein the strain continues to increase at near constant loads. On the the other hand, Fung's and Gent's models display strong \textit{strain hardening}, which makes it difficult to strain them further at progressively higher strains, and explains their observed FSI characteristics and renders them appropriate for modeling biological tissues with high collagen content.
    \item The strain hardening characteristics of Fung's and Gent's model are intimately coupled with their respective material parameters $\alpha$ and $\eta$ directly. However, for small deformation (when the tube is stiff, or when the flow rate is small, or near the clamped edge), the impact of the material parameters $\alpha, \eta$ on the FSI response diminishes to the point that the response becomes independent of the strain hardening parameter. In this regime, the Fung's and Gent's model behave like neo Hookean model.
    \item Tube laws, which relate the local pressure with the deformed area of the tube have been derived for all the five ($5$) hyperelastic models and cataloged in Table \ref{tb:1}. These tube laws act as constitutive equations for tube, and are independent of the nature of the flow being conveyed whether it is Newtonian, complex, compressible, incompressible, steady or otherwise.
\end{enumerate}
We also carried out a sensitivity analysis where the role of geometric nonlinearity was decoupled and a FSI analysis with only material nonlinearity was performed. The results show that geometric nonlinearity is indeed important in our problem and should be accounted for. This conclusion is in contrast to some other cases documented in literature \cite{Breslavsky_plate1} where the geometric nonlinearity did not influence the hyperelastic response of the structure appreciably. Geometric nonlinearity has a stronger influence on the inflation problem than it does on the FSI problem. 

This analysis has opened up new vistas for thought provoking research in the future. We may want to understand how thickness effects may be incorporated in this analysis in future \cite{AC18b}, which will significantly affect the structural mechanics of the system whilst enhancing the scope of application of our research. Transient effects and non Newtonian nature of the fluid are also worthy of being accounted for.

\appendix
\section{Appendix}
Pressure deflection curve for Gent's model for higher deformation is shown below in Fig.~\ref{fig:TubeLaw2_Gent}.
\begin{figure}[t]
\centering
\subfloat[]{\includegraphics[width =0.5\linewidth]{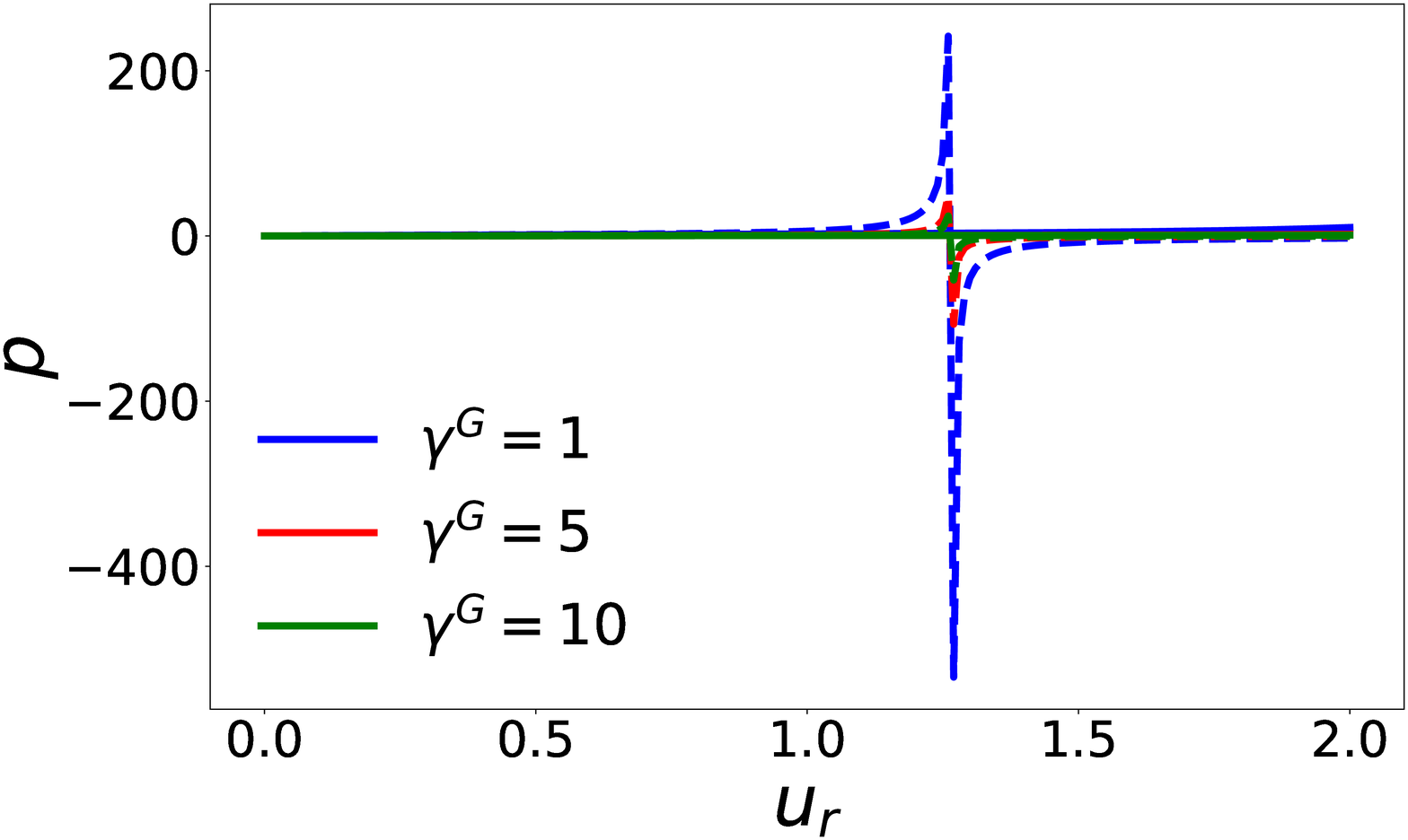}}\\
\caption{Pressure deflection curve for Gent's model at higher deformation}
\label{fig:TubeLaw2_Gent}
\end{figure}

\bibliographystyle{jfm.bst}
\bibliography{references}

\end{document}